\UseRawInputEncoding
\documentclass[12pt,a4paper]{article}
\pdfoutput=1
\usepackage{graphicx,epsfig}
\usepackage{dcolumn} 
\usepackage{slashed,color,amsmath,amssymb}
\usepackage{jheppub}
\usepackage{enumitem}

\usepackage{caption, subcaption}
\usepackage{multirow}
\usepackage{bigcenter}
\usepackage{verbatim}
\newcommand{\be}{\begin{equation}}
\newcommand{\ee}{\end{equation}}
\newcommand{\bea}{\begin{eqnarray}}
\newcommand{\eea}{\end{eqnarray}}
\newcommand{\bit}{\begin{itemize}}
\newcommand{\eit}{\end{itemize}}

\def\gsim{\lower0.5ex\hbox{$\:\buildrel >\over\sim\:$}}
\def\lsim{\lower0.5ex\hbox{$\:\buildrel <\over\sim\:$}}

\def\met{\rm E{\!\!\!/}_T}

\hypersetup{
   colorlinks=true,       
   linkcolor=blue,        
   citecolor=red,         
   filecolor=magenta,     
}

\preprint{IPPP/19/94, LAPTH-051/19}

\title{Determining the lifetime of long-lived particles at the HL-LHC}

\author{Shankha Banerjee$^{1}$, Biplob Bhattacherjee$^{2}$,  Andreas Goudelis$^{3}$,  Bj\"orn Herrmann$^{4}$, Dipan Sengupta$^{5}$, Rhitaja Sengupta$^{2}$}
\affiliation{\vspace*{0.1in}$^1$ Institute for Particle Physics Phenomenology, Department of Physics, Durham University, Durham DH1 3LE, United Kingdom}
\affiliation{\vspace*{0.1in}$^2$ Centre for High Energy Physics, Indian Institute of Science, Bangalore 560012, India}
\affiliation{\vspace*{0.1in}$^3$ Laboratoire de Physique de Clermont (UMR 6533), CNRS/IN2P3, Univ.\ Clermont Auvergne, 4 Av.\ Blaise Pascal, F-63178 Aubi\`ere Cedex, France}
\affiliation{\vspace*{0.1in}$^4$ Univ.\ Grenoble Alpes, Univ.\ Savoie Mont Blanc, CNRS, LAPTh, F-74000 Annecy, France}
\affiliation{\vspace*{0.1in}$^5$ Department of Physics and Astronomy, University of California San Diego, 9500 Gilman Drive, La Jolla, California, USA}

\emailAdd{shankha.banerjee@durham.ac.uk}
\emailAdd{biplob@iisc.ac.in}
\emailAdd{andreas.goudelis@clermont.in2p3.fr}
\emailAdd{herrmann@lapth.cnrs.fr}
\emailAdd{disengupta@physics.ucsd.edu}
\emailAdd{rhitaja@iisc.ac.in}

\abstract{
We examine the capacity of the Large Hadron Collider to determine the mean proper lifetime of long-lived particles assuming different decay final states. We mostly concentrate on the high luminosity runs of the LHC, and therefore, develop our discussion in light of the high amount of pile-up and the various upgrades for the HL-LHC runs.
We employ model-dependent and model-independent methods in order to reconstruct the proper lifetime of neutral long-lived particles decaying into displaced leptons, potentially accompanied by missing energy, as well as charged long-lived particles decaying ihnto leptons and missing energy. We also present a discussion for lifetime estimation of neutral long-lived particles decaying into displaced jets, along with the challenges in the high PU environment of HL-LHC. After a general discussion, we illustrate and discuss these methods using several new physics models. We conclude that the lifetime can indeed be reconstructed in many concrete cases. Finally, we discuss to which extent including timing information, which is an important addition in the Phase-II upgrade of CMS, can improve such an analysis.
} 

\begin{document}

\maketitle


\section{Introduction}
\label{sec:intro}

The lack of observation of new physics at the Large Hadron Collider (LHC) has prompted a re-evaluation of the strategies aiming to probe signals of physics beyond the standard model (BSM). Initial expectations had been that new physics would reveal itself in prompt searches involving leptons, jets, and missing energy, or, eventually, in the form of exotic resonances. However, no smoking gun signal has appeared in such searches so far. It is, therefore, only reasonable to entertain the possibility that new physics may manifest itself in unexpected ways, in the form of non-standard signatures. Although such a {\it terra incognita} can be daunting to explore, we can appeal to well-motivated theoretical scenarios for guidance. An attractive possibility is that some of the produced particles are long-lived, {\it i.e.}\ that the secondary vertices through which they decay are macroscopically displaced with respect to the primary interaction point at which they are produced. Such signatures appear in a large variety of new physics frameworks such as Supersymmetry \cite{Giudice:1998bp, Giudice:2004tc, Burdman:2006tz, Biswas:2009zp, Meade:2010ji, Biswas:2010yp, Ibe:2012hu, Graham:2012th, Bhattacherjee:2012ed, Arvanitaki:2012ps, Cerdeno:2013oya, Heisig:2013rya, Rolbiecki:2015gsa, Dercks:2018eua, Fukuda:2019kbp}, Twin Higgs models \cite{Chacko:2005pe}, gauge unification frameworks based on vector-like fermions \cite{Kowalska:2019qxm}, or Hidden Valley models \cite{Strassler:2006im, Strassler:2006ri, Strassler:2006qa}, as well as in frameworks including dark matter \cite{Hall:2009bx, TuckerSmith:2001hy, Co:2015pka, Hessler:2016kwm, Garny:2017rxs, Belanger:2018sti, Goudelis:2018xqi, Garny:2018ali, No:2019gvl} or baryogenesis \cite{Cui:2014twa}.

Searches for long-lived particles (LLPs) have already been pursued at previous experiments like CDF and D{\Large\o{}}, and are being pursued at ATLAS, CMS and LHCb in LHC, see, {\it e.g.}, Refs.\ \cite{Abulencia:2007ut, Gruenendahl:2008ew, Aaltonen:2009kea, Aad:2014gfa, Aaboud:2017mpt, Aaboud:2017iio, Aad:2019tua, Aaboud:2019opc, Aad:2019tcc, Khachatryan:2016sfv, Sirunyan:2017jdo, Sirunyan:2017sbs, Sirunyan:2019gut, Sirunyan:2019wau,Ilten:2015hya,Aaij:2015ica,Aaij:2016qsm,Aaij:2016xmb,Aaij:2016isa,Aaij:2017mic,Aaij:2019bvg} and will be one of the primary focus of new physics searches at the LHC in the coming years. For an overview of recent LLP searches, we refer the reader to Ref.\ \cite{Alimena:2019zri} and references therein. In addition to the multi-purpose experiments ATLAS and CMS, dedicated detectors like FASER~\cite{Ariga:2019ufm} and MATHUSLA \cite{Lubatti:2019vkf} have been proposed to probe long-lived particles \cite{Feng:2017uoz}. The range of new physics scenarios that such detectors can explore is both vast and very well-motivated \cite{Chou:2016lxi, Curtin:2018mvb}. Moreover, these proposals aim at filling in a ``lifetime gap'' between prompt collider searches and cosmological constraints such as Big Bang Nucleosynthesis (BBN), which is typically sensitive to lifetimes of the order of $0.1$ s or longer \cite{Kawasaki:2017bqm}.





In this paper we consider LLPs as states with a proper lifetime long enough such that they decay only after traversing some macroscopic distance (order of few cm) within the detector.
Such lifetimes can be induced either by rather small couplings or in specific kinematic configurations involving small mass splittings between the particles participating in the process or large propagator masses. Regardless of the underlying physics, LLPs introduce significant additional complications for experimental searches as compared to promptly produced particles. If the LLP is heavy and charged, it will leave a distinct track in the detector, making detection easier, while neutral LLPs are more difficult to detect.

We place ourselves in the hopeful scenario that Long-Lived Particles (LLPs) will be observed at the High Luminosity runs of the Large Hadron Collider (HL-LHC) and we examine its capacity to reconstruct the LLP lifetime.
Extracting the lifetime information can not only provide crucial information in order to, at least partly, reconstruct features of the underlying microscopic model, notably the coupling strength between the LLP and its decay products, but may also lead to the establishment of more unexpected connections between LHC observations and, \textit{e.g.}, cosmology, for an example see Ref.\ \cite{Belanger:2018sti}. Motivated by our simple analysis presented in Ref. \cite{Brooijmans:2018xbu}, we explore several ways to estimate the lifetimes of different kinds of LLPs with multifarious decay modes.

Several studies have appeared in the literature concerning different ways to estimate the lifetime of LLPs. Our analysis builds upon and expands this body of work in a number of ways:
\begin{itemize}

\item Many of the existing analyses have focused on the potential for lifetime determination in searches for charged LLPs \cite{Ambrosanio:2000ik,Ambrosanio:2000zu,Kawagoe:2003jv,Asai:2008sk,Ishiwata:2008tp,Kaneko:2008re}. The advantage of charged LLPs is that their boost and mass can be inferred from their tracks in the detector, while in the case of Charged Massive Stopped Particles (CHAMPs) different methods can be employed which have been explored, \textit{e.g.}, in \cite{Hamaguchi:2004df, Feng:2004yi, Arvanitaki:2005nq, DeRoeck:2005cur, Hamaguchi:2006vu, Asai:2009ka, Pinfold:2010aq}. The -- arguably, more challenging -- case of neutral LLPs, in which one has to reconstruct their decay products in order to estimate their lifetime has, on the other hand, received less attention. In this work, we attempt to fill in this void.

\item Many of the existing analyses have considered several thousands of LLP events in order to assess whether their lifetime can be reconstructed. Given the recent null results from numerous LHC searches, the current bounds only allow between a few hundreds to a few thousands of events. In this paper, we choose to work with cross-section \footnote{We mean the cross section of the full process, i.e., production of LLPs and their subsequent
decay into the observed final state, which is the production cross section of the LLPs times the branching fraction into the decay mode
analysed.} values that conform with these latest limits. At the same time, we place ourselves in the framework of the High-Luminosity LHC (HL-LHC), which is expected to accumulate a total luminosity of 3000 fb$^{-1}$, thus increasing the possibility of discovering and measuring the properties of long-lived particles even for more moderate cross-section values.

\item On the other hand, even though this luminosity increase will be extremely beneficial for LLP searches, with the accumulation of more data the pile-up (PU) events are going to negatively impact the prospects for observing LLPs and measuring their properties. Indeed, $\sim 140$ PU events per bunch crossing are expected in the HL-LHC \cite{Schmidt:2016jra}, compared to only 30-50 in the previous runs. However, the ATLAS and CMS detectors will also undergo major upgrades - both in the hardware and in the software fronts: increase in the rapidity range of the calorimeters, improved timing information, addition of timing layers which will help experimentalists to better understand and subtract pile-up events are but a few such examples. Although we are fully aware of the fact that, at the end of the day, these issues can only be analysed in full by the experimental collaborations themselves, we attempt to provide at least some preliminary discussion on how such factors may affect our capacity to reconstruct the lifetime of LLPs.

\item Most existing analyses have employed $\chi^2$ methods in order to reconstruct the LLP lifetime. In this work, we discuss how the inclusion of likelihoods and some machine learning algorithms fare against such canonical treatments.

\end{itemize}

In what follows we present ways to estimate the lifetime of LLPs considering various decay channels into different final states. We start with the simplest case of displaced leptons, continue with displaced jets and, eventually, study LLP decays involving invisible particles. Most of the analyses presented here can be applied to numerous LLP models involving such final states and also to several future colliders. Finally, we also study the prospect of the proposed MIP Timing Detector (MTD) \cite{MTD}, which will be included in the Phase-II upgrade of the CMS detector and will play a key role in improving the determination of the LLP lifetime during the HL-LHC runs. We should point out that at various stages we will allow for some leeway, in terms of the known quantities available to us, as well as potentially speculate on some uncertainties that will only be precisely estimated once the LHC resumes operation after the long shutdown 2 (LS2).
 
The paper is organised as follows: In Sec.\ \ref{sec:llp_lifetime}, we start by recalling basic formulae related to the LLP lifetime and discuss why we restrict ourselves to LLP decays within the tracker. In Sec.\ \ref{sec:diff_decay}, we discuss existing bounds on processes involving LLPs, we illustrate why the LLP lifetime cannot be reconstructed through a naive exponential fit, we propose alternative approaches and we study how these can be used to estimate the lifetime for different LLP decay modes. Sec.\ \ref{sec:timing} is dedicated to a discussion of the MTD and how adding timing information can improve the situation, not only concerning the lifetime estimation but also in order to identify the model by reconstructing the mass of the LLP in cases where the LLP decays involve invisible particles. Our conclusions are presented in Sec.\ \ref{sec:concl}.

\section{Long-lived particle lifetime reconstruction}
\label{sec:llp_lifetime}

We start this Section by reviewing some relations related to the lifetime of long-lived particles (LLPs).

\subsection{Kinematics of LLPs}
\label{ssec:lifetime}

In the laboratory frame, the decay length of a particle with mean proper decay time $\tau$ (as measured in its own rest frame) is given by
\begin{equation}
	d ~=~ \beta \gamma c \tau  \,,
	\label{eq:DL}
\end{equation}
where $\gamma = E/m = (1-\beta^2)^{-1/2}$ is the relativistic factor with $\beta = v/c = |\vec{p}|/E$, $v$ is the velocity of the decaying particle and $c$ denotes the speed of light. For simplicity, we will refer to the mean proper decay lifetime ($\tau$) or the mean proper decay length ($c\tau$) as just ``lifetime'' or ``decay length'', unless stated otherwise.
The decay probability of such particles follows the same distribution as the one encountered in radioactive decays. 
If we consider the production of a number $N_0$ of such unstable particles with mean proper lifetime $\tau$, the expected number of surviving particles $N(t)$ evolves as a function of time $t$ through the usual exponentially decreasing distribution
\begin{equation}
	N(t) ~=~ N_0 \, e^{-t/\tau} \,.
	\label{eq:tau_dist}
\end{equation}
By measuring the decay length $d_i$ of each event, together with the corresponding kinematical factor $\beta_i$, we can deduce the proper decay time associated to the event. Ideally, it is possible to infer the values of $N_0$ and $\tau$ by performing an exponential fit of the sample data, provided that enough statistics is available.  If the proper decay length is large, the number of LLP  decays  within the detector volume will be very small, and therefore we will require a large enough statistical sample to perform a faithful fit.

We  note that the geometrical acceptance probability for an LLP with a decay length $d$ as it traverses the detector is given by
\begin{equation}
    P_{\rm dec} ~=~ \frac{1}{4\pi}\int_{\Delta\Omega} {\rm d}\Omega~ \int_{L_{1}}^{L_{2}} {\rm d}L~ \frac{1}{d}e^{-L/d} \,, 
    \label{eq:Pdec}
\end{equation}
where $L_{1}$ and $L_{2}$ are the distances between the interaction point and the point where the LLP respectively enters and exits the decay volume, and  $\Delta\Omega$ is the geometric cross-section of the active detector volume \cite{Curtin:2018mvb}. Thus we clearly see that while the LHC can be sensitive to decays occurring within a certain displacement, the probability decreases if the displacement length is significant\footnote{Next generation dedicated LLP detectors like MATHUSLA\cite{Lubatti:2019vkf} and FASER\cite{Ariga:2019ufm} can therefore provide further coverage of the associated parameter space.}.


In the following, we will consider the production of a variety of long-lived particles, hereafter denoted by $X$, with different decay modes. The long-lived particle can be charged or neutral and therefore its identification efficiency depends on the tracker and the energy deposition in the detector, among various other factors.

\subsection{Restricting to decays within tracker}
\label{ssec:tracker}

In order to estimate the lifetime of the long-lived particle $X$, in the following we will restrict ourselves to decays occurring inside the tracker region of the detector. This not only allows access to the position of the secondary vertex (SV), but also helps in reconstructing the charged decay products of $X$ as well as in the measurement of the boost factor $\beta\gamma$. However, this restriction will limit the number of observed LLP decays, especially in the case of particles characterised by longer lifetimes leading to decays outside the tracker region. In this Section we quantify the fraction of decays we can expect within the tracker for given ranges of LLP masses and lifetimes.

\begin{figure}
    \centering
    \includegraphics[scale=0.09]{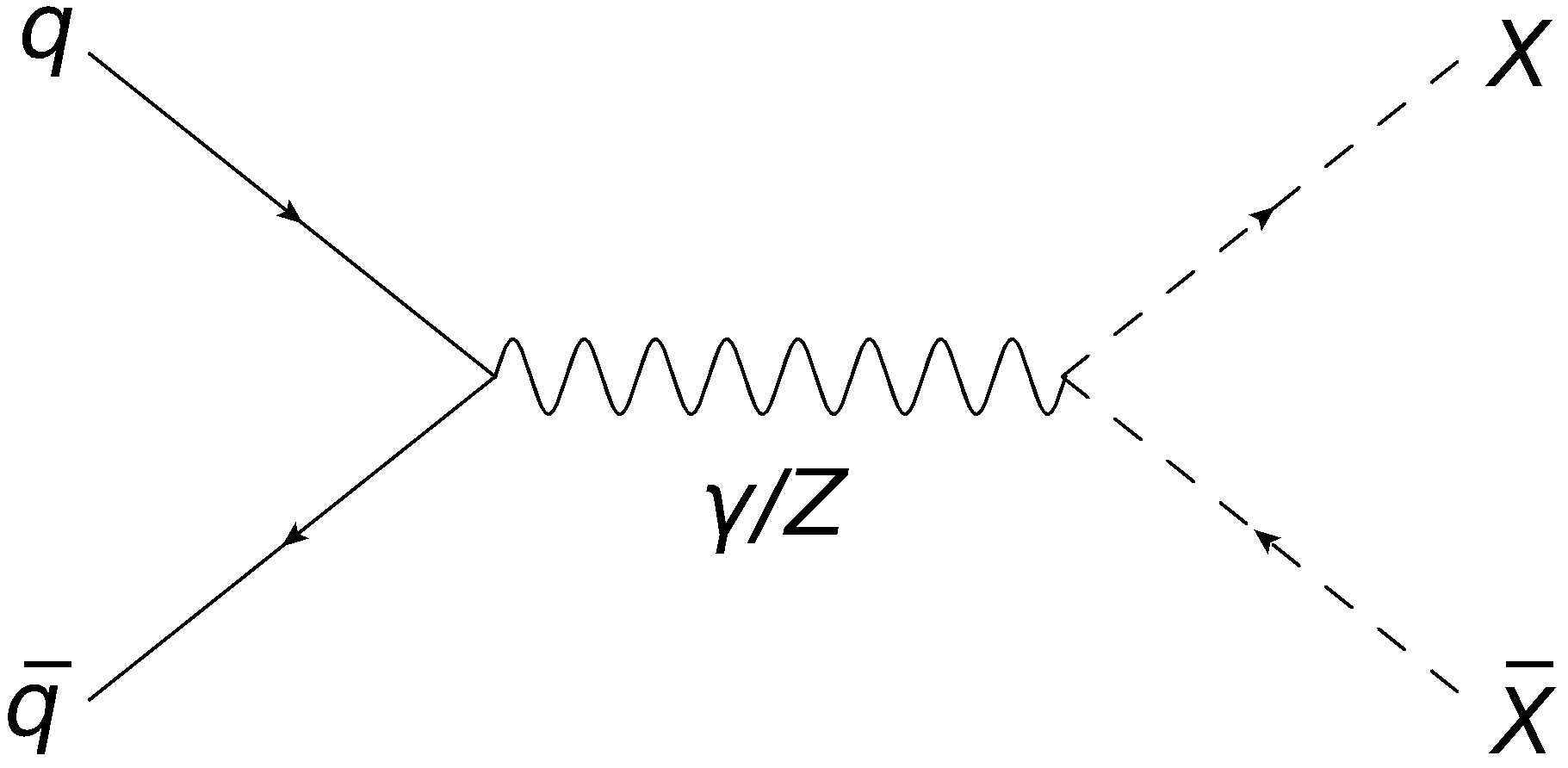}
    \caption{Quark-initiated $s$-channel pair production of scalar LLPs}
    \label{fig:feyn1}
\end{figure}

 Consider the production of a pair of long-lived particles at the LHC, $pp \to XX$, and their subsequent decays into Standard Model particles. Since the boost factor $\beta\gamma$ of the particles $X$ depends on their production mode, for illustration we focus on a supersymmetric (SUSY) model containing a $LLE$-type $R$-parity violating (RPV) coupling, for a review \textit{cf e.g.} \cite{Barbier:2004ez}. In this model a pair of sneutrino LLPs is produced through a quark-initiated $s$-channel process as in Figure \ref{fig:feyn1} and decays into two electron pairs, with a mean proper lifetime that is controlled by the LLP mass and the magnitude of the RPV coupling. Events have been simulated using \texttt{PYTHIA6} \cite{Sjostrand:2006za} at $\sqrt{s}=14$ TeV.

\begin{figure}
    \centering
    \includegraphics[width=0.49\textwidth]{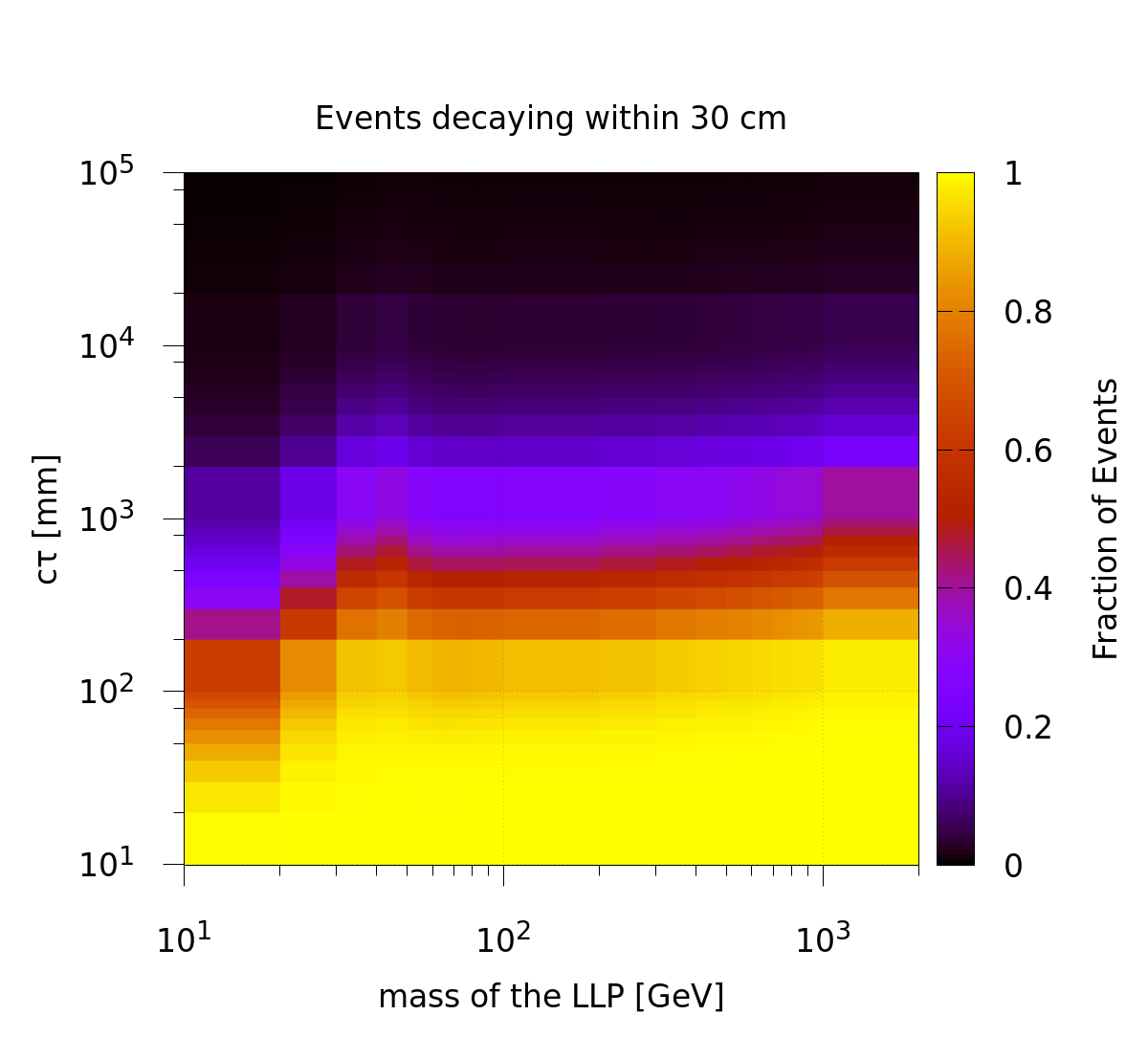}
    \includegraphics[width=0.49\textwidth]{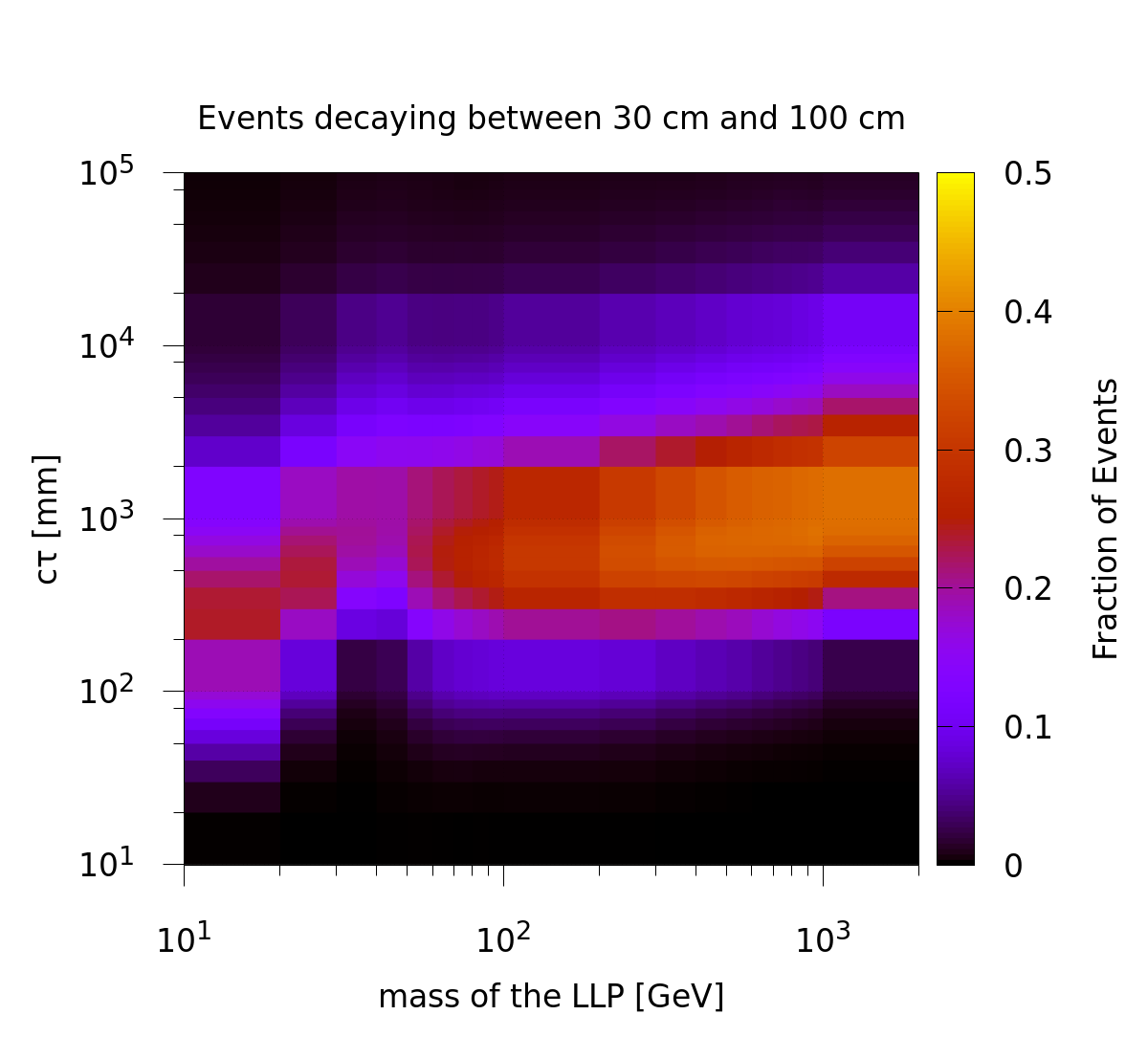}
    \caption{Fraction of LLPs decaying within 30\,cm (left) and between 30\,cm and 100\,cm (right) from the beam line as a function of their mass and proper decay length ($c\tau$).}
    \label{fig:fract_decay}
\end{figure}

In Figure \ref{fig:fract_decay} we show the variation of the fraction of LLP decays as a function of the LLP mass and proper decay length. We focus on two intervals for the decay length, namely a decay within 30\,cm (left panel) or between 30\,cm and 100\,cm (right panel) from the beam line. 
The former corresponds to the transverse dimension of the ATLAS large area tracking setup as given in \cite{ATL-PHYS-PUB-2017-014} and the latter to the rest of the tracker region \cite{Collaboration:2285585}. 

For particles with masses in the TeV range, we observe that even for proper lifetimes of the order of a few meters, about 40\% of the decays are expected to take place within 30\,cm from the beam line, {\it i.e.}\ within the silicon detector. For lighter particles having masses around 10\,GeV, this fraction turns to 20\%, provided the proper lifetime is $c\tau \lesssim 1$\,m. For even lighter particles, the sensitivity will be reduced, since they may have larger boost factors and consequently decay much later. Considering radial distances between 30\,cm and 100\,cm from the beam line, we find that we can expect at least $10\%$ decays within this region for a mass of about $\sim1{\rm~TeV}$ and a mean proper lifetime of $c\tau \sim 10$\,m. The same fraction of decays is expected for a particle mass of about 10\,GeV and a proper lifetime of $c\tau \sim 1$\,m.

The small increase in the fraction of decays within $30{\rm~cm}$ and a subsequent dip in the corresponding fraction in the region between $30{\rm~cm}$ and $100{\rm~cm}$, for LLP masses around $40-50{\rm~GeV}$ is due to the presence of the $Z$-pole in the LLP pair-production cross-section. At these mass values, the LLPs will preferentially be produced by the decay of an on-shell $Z$-boson and hence they will have very little boost and smaller decay lengths.

We also notice that in the right panel plot of Fig.~\ref{fig:fract_decay}, a red region is sandwiched between two purple regions, \textit{i.e.}, the decay fraction in this region first rises and then falls again. This is because as the lifetime increases up to a particular value ($c\tau\sim$ few $100{\rm~mm}$), we expect the decay fraction in this region to increase. However, when the lifetime becomes much higher, the decay length distribution becomes flatter and hence, the chances of the LLP to decay in regions of the detector outside the tracker also increase, making the decay fraction in this region decrease.

In summary, we find that for a wide range of proper decay length values $c\tau$, the probability of the LLP to decay within a distance of 30\,cm from the beam line is substantial. In the following, we will therefore restrict ourselves to decays taking place within this radius, {\it i.e.}\ within the region of the ATLAS tracker where large area tracking is possible, since it will serve well in identifying the position of the secondary vertex.

\section{Lifetime reconstruction for different LLP decay modes}
\label{sec:diff_decay}

After the previous preliminary remarks, let us now turn to our analysis. We will study to which extent the lifetime of a long-lived particle (LLP) can be reconstructed considering the following four decay scenarios:
\begin{itemize}
    \item Displaced leptons: In this case we will assume a neutral LLP decaying into a pair of leptons. We will generate our Monte Carlo data set using a supersymmetric model containing a $LLE$-type $R$-parity violating coupling \cite{Dreiner:1997uz,Barbier:2004ez,Graham:2012th,Evans:2016zau,Lee:2018pag,CMS:2014hka,Aad:2019tcc}. In this model the sneutrino is the LLP and it decays into two electrons. Given this final state, the position of the secondary vertex as well as the boost ($\beta\gamma$) of the LLP can be experimentally measured.
    \item Displaced leptons with missing transverse energy ($\met$): We will assume a neutral LLP decaying into a pair of leptons along with an invisible particle. Here we will employ a minimal Gauge-Mediated Supersymmetry Breaking (GMSB) model in which a long-lived lightest neutralino decays into a $Z$ and a nearly massless gravitino \cite{Giudice:1998bp,Meade:2008wd,Meade:2010ji,Lee:2018pag}. We will also discuss the feasibility of estimating the lifetime in a scenario with a heavier invisible particle in Sec.~\ref{sssec:3body_mass}. For this signature, although the position of the secondary vertex can be measured, the boost of the LLP cannot be reconstructed since part of the final state is invisible.
    \item Kinked (or disappearing) tracks: As a final case we will consider a charged LLP decaying into a lepton along with an invisible particle. The model that we will use for this analysis is again $LLE$-type RPV SUSY \cite{Bhattacherjee:2012ed,Aaboud:2017mpt,Sirunyan:2020pjd}, with a slepton LLP decaying into a charged lepton and a neutrino. In these signatures, the position of the kink or the position where the charged track disappears can provide the LLP decay position, and the $\beta\gamma$ of the LLP can be calculated from the charged LLP track itself.
    \item Displaced jets: Here we will consider a neutral LLP decaying into two jets. In this analysis, we will again use the R-parity-violating supersymmetric framework, but with a $LQD$-type coupling \cite{Barbier:2004ez,Graham:2012th,CMS:2020idp} inducing the decay of a long-lived sneutrino into a jet pair. For the displaced jets signature, the position of the secondary vertex can also be measured. It is possible to reconstruct the $\beta\gamma$ of the LLP from the jets, however, the reconstruction is plagued with important uncertainties due to the fact that the observed jets are quite different than the initial partons.  
\end{itemize}

Let us also note that this choice of models should not be taken to reflect any theoretical prejudice. They have been chosen simply for convenience, as they are already incorporated in the \texttt{PYTHIA6} framework and they can give rise to the experimental signatures that we will be studying in what follows. In the same spirit, we will not be concerned with the phenomenological viability of these models with regards to the full set of constraints that could be envisaged. Put simply, for the purposes of this work these models should be viewed as toy models.
Any model giving rise to such production modes will exhibit the results discussed below. The results are general and hold for most models exhibiting the respective topologies. 

\subsection{Typical cross-sections and triggering of LLPs}\label{sec:constraints}

We start by briefly discussing the cross-section upper limits that have been obtained in various experimental searches by CMS and ATLAS and which are relevant for our processs. This will guide us in chosing realistic values for these cross-sections which are allowed by current observations. 

\begin{itemize}
\item Displaced leptons (electrons or muons)

The CMS search for LLPs decaying to two leptons at $\sqrt{s}=8$ TeV with 19.6 fb$^{-1}$ of data sets an upper limit of few fb on the signal cross-section (production of LLPs from SM Higgs decays and branching of LLPs to electrons/muons) when the LLP mass is 20 GeV and it has a proper mean decay length ($c\tau$) of 1 cm \cite{CMS:2014hka}.

The search for displaced vertices of oppositely charged leptons in ATLAS at $\sqrt{s}=13$ TeV with 32.8 fb$^{-1}$ of data rules out signal production cross-sections up to $\sim 0.3$ fb for the signal model where a pair of squarks of mass 700 GeV is produced and decays into a $500$ GeV neutralino which is long-lived ($c\tau \sim$ 2-3 cm) and decays into two charged leptons (electron/muon) and a neutrino via an RPV coupling \cite{Aad:2019tcc}.



\item Displaced jets

The CMS search for displaced jets at $\sqrt{s}=13$ TeV with 132 fb$^{-1}$ data for the jet-jet model (where LLPs are pair produced and decay to two quarks) sets a cross-section upper limit of 1 fb for an LLP mass of 100 GeV having $c\tau$ of 1 cm \cite{Sirunyan:2020cao}. This study has imposed a bound of 0.07 fb at 95\% confidence level for simplified models with pair-produced neutral LLPs decaying into quark-antiquark pairs for long-lived particle masses larger than 500 GeV and mean proper decay lengths between 2 and 250 mm.
In Ref.~\cite{CMS-PAS-EXO-19-013}, LLPs with mean proper decay length ranging between 0.1 mm and 100 mm have been studied. In the context of $R$-parity violating SUSY scenarios, a cross-section bound of 0.08 fb at 95\% confidence level for neutralino and gluino masses between 800 GeV and 3 TeV, has been obtained. The mean proper decay length has been constrained between 1 mm and 25 mm for such pair-produced particles.

For ATLAS, the search for displaced hadronic jets in the inner detector and muon spectrometer at $\sqrt{s}=13$ TeV with 33.0 fb$^{-1}$ of data sets an upper limit of $\sim 10^{-2}$ on the branching of SM Higgs to scalar LLPs which in turn decay to give displaced jets when the LLP has a mass of 25 GeV and $c\tau$ of 60-70 cm \cite{Aad:2019xav}.

\item Disappearing tracks

In the recent CMS search for disappearing tracks \cite{Sirunyan:2020pjd} at $\sqrt{s}=13$ TeV with 140 fb$^{-1}$ of data, the upper limit on cross-section times branching of a chargino decaying into a wino-like neutralino is found to be $\sim5$ fb, for a 500 GeV chargino with $c\tau = 100$ cm. 

The disappearing tracks search at $\sqrt{s}=13$ TeV with 36.1 fb$^{-1}$ of data in ATLAS \cite{Aaboud:2017mpt} excludes charginos with masses up to $\sim$ 560 GeV having $c\tau$ of 30 cm when tan$\beta$=5 and $\mu>0$.

\end{itemize}

From this discussion on the various cross-section upper limits we conclude that the strongest bounds until now are obtained by ATLAS in the displaced leptons search and the corresponding upper limit of $\mathcal{O}(0.1)$ fb applies for an LLP of mass 500 GeV having $c\tau$ 2-3 cm, with the LLP stemming from the decay of a 700 GeV particle. All other limits are mostly in the ballpark of at least a few fb for the $c\tau$ values for which they are the most sensitive. 
In this work, we therefore choose typical cross-sections of 1 fb, 0.1 fb and 0.05 fb for the various LLP processes. 
These amount to 3000, 300 and 150 pairs of LLPs produced at the HL-LHC for 3000 fb$^{-1}$ of luminosity respectively. Although these are conservative numbers for some of the scenarios in which the limits are weaker, which implies that the correspondng cross-sections can still be greater than 1 fb, they will provide us with an idea of how well the lifetime can be estimated even with a moderate number of observed events.

The next step is to ensure that the LLP process can be triggered upon. If the event does not pass triggering criteria, it will be lost forever and this will severely hamper the prospects of discovering LLPs at the LHC \footnote{LHCb is proposed to have a trigger-less readout system after the Long Shutdown 2 (LS2, between 2018 and 2019) \cite{LHCbCollaboration:2014vzo}}. 
Usually the first level of triggers are hardware-based and have very limited information available at a coarser resolution. These are followed by software-based triggers, which have access to the full detector granularity.

Until now no tracking information was available at the first level triggers, however, there is a proposal to include tracking at the level-1 (L1) in the Phase-II upgrade. This opens up the possibility to use the L1 tracking for designing dedicated LLP triggers. 
Several attempts have been made to develop dedicated triggers for LLPs, both by experimental collaborations \cite{Aad:2013txa,Aaboud:2018aqj} and phenomenological studies \cite{Bhattacherjee:2020nno,Gershtein:2020mwi}.
The Phase-II upgrade of the CMS L1 trigger \cite{CERN-LHCC-2020-004} also has some discussions on how specific triggers can be developed for triggering events with displaced objects. Ref. \cite{CERN-LHCC-2020-004} has also discussed the prospect of using the ECAL timing for triggering on displaced jets, having a $p_T$ as low as 20 GeV. In summary, even in the high PU environment of HL-LHC, triggering of LLPs will be possible making use of the Phase-II upgrades along with the unique features of processes involving LLPs. 


\subsection{A note on backgrounds}

For any discovery at the LHC, one has to carefully consider the backgrounds. Despite the numerous challenges characterising long-lived particles, they generically tend to give rise to fairly clean channels with relatively low background rates. Therefore, throughout our analysis we have neglected processes and instrumental effects that can act as backgrounds to the LLP signatures that we consider.

As an example, the authors of \cite{Mason:2019okp} followed one CMS \cite{Khachatryan:2016mlc} and one ATLAS \cite{Aad:2016xcr} analysis to estimate the double differential inclusive jet cross-section and the inclusive prompt photon cross-section, respectively. They found that if the probability to measure an object with a given lifetime is modelled as a gaussian smear with a time resolution of $30$ ps (the standard deviation of the gaussian distribution), the number of background events mis-measured to have a time-delay of more than $1$ ns was negligible, as $1$ ns lies more than 30 standard deviations away from the central value of the distribution. They further showed that for objects with a worse resolution, such as electrons, a 60 ps time resolution would, similarly, lead to negligible backgrounds. Moreover, the authors estimated the number of pile-up events, which can also act as backgrounds. They computed the number of pile-up backgrounds to be around $10^7$, considering the fake rate of jet $\rightarrow$ photon $\sim 10^{-4}$, fraction of track-less jets $\sim 10^{-3}$, and the inclusive cross-section of $\sigma_{\rm inc} \sim 80$ mb. Applying a gaussian smear with a slightly larger resolution of $190$ ps, they found that the number of background events for a time delay of $\Delta t > 1$ ns ($\Delta t > 2$ ns) is 0.7 (0). A recent CMS paper \cite{Sirunyan:2019gut} and the LLP community white paper \cite{Alimena:2019zri} further categorised various other sources of backgrounds. The respective number of background events from beam halo, satellite bunch crossings and cosmic rays was found to be of the order of $0.5$, $1$ and $1$ respectively, after imposing $p_T$ cuts on the objects. Other sources of backgrounds can arise from fake-particle signatures that mimic real particles. These usually arise from spurious detector noise and are very hard to model with Monte Carlo simulations. 

All in all, the low level of background rates that is expected in LLP searches allows us to neglect their impact, at least in this preliminary study. Needless to say that, in case one of the LHC experiments does observe an excess, they will have to be carefully studied.

\subsection{Displaced leptons}
\label{ssec:disp_leps}

We start with the experimentally simplest case in which a long-lived particle $X$ decays into two leptons within the inner tracker of ATLAS. In this case, the position of the secondary vertex can be identified precisely by observing the lepton tracks in the tracker~\footnote{Experimentally, the procedure to reconstruct a secondary vertex is generally done in two steps as detailed in~\cite{ATL-PHYS-PUB-2019-013}. The efficiency depends on the process/model in question, and on the associated number of charged particles/tracks that can be reconstructed first at the truth level and then at the detector level. The final efficiency can vary between as much as 100\% for models with a large number of tracks to about 20\% for models with leptons and missing energy.}. In general, the larger the number of tracks, the greater is the efficiency of reconstructing the secondary vertex, \textit{cf} Ref. \cite{Chatrchyan:2014fea}. The position of the secondary vertex for a pair of displaced leptons can be reconstructed with a precision of few ($\mathcal{O}(10)$) $\mu\text{m}$ in the transverse direction if the decay occurs towards the inner tracker region, and becomes more uncertain for longer transverse displacements, \textit{cf} Refs. \cite{Chatrchyan:2014fea,Alimena:2019zri}.  The mass of the decaying particle $X$ can be inferred from the dilepton invariant mass distribution. Finally, the boost factor of the decaying particle $X$ can be determined as $(\beta\gamma)_{\text{LLP}} = p/m$, where $p$ and $m$ are the absolute momentum and the invariant mass of the dilepton system, respectively. We present our analyses assuming two displaced electrons in the final state. In the case of muons, we need not restrict to the inner tracker and can rather consider decays up to the muon spectrometer \cite{Aaboud:2018jbr}. In this case, we could also use the muon tracks in the muon spectrometer to reconstruct the secondary vertex as well as the four-momentum of the LLP.

\subsubsection{Lessons from a naive exponential fit}
\label{ssec:naive}

Let us first attempt to reconstruct the LLP lifetime through a simple exponential fit \footnote{Note that part of this discussion has already been presented in Ref.\  \cite{Brooijmans:2018xbu}. In order to keep the presentation self-contained, we recapitulate the procedure here and expand upon it wherever necessary.}. We will see that experimental cuts introduce a bias on the sample and, hence, hamper the lifetime estimation. Solutions to this issue will be suggested in the following Sec.~\ref{ssec:realistic} and further elaborated upon in the subsequent Sections.

For this introductory exercise, we consider an ideal situation in which initial and final state radiation as well as smearing effects are absent, and the four-momenta of the long-lived particles $X$ can be measured with infinite precision. We generate our data sample of parton-level events, $pp \to XX$, at $\sqrt{s}=14$ TeV using {\tt PYTHIA\,6} for different masses and lifetimes of the particle $X$ \footnote{For this illustration we have used the $R$-parity violating supersymmetric (SUSY) model containing a $LLE$-type coupling.}. Note that the $\beta\gamma$ distribution of the LLPs will vary depending on their production mode and is, therefore, a model-dependent quantity.

We demand that both the electrons coming from the decay of the long-lived particle $X$ have transverse momentum $p_T > 20$ GeV and, in order to illustrate one key difference between the previous LHC runs and the HL-LHC, we impose two alternative pseudorapidity cuts, namely $|\eta| < 2.4$ (previous Runs) or $|\eta|<4.0$ (HL-LHC) \cite{Backhaus:2683265}. The samples with only the $p_T$ and $\eta$ cuts applied will be referred to as ``basic cuts'' (BC). Since the reconstruction of the secondary vertex becomes more difficult and less precise as the latter approaches the outer surface of the tracker, we impose an additional condition on the displacement of the secondary vertex with respect to the interaction point. We restrict ourselves to events for which the transverse decay length $d_T$ of $X$ lies within the region of the ATLAS tracker where large area tracking is possible (which extends out to 30 cm in the radial direction) and the displacement $|d_z|$ in the longitudinal direction is within 150\,cm which corresponds to the half-length of the ATLAS tracker. The samples with these additional cuts on the decay length along with the basic cuts applied will be hereafter referred to as ``extra cuts'' (EC). All used cuts can be summarized as follows:
\begin{itemize}[leftmargin=1.5cm]
    \item[{\bf (BC)}] $p_T>20{\rm~GeV}$ and $|\eta|<2.4$ (previous LHC Runs) or $|\eta|<4.0$ (HL-LHC) for both electrons,
    \item[{\bf (EC)}] $0.1{\rm~cm} < d_T < 30{\rm~cm}$ and $d_z < 150{\rm~cm}$ for displacement of secondary vertex, in addition to (BC).
\end{itemize}

\begin{figure}
    \centering
    \includegraphics[width=0.49\textwidth]{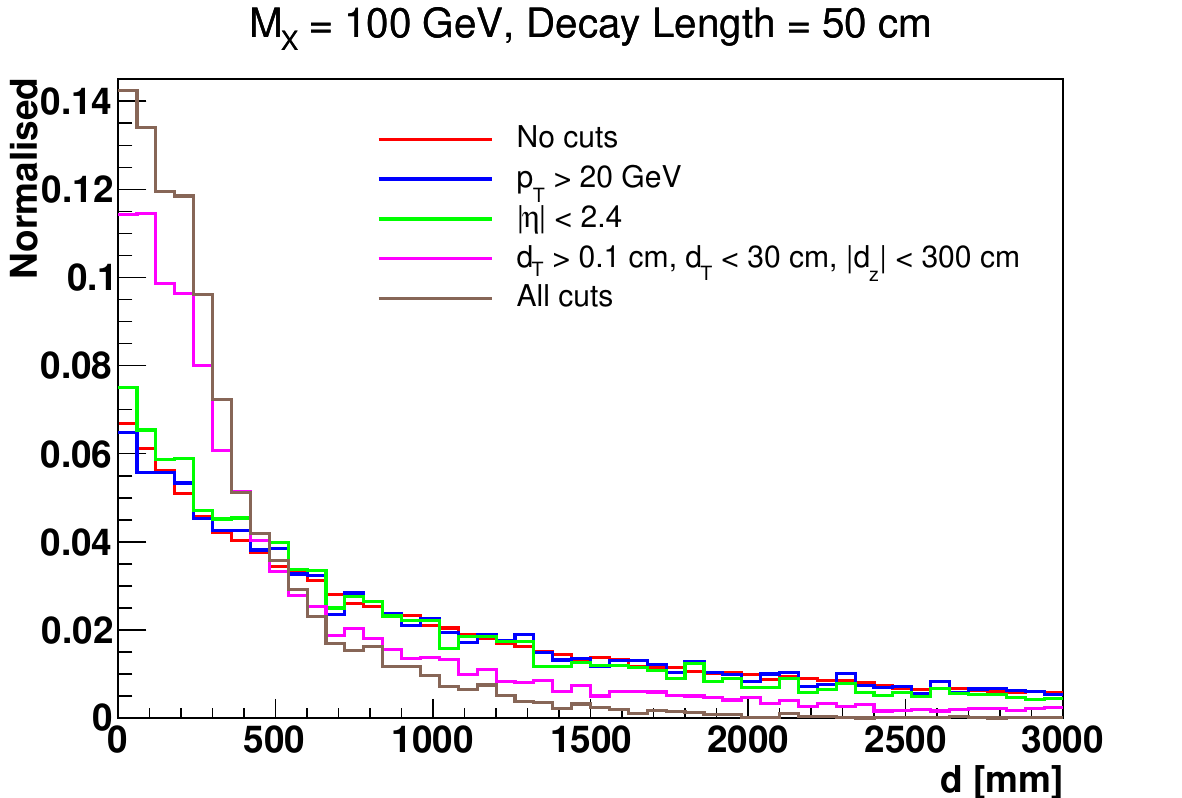}
    \includegraphics[width=0.49\textwidth]{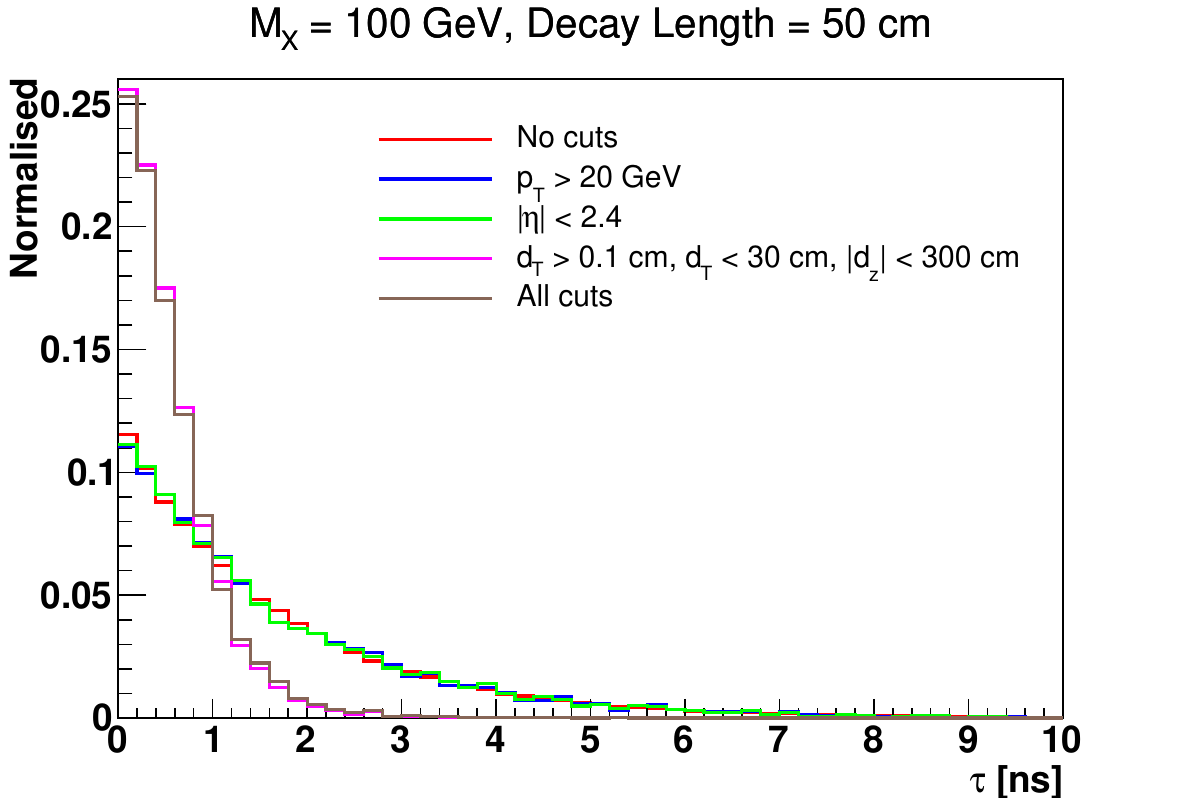} \\[2mm]
    \includegraphics[width=0.49\textwidth]{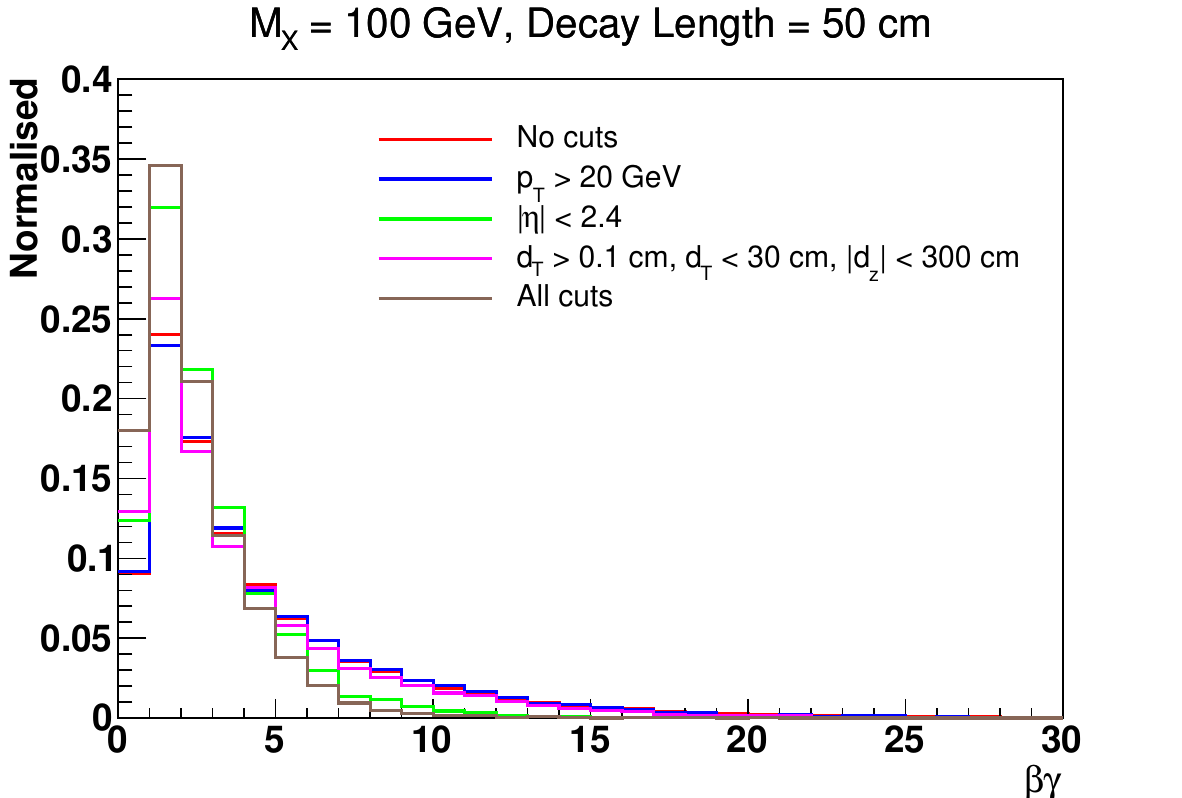} 
    \includegraphics[width=0.49\textwidth]{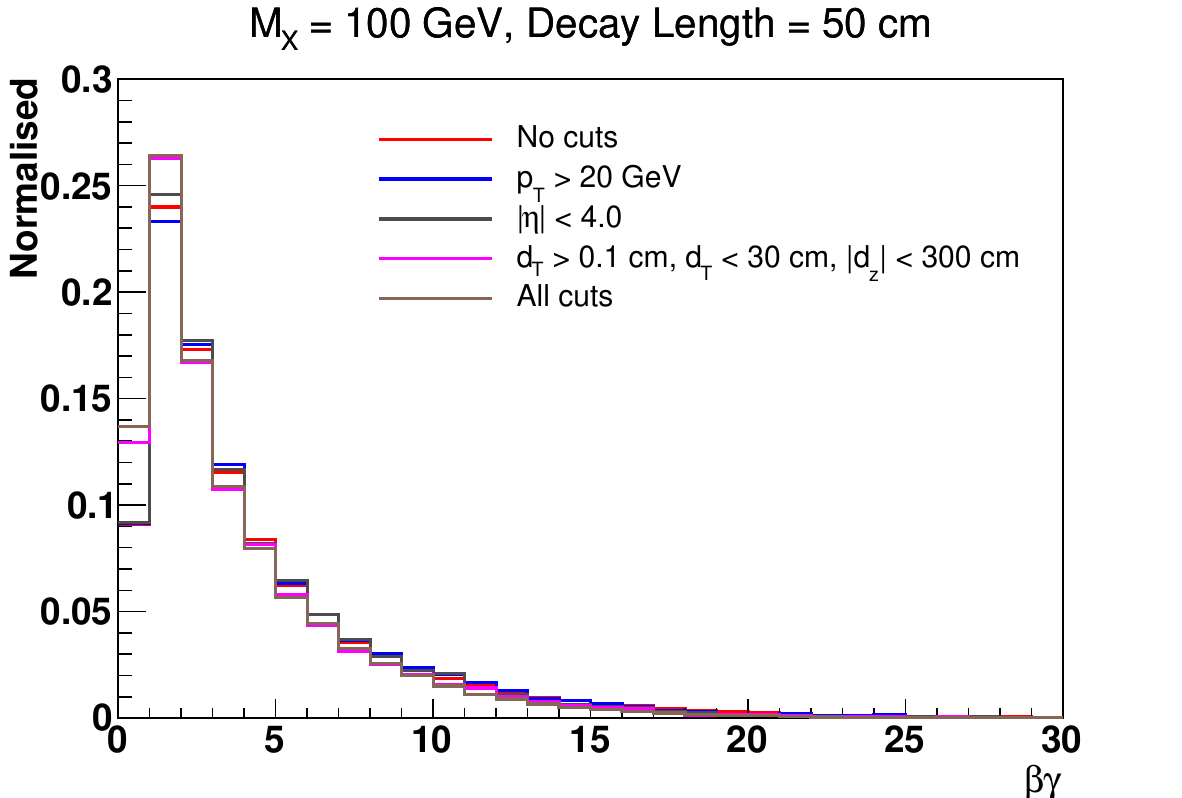}
    \caption{Effect of the cuts on the distribution of the reconstructed decay length $d$ (upper left), the proper lifetime $\tau$ (upper right), and the the boost factor $\beta\gamma$ (bottom left and right for two different pseudorapidity cuts $|\eta|<2.4$ and $|\eta|<4.0$ respectively). 
    The reconstruction here is actually a pseudo-reconstruction where all the detector effects and inefficiencies have not been taken into account.
    The distributions have been obtained assuming a mass of 100\,GeV and a proper decay length of 50\,cm. In each panel, we indicate the specific cuts that have been applied.}
    \label{fig:cuts_effect}
\end{figure}

For the lifetime estimate we use all the selected LLPs, and not specifically one or two LLPs per event.
In Figure \ref{fig:cuts_effect} we show the impact of these cuts on the decay length $d = \beta \gamma c \tau$, proper lifetime $\tau$ and $\beta\gamma$ distributions (top-left, top-right and bottom-left panels respectively), when applied individually as well as all combined. 

We first focus on the case in which $|\eta|<2.4$. We observe that the pseudorapidity cut on the electrons introduces a bias on the $\beta\gamma$ distribution towards smaller values. 
This is explained by the fact that  $|\eta| < 2.4 $  restricts events to the central region, characterised by high $p_{T}$ and low $p_{z}$. The fraction of events with high values of $\beta\gamma = p/m = \sqrt{p_x^2+p_y^2+p_z^2}/m$ in the central region is much smaller compared to the forward region\footnote{By the optical theorem, the cross-section in a $2\to2$ scattering in the centre of mass frame is  peaked in the forward direction, i.e implying more events with large $p_{z}$.}, and therefore the rapidity cut rejects a large fraction of such events.
On the other hand, we see that the cuts on the $p_T$ of the electrons and on the transverse decay length of $X$ affect the $\beta\gamma$ distribution only slightly. 

For the decay length $d$, we observe that this distribution is mostly affected by the cuts on the transverse decay length of $X$, pushing the spectrum towards lower values of $d$. Similarly, these cuts also shift the proper lifetime $\tau$ distribution towards lower values, hence, biasing our samples in favour of events characterised by smaller proper decay lengths. This implies that the observed distribution is overall skewed with respect to the underlying one, and our estimate for the lifetime will also be biased towards smaller values.

Based on the discussion presented in Sec.~\ref{sec:llp_lifetime}, we now attempt to estimate the lifetime of the particle $X$ through a simple exponential fit using the \texttt{TF1} class integrated in the \texttt{ROOT} environment \cite{ROOT}. The performance of this estimation depends both on the mean proper lifetime of the LLP and on its mass. To illustrate this dependence, we perform our fit assuming two different LLP masses, 100\,GeV and 1\,TeV, and three different mean decay lengths, 10\,cm, 50\,cm, and 1\,m. The results of this exercise are summarised in Table \ref{tab:reco_life_1}, which shows the estimated lifetimes based on the samples without any cuts and with the various cuts applied for all six cases. For completeness, we also quote the number of LLPs remaining after each set of cuts is applied, starting with a sample of 10\,000 LLPs (i.e., 5\,000 events since in each event a pair of LLPs is produced)~\footnote{In this Section we choose to work with a relatively large number number of events simply for illustration reasons. In the following Sections, our choices will be guided by the constraints discussed in Sec.~\ref{sec:constraints}.}.

\begin{table}
    \centering 
    \begin{tabular}{|c|c||c|c|c|}
        \hline
        $M_X$ & DL & \multicolumn{3}{c|}{Reconstructed DL}      \\
        \cline{3-5}
             [GeV] & [cm] & without cuts [cm] & with BC [cm] & with EC [cm] \\
        \hline
        \hline
        100 & 10 & 9.95 $\pm$ 0.10 & 9.91 $\pm$ 0.13 ~[6117] & 7.81 $\pm$ 0.10 ~[5375]\\
        100 & 50 & 49.74 $\pm$ 0.51 & 49.55 $\pm$ 0.64 ~[6117] & 15.25 $\pm$ 0.29 ~[2590]\\
        100 & 100 & 99.48 $\pm$ 1.01 & 98.82 $\pm$ 1.29 ~[6117] & 18.12 $\pm$ 0.45 ~[1539]\\
        \hline
        \hline
        1000 & 10 & 9.97 $\pm$ 0.10 & 10.02 $\pm$ 0.10 ~[9546] & 8.88 $\pm$ 0.08 ~[9050]\\
        1000 & 50 & 49.88 $\pm$ 0.49 & 50.09 $\pm$ 0.51 ~[9546] & 19.61 $\pm$ 0.26 ~[5227]\\
        ~1000~ & ~100~ & ~99.85 $\pm$ 0.99~ & ~100.15 $\pm$ 1.02 ~[9546]~ & ~22.37 $\pm$ 0.35 ~[3233]~\\
        \hline
    \end{tabular}
    \caption{Lifetime estimates obtained from exponential fitting of the $\tau$ distribution for six combinations of LLP mass ($M_X$, in GeV) and decay length (DL, in cm) based on an initial sample (without cuts) of 5\,000 events (10\,000 LLPs). We indicate the reconstructed decay lengths $d = c\tau$ (in cm) together with the number of events (in brackets) remaining after each set of cuts (BC or EC) is applied.}
    \label{tab:reco_life_1}
\end{table}

We observe that although the LLP mean decay length can be accurately reconstructed when the BC sample is used, once the (necessary) Extra Cuts are applied the result of the fitting procedure becomes incompatible with the actual underlying value. The situation becomes worse when the true decay length is large, since a larger fraction of the decays occurs beyond the limit of $d_T < 30$\,cm. The induced bias leads to results which can deviate from the actual decay length by almost one order of magnitude. The situation becomes marginally better when the mass of the LLP increases, since heavier LLPs are characterised by smaller $\beta\gamma$ values and are more centrally produced, but it is clear that a naive exponential fit to the data does not constitute a viable option to reconstruct the LLP lifetime.

Moving to the case of the HL-LHC ($|\eta| \lesssim 4.0$), which will be our focus in everything that follows, in the bottom-right panel of Figure \ref{fig:cuts_effect}, we also show the effect of the looser $\eta$ cut on the $\beta\gamma$ distribution (bottom-right panel). We observe that in this case the $\beta\gamma$ distribution is affected substantially less, implying that the HL-LHC upgrade may perform better in estimating the LLP lifetime. However, the effects of the Extra Cuts which concern the position of the SV, remain. Moreover, as we already mentioned Pile-Up is expected to become an important issue in the HL-LHC environment. In the next Section, we discuss how the addition of 140 PU vertices per hard interaction affects the situation and it can be remedied.

\subsubsection{The high PU environment of HL-LHC}
\label{ssec:PU}

To simulate the high PU environment of HL-LHC, we generate 1 million soft QCD events using \texttt{PYTHIA8} \cite{Sjostrand:2007gs} and use these as the pile-up.  We merge this PU with the hard process using the \texttt{PileUpMerger} of \texttt{Delphes-3.4.2} \cite{deFavereau:2013fsa}, assigning to each event a number of PU events which drawn from a Poisson distribution with an average value of 140. The average number of PU vertices follows from the peak luminosity of HL-LHC, which is proposed to be $\sim 5\times 10^{34}{\rm cm}^{-2}{\rm s}^{-1}$. 
The vertices follow a two-dimensional Gaussian distribution, with $z=0$ and $t=0$ having the maximum probability of having a vertex, with $\sigma_z= 5.3$ cm and $\sigma_t$= 160 ps respectively. The total spread is of 25 cm and 800 ps respectively in $z$ and $t$ directions.
We use \texttt{PYTHIA6} to generate the hard process at $\sqrt{s}=14$ TeV and \texttt{Delphes-3.4.2} for detector simulation. The total integrated luminosity used in this work is 3000 fb$^{-1}$. 

As we saw, for the lifetime estimation of any particle we need the information of its decay position in the detector and the boost of the particle, where the latter can be reconstructed from its decay products. The distances are mostly measured with respect to the primary vertex, which is the vertex which has the maximum value of $\sum_{n_{trk}} p_T^2$ or $\sum_{n_{trk}} p_T^2/n_{trk}$, where $n_{trk}$ is the number of tracks starting from that vertex. Then, in processes with LLPs which are not accompanied by many high-energy prompt particles, and in the presence of $\sim 140$ other such vertices, the chances of misidentifying any PU vertex as being the PV will be higher (for a quantitative discussion \textit{cf} Ref. \cite{Bhattacherjee:2020nno}). In such a scenario, the $d_z$ of the secondary vertex calculated from the reconstructed PV will not be the same as that calculated from the vertex where the LLP is actually produced, which will affect the overall measured LLLP decay length ($d$) in the lab frame. However, in all cases, the \textit{transverse} distance of the SV ($d_T$) remains the same, and therefore, it is more convenient to use $d_T$ in all further calculations, instead of $d$. 

The high amount of PU at HL-LHC, among other problems, will affect the electron isolation adversely, since a large amount of tracks now enter the electron isolation cone. However, for displaced leptons we can only use the tracks which start close to the identified secondary vertex. We understand, then, that using such an isolation technique will not constitute a viable option at the trigger level, notably the L1 trigger. However, since here we are mostly concerned with the offline analysis, where we have fully reconstructed tracks, such an isolation can be applied and is helpful in identifying the displaced leptons even in the high PU environment of the HL-LHC.

\begin{figure}[hbt!]
\centering
\includegraphics[width=0.85\textwidth]{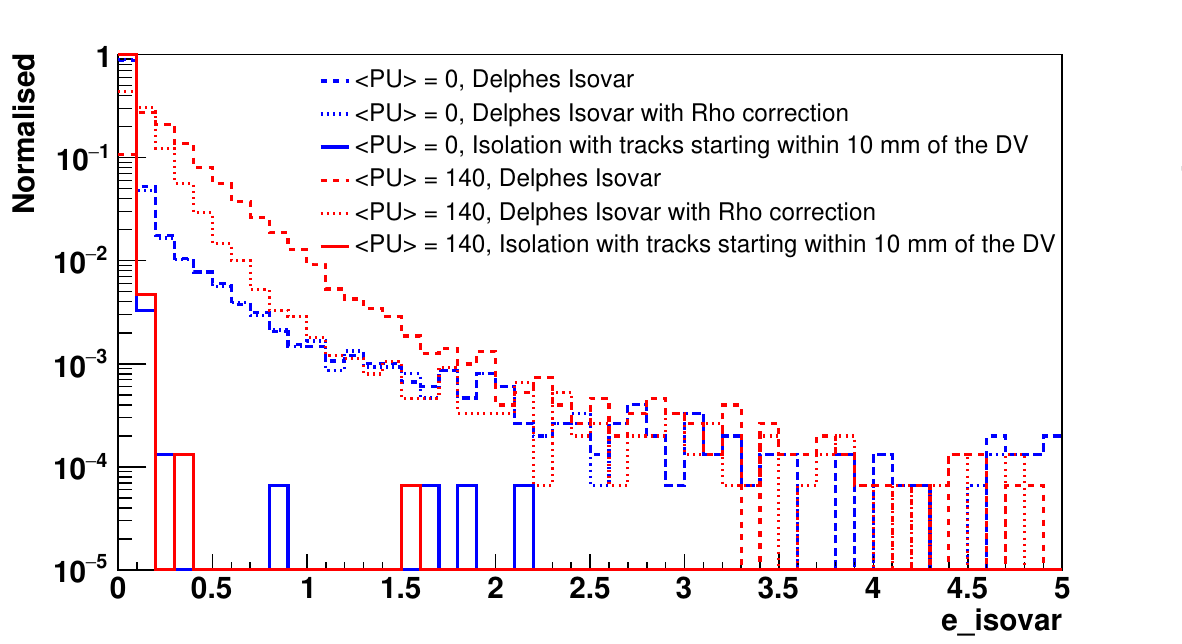}\\
\includegraphics[width=0.85\textwidth]{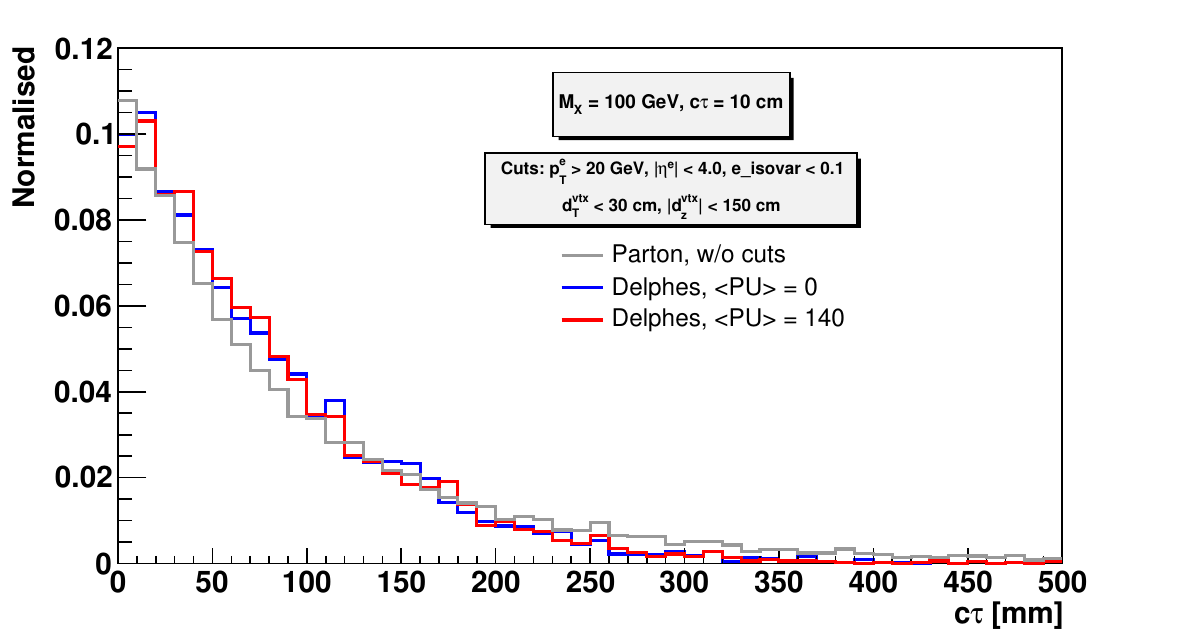}
\caption{Isolation of electrons}
\label{fig:e_isovar}
\end{figure}

Fig.\ref{fig:e_isovar} ({\it top panel}) shows the isolation variable for electrons calculated from Delphes with an isolation cone of 0.3 and the isolation variable calculated using tracks starting within 10 mm of the identified secondary vertex of the electrons. We find that taking only displaced tracks improves the isolation drastically as expected. Also, we can see that for $\langle {\rm PU}\rangle=140$, the \texttt{RhoCorrected} isolation is slightly better than the usual isolation variable, since it has some PU mitigation already present. When only displaced tracks starting near the SV are used for isolation, the amount of PU really does not affect the distributions much. As we can see from the {\it bottom panel} of fig.\ref{fig:e_isovar}, the $c\tau$ distribution for the low and high PU scenarios are very similar, with no bias in any preferred direction.
Also, since we are using $d_T$ of the vertex, which remains the same irrespective of whether we identify the LLP production vertex as the PV, it does not depend on whether the added number of PU is high (around 140) or low (close to zero). Therefore, for simplicity, we merge our hard process with low PU.

\subsubsection{Systematic uncertainties on the $d_T$ and $\beta_T\gamma$ distributions}
\label{ssec:PDF_scale}

\begin{figure}[hbt!]
\centering
\includegraphics[width=0.5\textwidth]{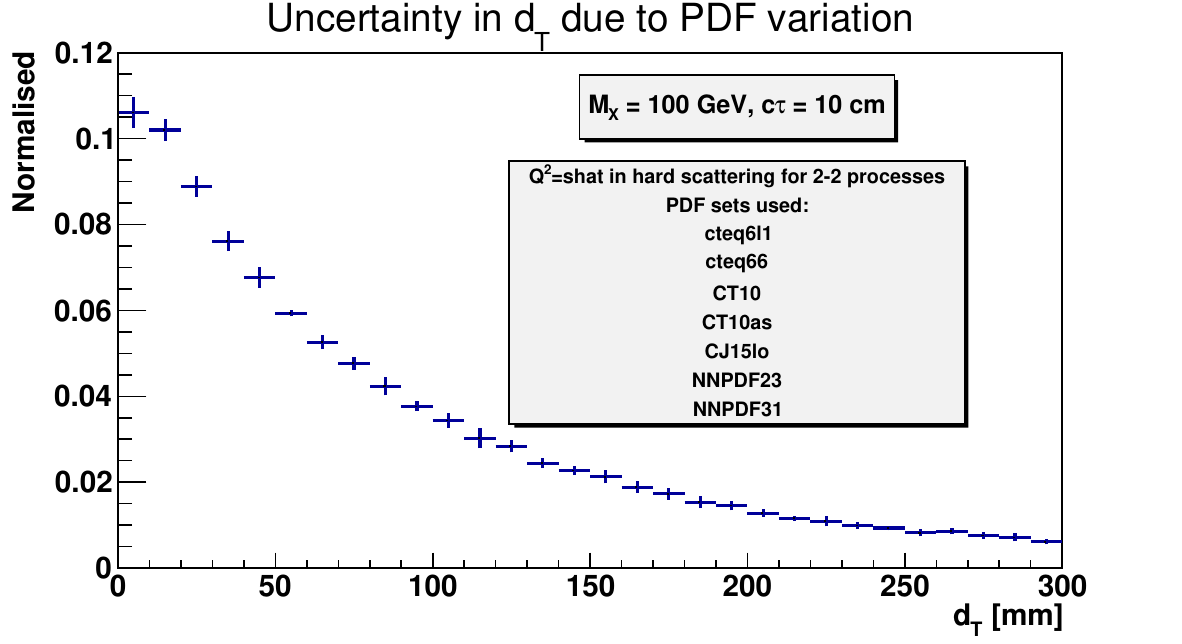}~
\includegraphics[width=0.5\textwidth]{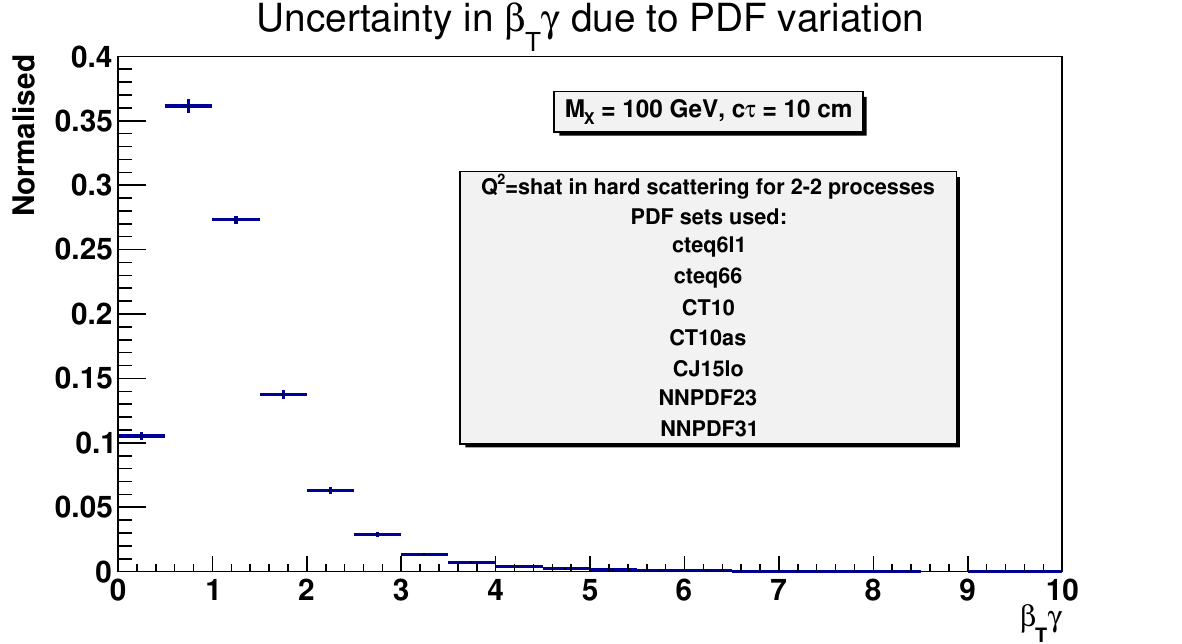}\\
\includegraphics[width=0.5\textwidth]{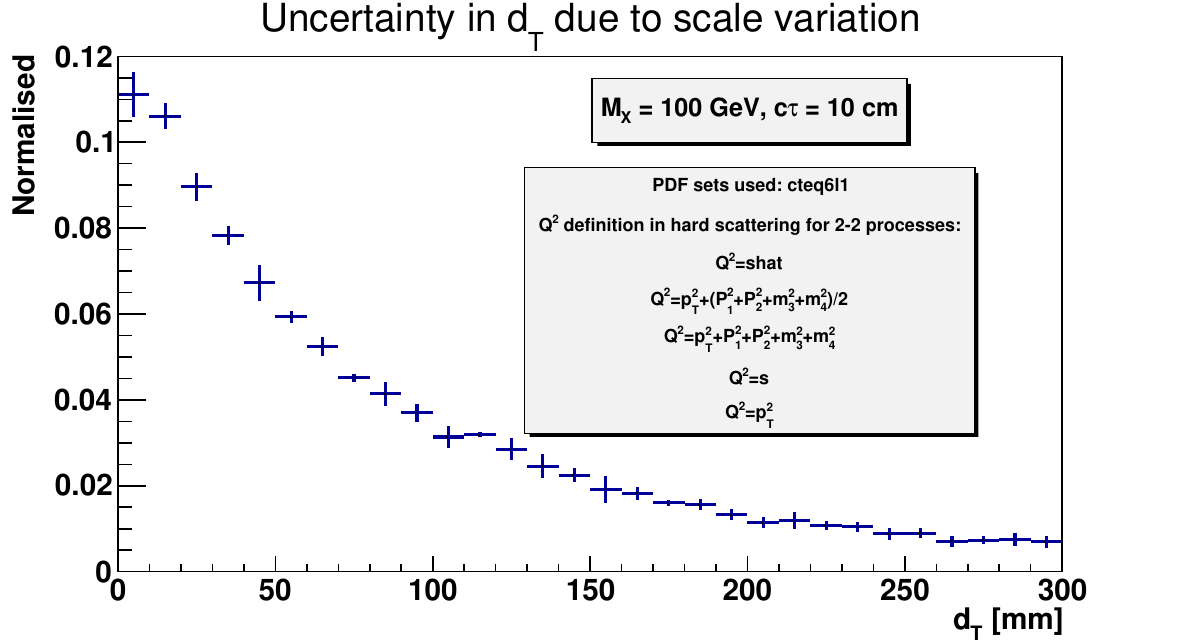}~
\includegraphics[width=0.5\textwidth]{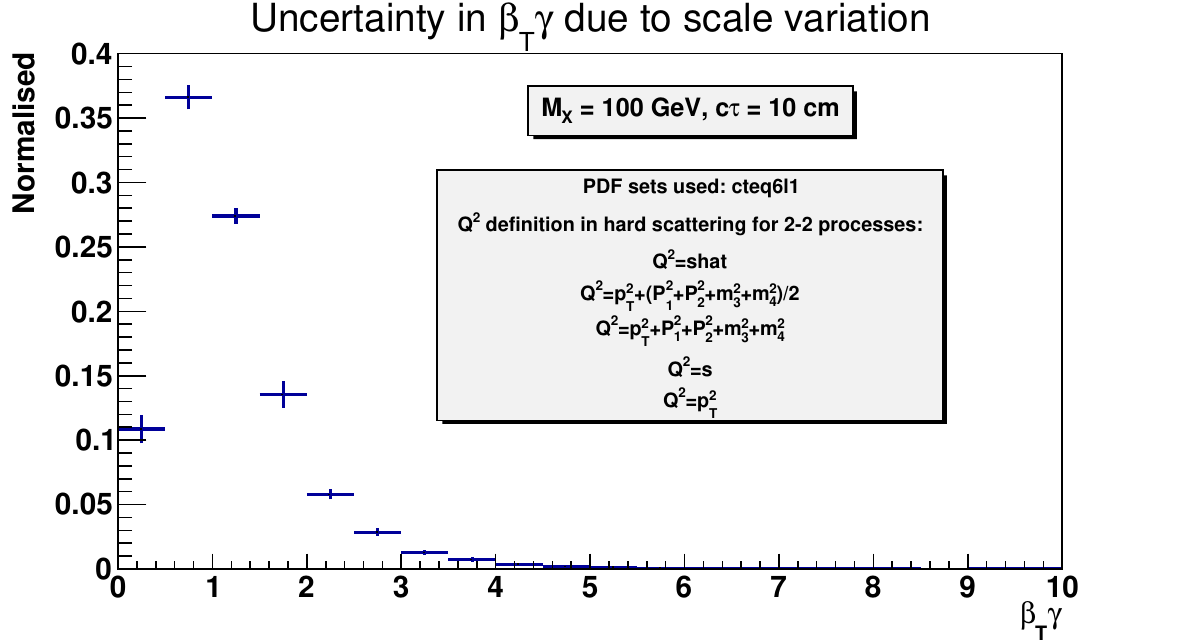}
\caption{Uncertainties in the $d_T$ ({\it left}) and $\beta_T\gamma$ ({\it right}) distributions with variation of PDF ({\it top}) and scale for the $2\rightarrow 2$ hard process ({\it bottom}) for an LLP of mass 100 GeV and $c\tau$ of 10 cm. The error is not more than 0.5\% in both the cases, even for varying $c\tau$ values.}
\label{fig:PDF_scale}
\end{figure}

In this Section we study the effect of varying the parton distribution functions (PDF sets) and the scale of the $2\rightarrow 2$ hard process on the $d_T$ and $\beta_T\gamma$ distributions. Since these two are the most important quantities required for the lifetime estimation of a LLP, we want to study the sources of systematic uncertainties and the ensuing uncertainty in their distributions. These effects depend on the decay length and mass of the LLP, and we perform this exercise for the same benchmark points as were chosen in Sec.~\ref{ssec:naive}. The results for an LLP of mass 100 GeV and decay length 10 cm are shown in fig. \ref{fig:PDF_scale}. Overall, the uncertainties due to the scale variation are slightly greater than those due to PDF variation. However, for both the $d_T$ and $\beta_T\gamma$ distributions, the important observation is that changing the PDF set or the scale of the hard interaction does not drastically alter the shape of the distributions. For simplicity, we will demonstrate the impact of such systematic effects on the lifetime estimation by assuming a flat uncertainty of 10\% and 20\%, however, one might use the uncertainties of different bins for different lifetimes separately if they have been computed beforehand.

\subsubsection{Towards more realistic assessments}
\label{ssec:realistic}

From the previous discussion we see that it is not possible to reconstruct the LLP lifetime in a fully model-independent manner based only on the measurement of its decay position, its mass, and its boost. The extra cuts (EC) introduce a bias towards smaller lifetimes in the sample because they restrict it to decays within a transverse distance of 30\,cm from the beam line and hence reject decays characterised by larger proper lifetimes. We discuss few approaches that can be envisaged in order to circumvent this problem, which we will further develop in the subsequent Sections. 

The first approach requires minimal experimental information but incorporates theoretical bias, as it relies on a model-dependent $\chi^2$ fitting. If, from experiment, we have the distribution of the transverse decay length $d_T$ and the mass $M_{X}$ of the LLP $X$, then we can simulate the process for that particular mass within the framework of a concrete model. Simulating events for different lifetimes and performing a binned $\chi^2$ analysis of the $d_T$ distribution, we can obtain an estimate for the actual lifetime of $X$. The results that can be obtained in this way will be presented in Sec.~\ref{sssec:chi2_dep}. 

The second approach is also based on a $\chi^2$ (or likelihood) analysis. In contrast to the previous method it is fairly model-independent, but requires additional experimental information. To be more precise, along with the $d_T$ distribution, we will assume that we can obtain the transverse boost factor ($\beta_T\gamma$) distribution of $X$ from experiment. It is then possible to fit the (normalised) $\beta_T\gamma$ distribution by an appropriate function and use the latter as a probability density function to generate random numbers. As a second step, random numbers will also be generated for each $c\tau$ distribution. Then, multiplying the two sets of random numbers leads to the $d_T$ distribution for that particular lifetime. Based on this, we can vary the lifetime and perform a $\chi^2$ analysis comparing the experimental $d_T$ distribution and the one generated using the procedure we just described to estimate the actual lifetime of $X$. This method will be discussed in Sec.~\ref{sssec:chi2_indep}.

\subsubsection{$\chi^2$ fitting of $\beta_T\gamma c\tau$ distribution: model-dependent analysis}
\label{sssec:chi2_dep}

Let us start with the model-dependent approach. For the case in which the LLP decays into lepton pairs, the transverse decay length distribution and the LLP mass $M_X$ can be experimentally measured. Then, within the framework of a concrete model, we can simulate the process assuming different lifetimes and perform a $\chi^2$ analysis in a straightforward manner. The minimum of the resulting $\chi^2$ distribution provides an estimate for the lifetime of the LLP. The reason why some knowledge about the underlying model is useful is because, as already mentioned above, the $\beta\gamma$ distribution depends on the production mechanism and the decay length distribution is generated at the Monte-Carlo level by multiplication of the former with the lifetime distribution. 

From now on we restrict our analysis to four out of the six benchmarks given in Table \ref{tab:reco_life_1}, which are characterised by mean proper decay lengths of 10\,cm or 50\,cm. For each configuration, we generate 3000, 300 and 150 events, corresponding to cross-sections of 1 fb, 0.1 fb and 0.05 fb respectively, and apply the EC on the obtained samples. The events passing these cuts constitute the ``experimental'' $d_T$ distributions that we wish to fit. To this end, we generate the same process with the LLP mass set equal to the invariant mass of the two final state electrons and varying the lifetime. For each lifetime, we generate 5\,000 events and then apply the EC. We construct $d_T$ distributions for LLPs simulated for different lifetimes with the same bin size as the $d_T$ distribution of the discovered LLP that we obtain from the experiment. Here, we have set the bin size to 1\,cm. We then calculate the $\chi^2$ value between these two distributions as
\begin{equation}
    \chi^2 ~=~ \sum_{i=1}^N \frac{(n_{\rm th}-n_{\rm exp})^2}{\sigma^2} 
           ~=~ \sum_{i=1}^N \frac{(n_{\rm th}-n_{\rm exp})^2}{(b-a)^2} \,,
    \label{eq:chi2}
\end{equation}
where $N$ is the total number of bins in the distribution, and $n_{\rm exp}$ and $n_{\rm th}$ are the experimentally observed and theoretically expected number of events in each bin. The expected number of events in each bin ($n_{\rm th}$) is normalized to the total number of events observed in experiment. The denominator is the square of the 68\% confidence level uncertainty in the observed number ($n_{\rm exp}$), which is equal to the difference between the $68\%$ upper ($b$) and lower ($a$) limits on the value of $n_{\rm exp}$. The latter are given by
\begin{align}
    \begin{split}
    a ~&=~ \frac{1}{2}F^{-1}_{\chi^2}\Big[ 0.32, \, n_d=2n_{\rm exp}\Big] \,, \\
    b ~&=~ \frac{1}{2}F^{-1}_{\chi^2}\Big[ 0.68, \, n_d=2\big(n_{\rm exp}+1\big)\Big] \,,
    \end{split}
    \label{eq:sigma}
\end{align}
where $F^{-1}_{\chi^2}$ is the quantile of the $\chi^2$ distribution with the number of degrees of freedom for the $\chi^2$ distribution given by $n_d$ \cite[Eq. 9.18]{chi_square}. The uncertainty taken here is exact and equals to $\sqrt{n}$ for large values of $n$. 

\begin{figure}
    \centering
    \includegraphics[width=0.49\textwidth]{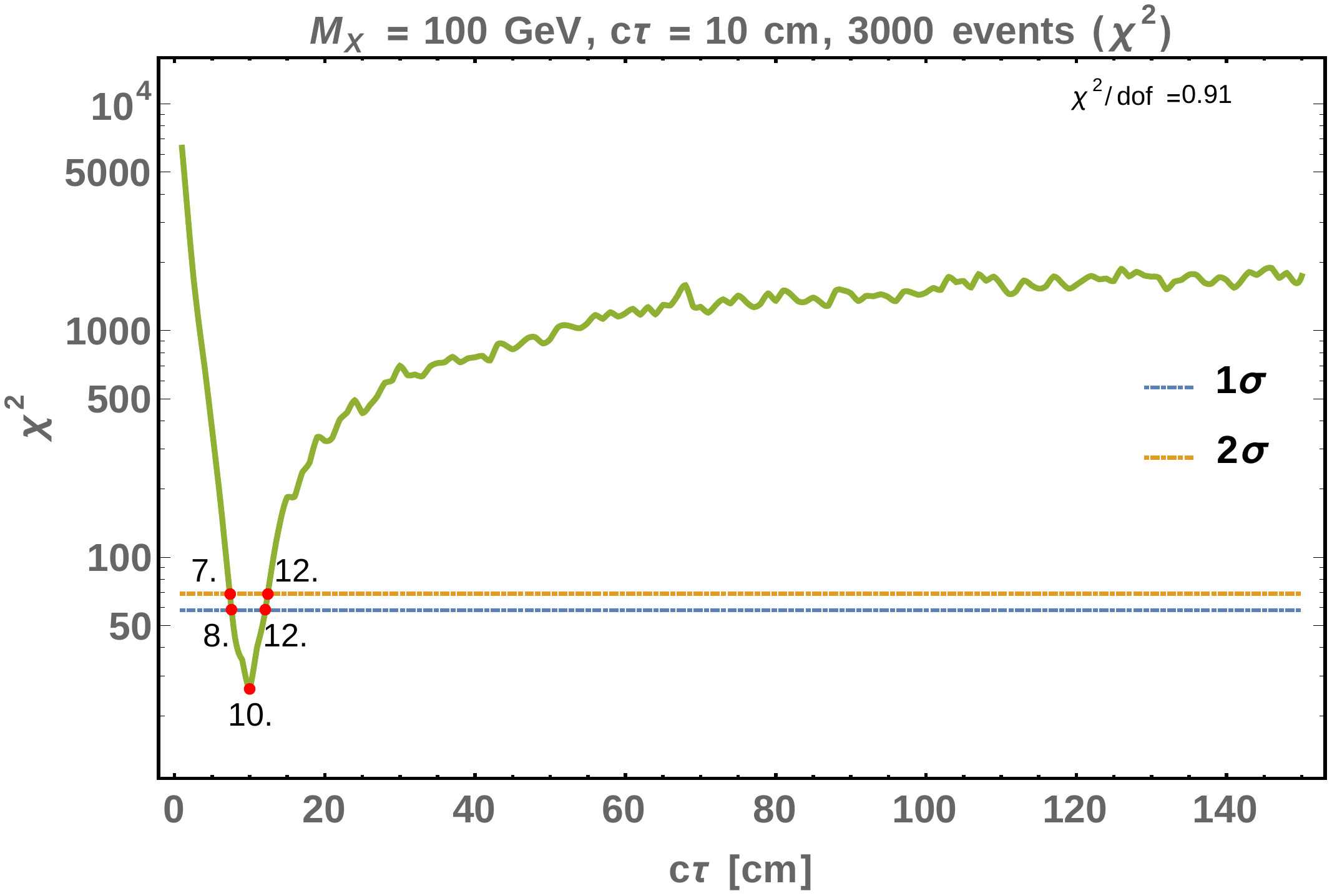}~
    \includegraphics[width=0.49\textwidth]{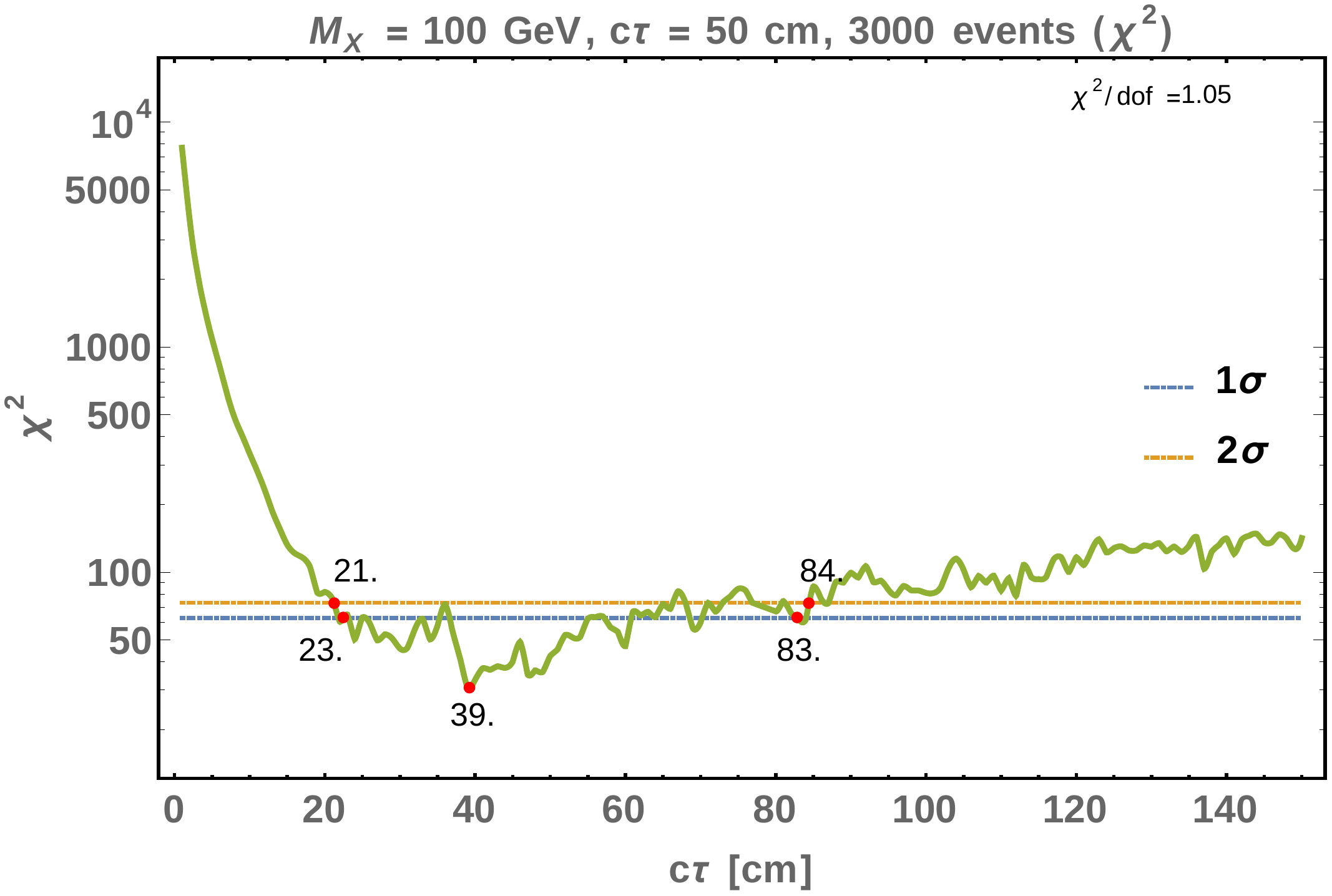} \\[2mm]
    \includegraphics[width=0.49\textwidth]{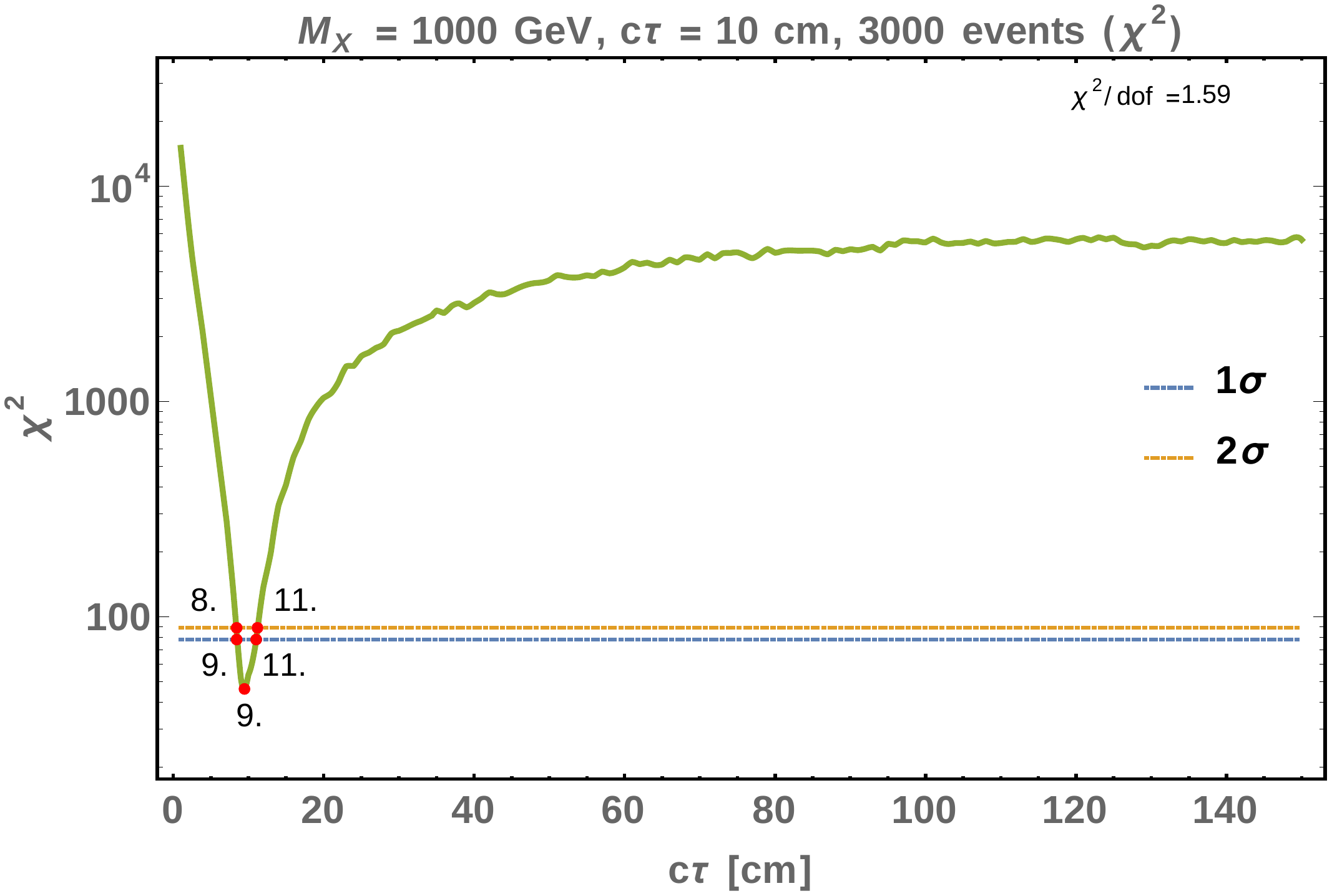}~
    \includegraphics[width=0.49\textwidth]{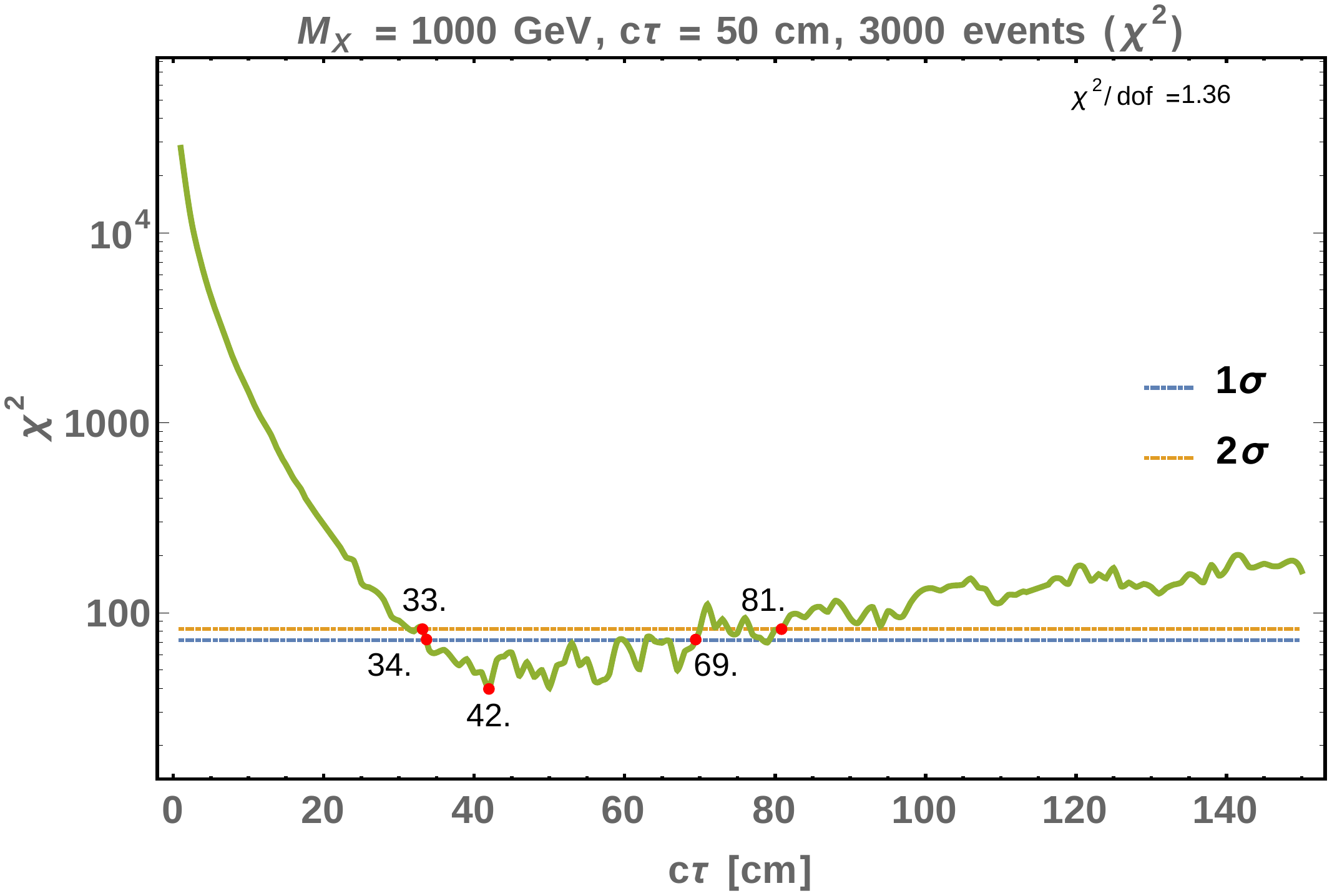}
    \caption{Model-dependent $\chi^2$ as a function of the reconstructed decay length $c\tau$ for LLP masses of 100 GeV (top) and 1000 GeV (bottom) and decay lengths of 10\,cm (left) and 50\,cm (right). The $\chi^2$ is based on the data samples after applying the EC for the displaced lepton signature.
    }
    \label{fig:chi2_dep}
\end{figure}

In our simulation, we vary the mean proper decay length between 1\,cm and 150\,cm with a step size of 1\,cm. Figure \ref{fig:chi2_dep} shows the resulting $\chi^2$ values as a function of the decay length for the four cases, with 1 fb cross-section for each. The reconstructed lifetimes, corresponding to the $\chi^2$ minimum for all the three assumed cross-sections are summarized in Table \ref{tab:chi2_dep} together with the $1\sigma$ and $2\sigma$ lower and upper limits on the lifetime in each case \footnote{Note that for some of our benchmarks the reconstructed decay length entry in Table \ref{tab:chi2_dep} is absent. This is due to the fact that for these cases, no discernible minimum could be identified for the $\chi^2$ distribution within the lifetime interval that we have considered. A similar notation will be followed in the subsequent Sections, whenever necessary.}. Here, the number of degrees of freedom ($dof$) for the $\chi^2$ analysis has been taken to be one less than the number of bins over which the sum in Eq.\ \eqref{eq:chi2} has been carried out ($N = 30$, $dof=29$).

We observe that in our best-case scenario ($\sigma = 1$ fb) the mean proper lifetime can be reconstructed with a precision of $\sim 20\%$ for decays lengths of the order of 10 cm, whereas for 50 cm we obtain the lifetime within roughly a factor of 2. Note also that, again, with increasing mass the $\chi^2$ minimum becomes more prominent. We find that the 1$\sigma$ and 2$\sigma$ lower and upper limits often have the exact same values. This is just due to the fact that the results here have been quoted with a precision of 1 cm, which corresponds to the bin size and scan interval. Even with less statistics, the lower decay length can be reconstructed, although the error band increases, and for the higher decay length benchmark, with 0.1 or 0.05 fb cross-section, only a weak lower bound can be placed on the lifetime. However, from our discussion on the typical allowed cross-sections of LLP processes, we found that the upper limits usually get weaker for LLPs with high $c\tau$ values and, therefore, our optimistic scenarios are, indeed, viable.

\begin{table}
    \centering
    \begin{tabular}{|c|c|c||c||c|c||c|c|}
    \hline
    $M_X$ & DL & cross-section & Rec.\ DL & $1\sigma$ LL & $1\sigma$ UL & $2\sigma$ LL & $2\sigma$ UL \\
    (GeV) & (cm) & (fb) & (cm) & (cm) & (cm) & (cm) & (cm) \\
    \hline\hline
    \multirow{3}{*}{100} & \multirow{3}{*}{10} & 1 & 10 & 8 & 12 & 7 & 12\\
    & & 0.1 & 11 & 4 & 23 & 4 & 27\\
    & & 0.05 & 7 & 3 & 30 & 2 & 35\\    
    \hline
    \multirow{3}{*}{100} & \multirow{3}{*}{50} & 1 & 39 & 23 & 83 & 21 & 84\\
    & & 0.1 & - & 9 & $>$150 & 8 & $>$150\\
    & & 0.05 & - & 5 & $>$150 & 4 & $>$150\\
    \hline\hline
    \multirow{3}{*}{1000} & \multirow{3}{*}{10} & 1 & 9 & 9 & 11 & 8 & 11\\
    & & 0.1 & 10 & 7 & 16 & 6 & 17\\
    & & 0.05 & 11 & 6 & 22 & 5 & 25\\    
    \hline
    \multirow{3}{*}{1000} & \multirow{3}{*}{50} & 1 & 42 & 34 & 69 & 33 & 81\\
    & & 0.1 & - & 19 & $>$150 & 17 & $>$150\\
    & & 0.05 & - & 10 & $>$150 & 9 & $>$150\\    
    \hline
    \end{tabular}
    \caption{Lifetime estimates by model-dependent $\chi^2$ fitting of the $d_T$ distribution (as shown in Figure \ref{fig:chi2_dep}) for the displaced leptons signature assuming different combinations of the LLP mass $M_X$ and decay length (DL). We display the reconstructed decay length as well as the corresponding lower (LL) and upper (UL) limits at the $1\sigma$ and $2\sigma$ confidence level, respectively. }
    \label{tab:chi2_dep}
\end{table}

\begin{table}
    \centering
    \begin{tabular}{|c|c||c||c|c||c|c|}
    \hline
    cross-section & Syst. uncertainty & Rec.\ DL & $1\sigma$ LL & $1\sigma$ UL & $2\sigma$ LL & $2\sigma$ UL \\
    (fb) &  & (cm) & (cm) & (cm) & (cm) & (cm) \\
    \hline\hline
    \multirow{2}{*}{1} & 10\% & 10 & 7 & 13 & 7 & 13\\
    & 20\% & 10 & 6 & 13 & 6 & 14\\    
    \hline
    \multirow{2}{*}{0.05} & 10\% & 7 & 3 & 30 & 2 & 36\\
    & 20\% & 7 & 2 & 30 & 2 & 36\\
    \hline
    \end{tabular}
    \caption{Effect of including PDF/scale systematic uncertainties to the lifetime estimation for the benchmark point $M_X$=100 GeV, $c\tau$=10 cm for two different cross-sections.}
    \label{tab:chi2_dep_syst}
\end{table}

\begin{table}
    \centering
    \begin{tabular}{|c|c||c||c|c||c|c|}
    \hline
    Smearing on $d_T$ & cross-section & Rec.\ DL & $1\sigma$ LL & $1\sigma$ UL & $2\sigma$ LL & $2\sigma$ UL \\
    (mm) & (fb) & (cm) & (cm) & (cm) & (cm) & (cm) \\
    \hline\hline
    \multirow{2}{*}{10} & 10 & 10 & 8 & 11 & 8 & 11\\
    & 0.05 & 7 & 3 & 39 & 2 & 45\\
    \hline
    \end{tabular}
    \caption{Effect of adding smearing-related systematic uncertainties to the lifetime estimation for the benchmark point $M_X$=100 GeV, $c\tau$=10 cm for two different cross-sections.}
    \label{tab:chi2_dep_smear}
\end{table}

Table \ref{tab:chi2_dep_syst} shows the effect of adding a flat systematic uncertainty due to the choice of PDF or scale of the hard interaction, on the number of events that is expected in a bin of the $d_T$ histogram of the benchmark with $M_X$=100 GeV and $c\tau$=10 cm with cross-sections 1 fb (0.05 fb), as has been discussed in Sec.~\ref{ssec:PDF_scale}. We have ignored here the variation of this error with the decay length of the LLP. We observe that even with a systematic uncertainty of 20\%, the results are not affected substantially. 
We have mentioned earlier that the error in identifying the secondary vertex is very small in the current LHC runs, where the PU is low. 
For HL-LHC, it might be difficult to correctly identify the position of the SV due to the high amount of PU which leads to large track multiplicity. Therefore, we add a 10 mm smearing on the position of the SV to see how it affects the lifetime estimation. Table \ref{tab:chi2_dep_smear} shows our results for an LLP of mass 100 GeV and decay length 10 cm decaying into electrons and having a cross-section of 1 fb (0.05 fb), whose $d_T$ position has a Gaussian error of 10 mm. The error band on the estimated lifetime increases by about $\sim$10 cm for the low cross-section scenario, and for the higher cross-section, there is no change.

The results shown in Table \ref{tab:chi2_dep} have been obtained using the same model as the one that was used in order to generate our pseudo-experimental sample. A reasonable question would be to ask how would these results change if we assumed the wrong model, like a different production mode of the LLP $X$. In this case, we will obtain the wrong lifetime estimate but with comparable error bands. For example, if the actual underlying process was production of $X$ from the decay of an on-shell resonance, and we had assumed non-resonant production, then the $\chi^2$ analysis would give minima at a wrong decay length due to differences in the boost factor  distributions of the two processes. However, it is possible to identify such resonant LLP production and the mass of the intermediate resonance, if any, from the total invariant mass distribution of the two LLPs' decay products. Moreover, the spin information of LLP $X$ can also be inferred from the angular distributions of its decay products as discussed in Ref.~\cite{Barr:2005dz} for sleptons and in Refs.~\cite{Barr:2005dz,Aad:2013xqa,Khachatryan:2014kca} for the Higgs boson. Hence, it is possible to deduce several key features of the underlying model if we can reconstruct all the decay products of the LLP pair. Even for decays of LLPs involving invisible particles, there are methods to identify the model and its parameters, like the LLP mass, as we will discuss later.

\subsubsection{Unbinned and binned likelihood estimators}
\label{sssec:likelihood}

When the number of observed events is lower the $\chi^2$ analysis becomes less reliable, since the $\chi^2$ value as defined in Eq.~\eqref{eq:chi2} does not follow the $\chi^2$ distribution if the errors cannot be treated as Gaussian. In this case we can envisage other estimators such as the likelihood, the unbinned version of which is defined as:
\begin{equation}
L_j = \prod_{i=1}^{N_{obs}} f(ct_{i};c\tau_j,W)
\label{eq:unbinned_L}
\end{equation}  
where $N_{obs}$ is the total number of observed events and $f(ct_i;c\tau_j)$ is the probability of observing an LLP with decay length $ct_{i}$ if the mean decay length is $c\tau_j$. Therefore, the $c\tau_j$ value which maximises $L_j$ will provide an estimate of the lifetime of the LLP. However, one has to correctly identify the probability distribution function $f(ct_{i};c\tau_j,W)$. If we had no cuts imposed on our sample, this would just be $f(ct_{i};c\tau_j) \sim {\rm exp}(-ct_i/c\tau_j)$. Application of the cuts biases the sample as we have seen in Sec.~\ref{ssec:naive}, and the bias is dependent on $c\tau_j$ as well as other factors like the mass of the LLP and its production mode, all of which we have denoted by $W$. The prescription for performing an unbinned likelihood analysis is to first fit the $d_T$ or $c\tau$ distributions for different $c\tau_j$ within a particular model and for the mass of the LLP as obtained from experiment, and find suitable PDFs corresponding to each $c\tau_j$. If all of them can be fitted well with a single function, say, $f(ct_i;\vec{\theta})$, where $\vec{\theta}$ is the set of parameters, then one might try fitting these parameters as functions of $c\tau_j$, $\vec{\Theta}(c\tau_j)$. Then Eq.~\eqref{eq:unbinned_L} can be written as:
\begin{equation}
L_j = \prod_{i=1}^{N_{obs}} f(ct_{i};\vec{\Theta}(c\tau_j))
\label{eq:unbinned_L_2}
\end{equation}
where the functions $\vec{\Theta}(c\tau_j)$ and $f(ct_{i};\vec{\Theta}(c\tau_j))$ are predetermined for a given model and LLP mass, and therefore the unbinned maximum likelihood estimator can be used in this case.
In order to check the robustness of our estimation against different choices of distribution functions, we fitted 80 different functional forms from the \texttt{Python} \texttt{fitter} package \cite{fitter}, assuming different decay lengths of a 100 GeV LLP, produced as discussed in Sec.~\ref{ssec:tracker}. 
While performing this exercise we found that for different $c\tau$ values, different functions were providing the best fit, for example, for $c\tau$ = 5 cm the $\beta$ or $\chi^2$ functions appear to be optimal whereas for $c\tau$ = 20 cm the generalised exponential function \footnote{$f(x;a,b,c)=(a+b(1-{\rm exp}(-cx))){\rm exp}(-ax-bx+\frac{b}{c}(1-{\rm exp}(-cx)))$, for $x>0$, $a,b,c>0$.} provides the best fit.
It is, therefore, difficult to generalise this procedure for all $c\tau$ values as well as for LLPs with different mass and production modes.

\begin{figure}
    \centering
    \includegraphics[width=0.49\textwidth]{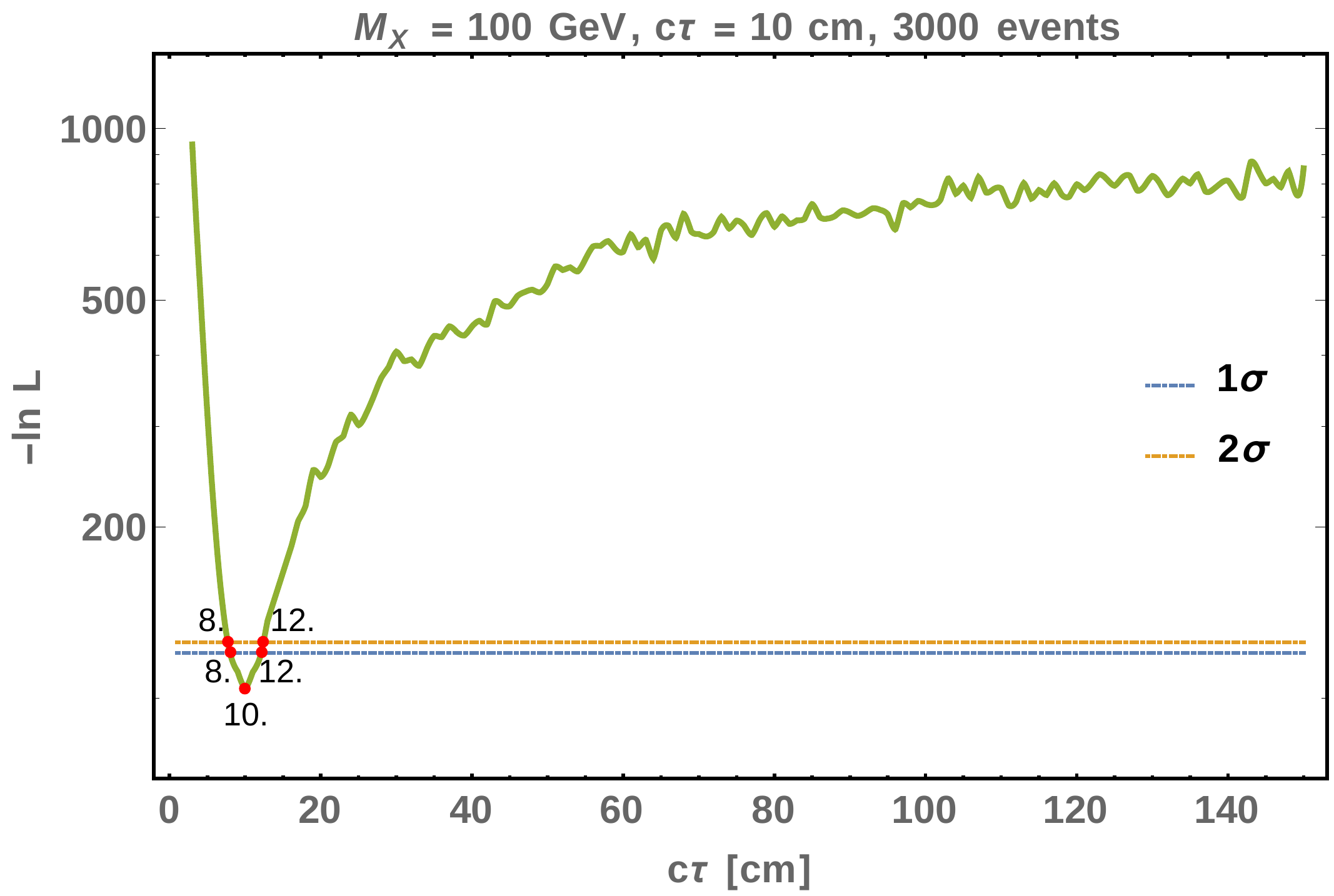}~
    \includegraphics[width=0.49\textwidth]{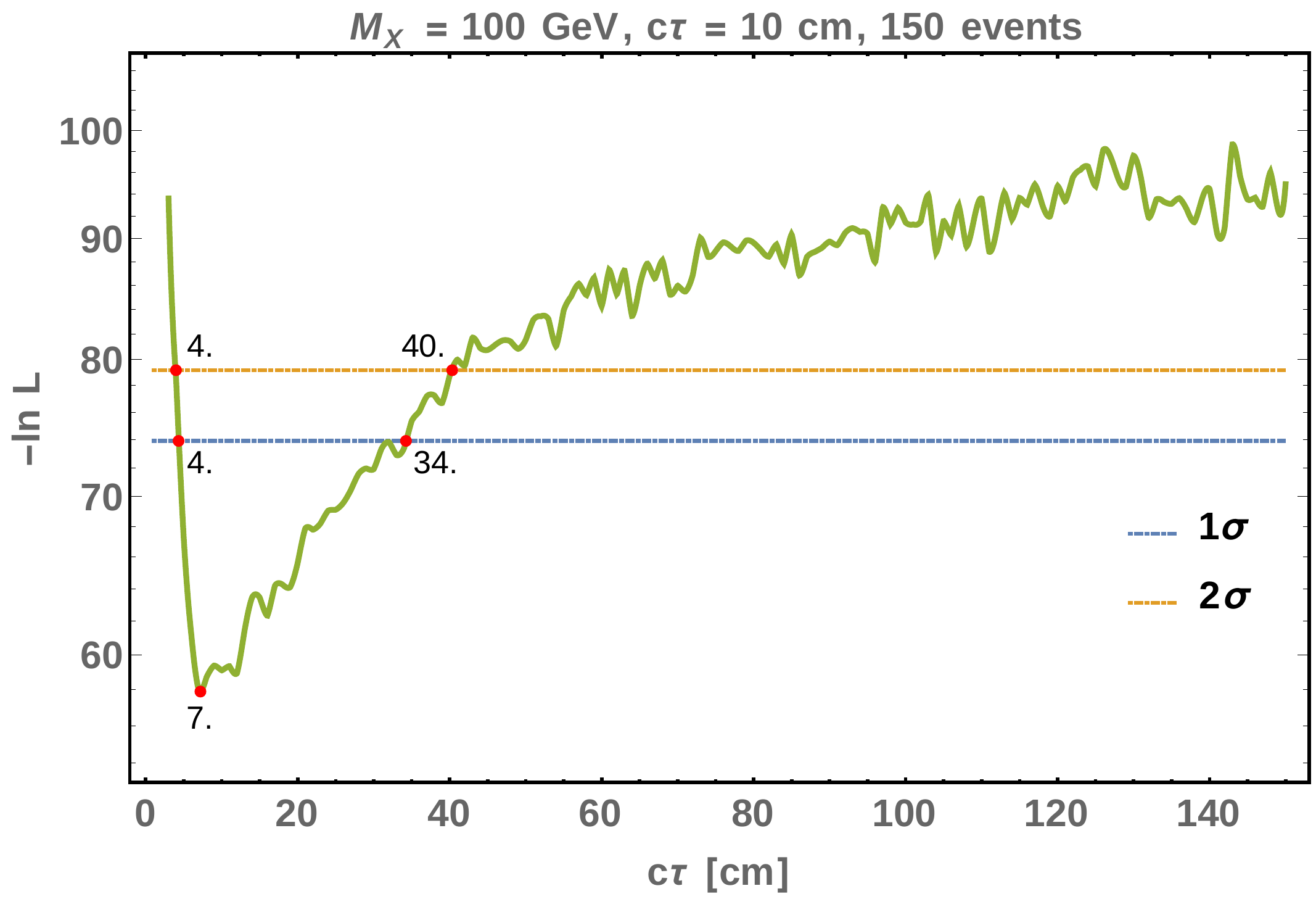}
    \caption{Model-dependent $-{\rm ln}L$ as a function of the reconstructed decay length $c\tau$ for LLP mass of 100 GeV and decay length 10\,cm with 3000 events ({\it left}) and 150 events ({\it right}), for the displaced lepton signature.
    }
    \label{fig:L_dep}
\end{figure}

Since the probability distribution function is unknown, we turn naturally to the binned likelihood analysis, where we can treat the number of events observed in each bin of the $d_T$ distribution as coming from a Poisson distribution with $\mu$ being the number of events expected from the theoretical simulation that we perform for different $c\tau$ values assuming a specific model and mass of LLP. It is defined as:
\begin{equation}
L = \prod_{i=1}^{N} \frac{n_{{\rm th},i}^{n_{{\rm exp},i}}}{n_{{\rm exp},i}!}{\rm exp}(-n_{{\rm th},i})
\label{eq:binned_L}
\end{equation}
where the symbols have the same meaning as described in Sec.~\ref{sssec:chi2_dep}. In Eq.~\eqref{eq:binned_L}, we are therefore using $N$ different Poisson PDFs with different expectation values coming from the theoretical simulations. As discussed in the case of unbinned likelihood, the value of $c\tau$ which maximises $L$ is the estimated lifetime of the LLP. One can also use $-2{\rm ln} L$ as the estimator, and then one has to minimise this quantity. For data following a Gaussian distribution, $\chi^2 = -2{\rm ln} L$ and $\Delta\chi^2 = 2\Delta {\rm ln}L= F^{-1}_{\chi^2 m}(1-\alpha) \Rightarrow \Delta {\rm ln}L = F^{-1}_{\chi^2 m}(1-\alpha)/2$ \cite[Chapter 40]{10.1093/ptep/ptaa104}, where $F^{-1}_{\chi^2 m}$ is the chi-square quantile for $m$ degrees of freedom. We use this to quote the 1$\sigma$ and 2$\sigma$ error bands on the lifetime estimated using the $-{\rm ln}L$ method. Fig.\ref{fig:L_dep} shows $-{\rm ln} L$ as a function of $c\tau$ for the benchmark with $M_X$=100 GeV and $c\tau$=10 cm, for two different cross-sections $-$ 1 fb (3000 events) and 0.05 fb (150 events). We find that the results are comparable to the estimates that we get from the $\chi^2$ analysis. However, as we discussed earlier, for a smaller number of observed events, the likelihood is probably a more reliable estimator.

\subsubsection{Lifetime estimation using machine learning based regression}

Instead of using a $\chi^2$ (or likelihood) analysis to estimate the lifetime, one can also employ machine learning-based (ML) regression techniques. To illustrate how these -- more sophisticated -- techniques work, we use the DNN based \texttt{KerasRegressor} and the \texttt{XGBRegressor}, both from \texttt{Scikit-learn} \cite{sklearn} to compare their outcome with the one obtained from our $\chi^2$ analysis for some of our benchmarks. We use the same format of input as for the model-dependent $\chi^2$ discussed above. We train the networks with the simulated $d_T$ distributions which are generated for a particular LLP mass within a specific model and for $c\tau$ values between 1 to 150 cm. The inputs are, therefore, the bin entries for 30 bins from 1-30 cm of the $d_T$ distribution normalised to the total number of observed events in the experiment. This set of 150 histograms from simulations are randomly split into training and validation sets, with 100 and 50 $c\tau$ distributions respectively, and we check that the network is not overfitting.
The trained network is then used to predict the decay length of the benchmark points from their $d_T$ distributions.
We have seen in the previous Sections that addition of PU does not significantly affect our results for the case of displaced leptons. In order to train ML-based regressors it is better to use a larger sample size in the simulations in order to ensure that the histograms that are used for training are smooth, and since generating the process with PU is both resource and time consuming, we ignore PU for a simple illustration of the ML techniques for lifetime estimation, and generate 50 000 events (100 000 pairs of LLPs) to obtain the training data set. The testing is done with LLP samples having a cross-section of 1.67 fb (around 5000 events by the end of HL-LHC).

In the \texttt{XGBRegressor} case, we employ a \texttt{GradientBoostingRegressor} with the following parameters:

\begin{center}
\texttt{n\_estimators=2500, max\_depth=8, learning\_rate=0.1,} \\
\texttt{min\_samples\_leaf=5, min\_samples\_split=5}.
\end{center}

Using the quantile regression presented in \cite{grad-boost}, and taking as an example our benchmark of a 100 GeV LLP with $c\tau=50$ cm, the reconstructed central value is $63$, the $2\sigma$ lower limit is $47$, and the 2$\sigma$ upper limit is $116$ from the \texttt{XGBRegressor}. We can, hence, see that the performance is reasonably comparable to the $\chi^2$ analysis. 

For the DNN based regressor (\texttt{KerasRegressor} \cite{keras}), we have used two fully connected hidden layers with 30 and 32 nodes respectively, with RELU activation function. The loss of the network is taken to be the mean squared error and ADAM optimiser has been used with its default learning rate of 0.001. We find that the lifetime estimates are $10$ and $57$ for the LLP benchmarks with $M_X=100$ and $c\tau=$10 cm and 50 cm respectively. 

In both cases, we should point out that we have performed a naive optimization of the hyperparameters based on trial-and-error and further optimization may be possible. Such a study is, however, clearly beyond the scope of the present work.

The size of the training and validations samples might seem small compared to those used in other problems, like jet tagging. However, even with this smaller training sample, we observed that the ML-based regressors could perform a very good fit, and the predictions are close to the expected results. This is due to the fact that the lifetime depends on very few parameters and, therefore, the regressors can learn the variation in the $d_T$ distribution with $c\tau$ even with a smaller sample.

\subsubsection{$\chi^2$ fitting of $\beta_T\gamma c\tau$ distribution: model-independent analysis}
\label{sssec:chi2_indep}

\begin{figure}
    \centering
    \includegraphics[width=0.6\textwidth]{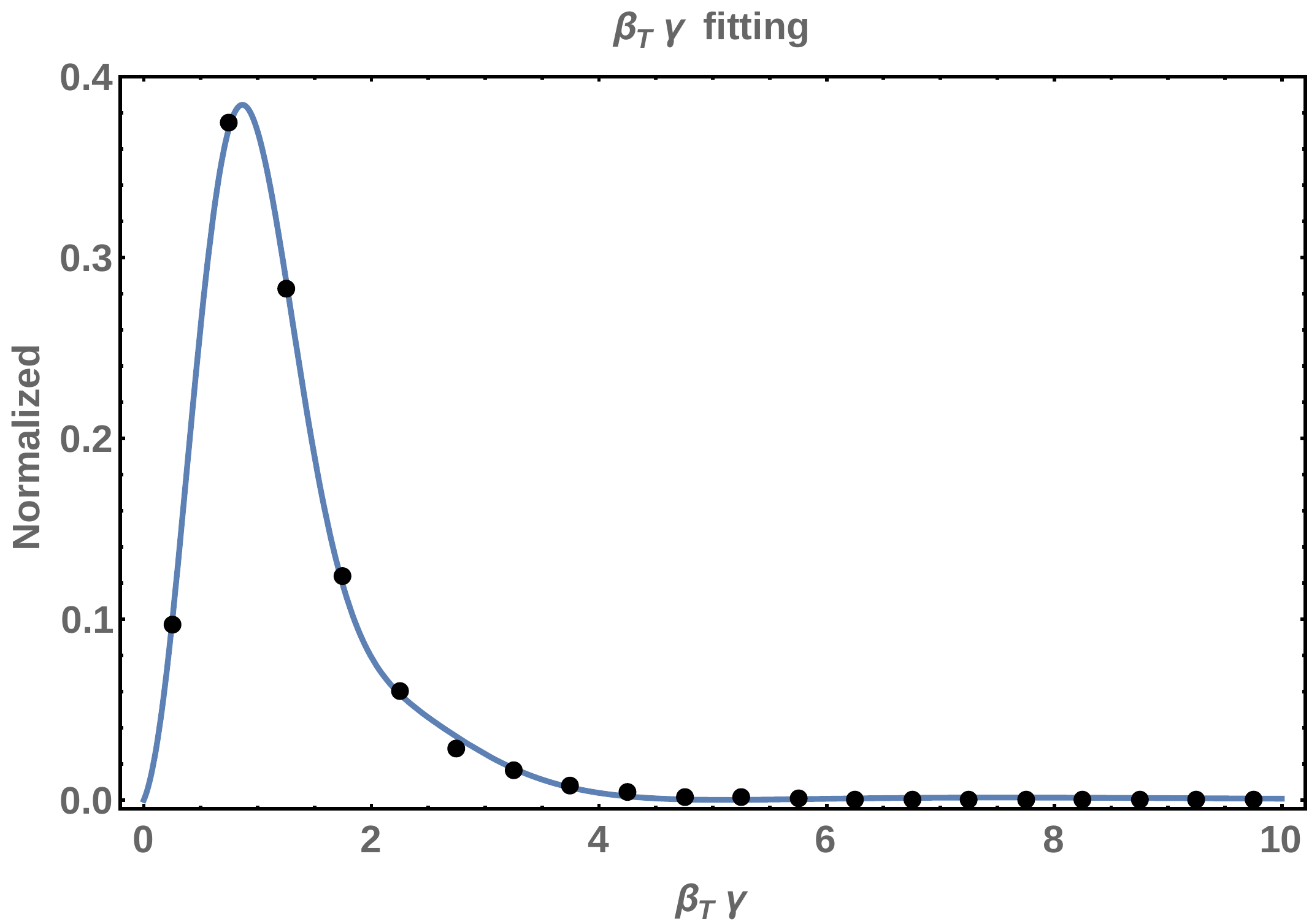}
    \caption{Fit of the normalized $\beta_T\gamma$ distribution for the benchmark $(m_X, c\tau) = (100{\rm~GeV},10{\rm~cm})$ using a function of the form: \small{$A~e^{-x}x-B~e^{-x^2}x-C~e^{-x}x^2+D~e^{-x^2}x^2-E~e^{-x^2}x^3+F~e^{-x}x^4$}}
    \label{fig:bgfit1}
\end{figure}

Let us now turn to our model-independent method which, however, requires certain additional experimental information. As we already mentioned in Sec.~\ref{ssec:realistic}, here we will assume that the $\beta_T \gamma$ and $d_T$ distribution can be extracted from experiment. This distribution will be fitted with a suitable function which, if treated as a probability density function, can be used to generate a large number of random values for $\beta_T\gamma$. In Figure \ref{fig:bgfit1} we show the normalised $\beta_T\gamma$ distribution we obtain for benchmark $(m_X, c\tau) = (100{\rm~GeV},10{\rm~cm})$, along with the corresponding fit. Moreover, we employ an additional function of the form $-c\tau \ln U[r] $, where $U[r]$ generates a random number distributed uniformly between $0$ and $1$, to generate exponential lifetime distributions with different values for $c\tau$. Multiplying the two sets of random numbers, we obtain a $d_T$ distribution for various combinations of $\beta_T\gamma$ and $c\tau$. This $d_T$ distribution can then be used to perform a $\chi^2$ analysis similar to the one described in Sec.~\ref{sssec:chi2_dep} and obtain an estimate of the LLP lifetime. In this case, no specific model assumption is needed, since the knowledge of the $\beta_T\gamma$ distribution encapsulates all the necessary model information. Therefore, this is indeed a model-independent approach, provided that the information on $\beta_T \gamma$ is experimentally accessible. 

\begin{figure}
    \centering
    \includegraphics[width=0.49\textwidth]{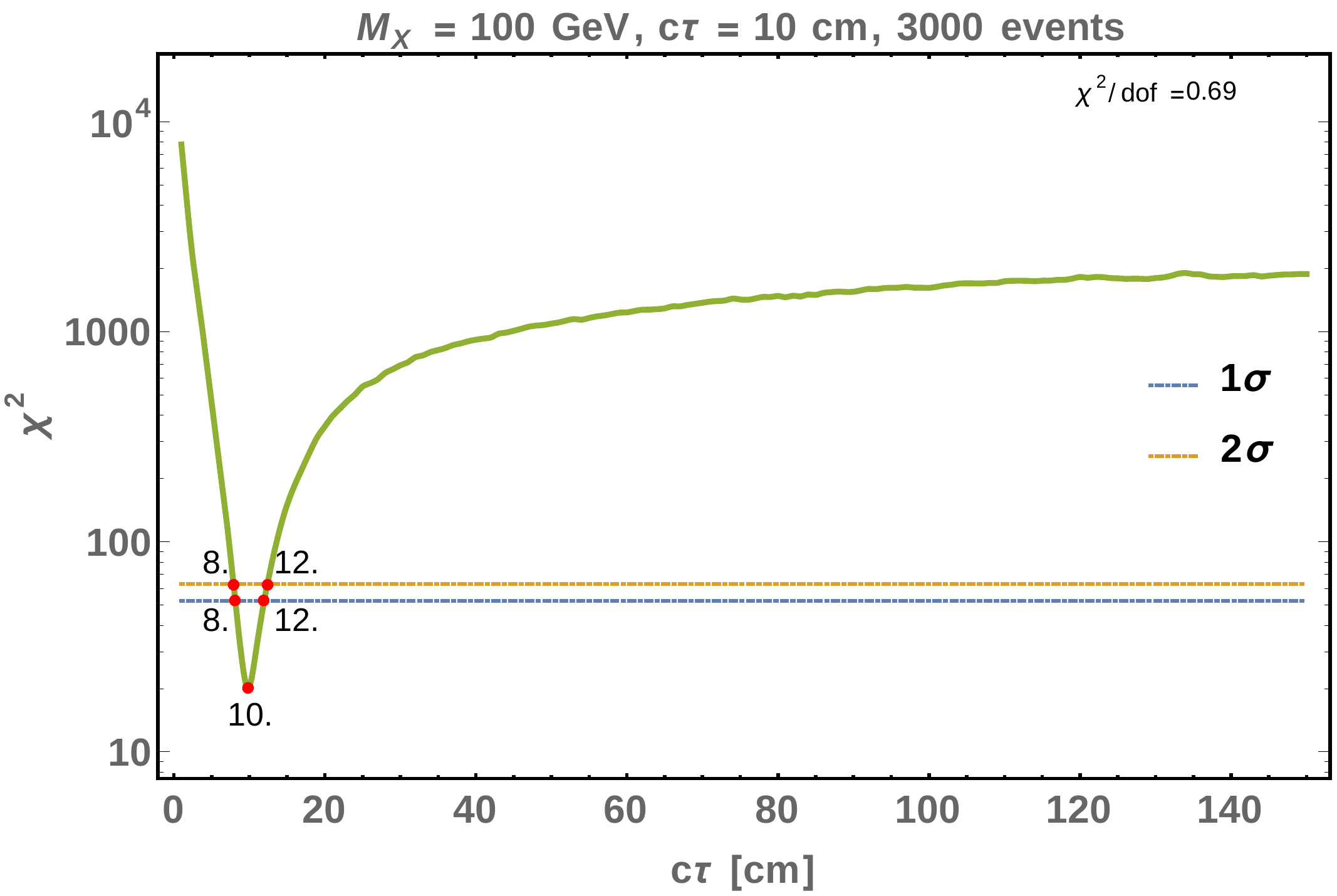}~
    \includegraphics[width=0.49\textwidth]{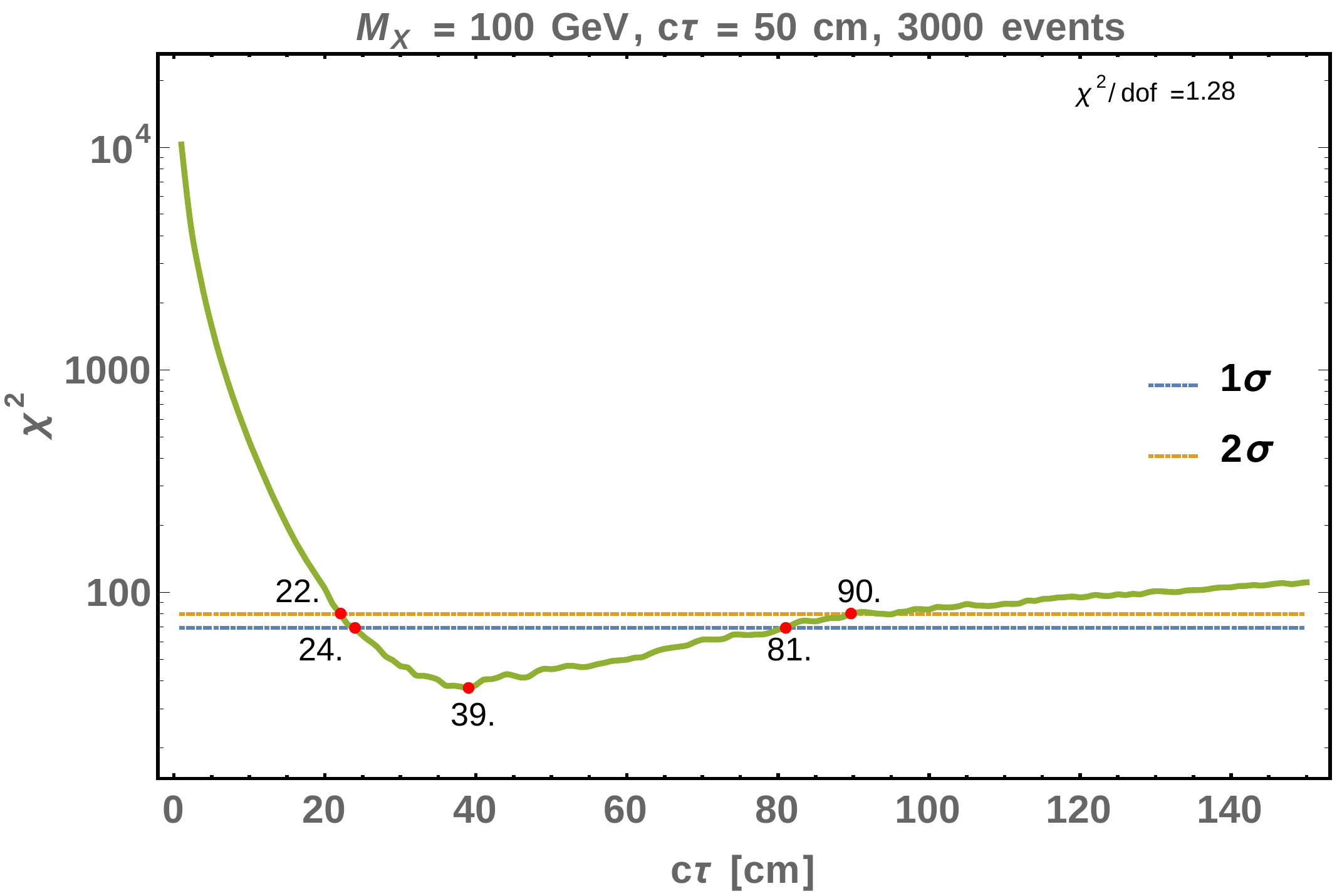} \\[2mm]
    \includegraphics[width=0.49\textwidth]{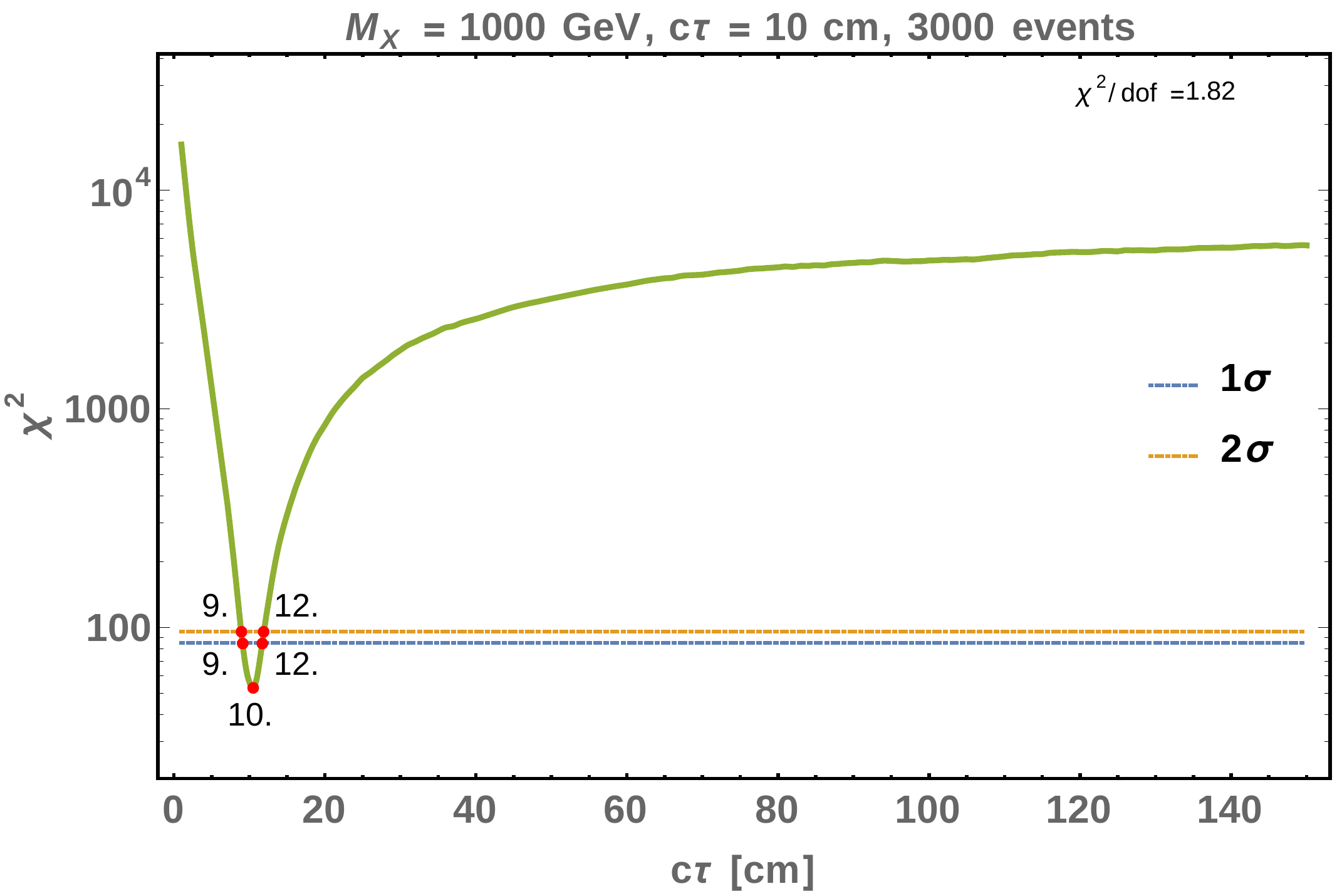}~
    \includegraphics[width=0.49\textwidth]{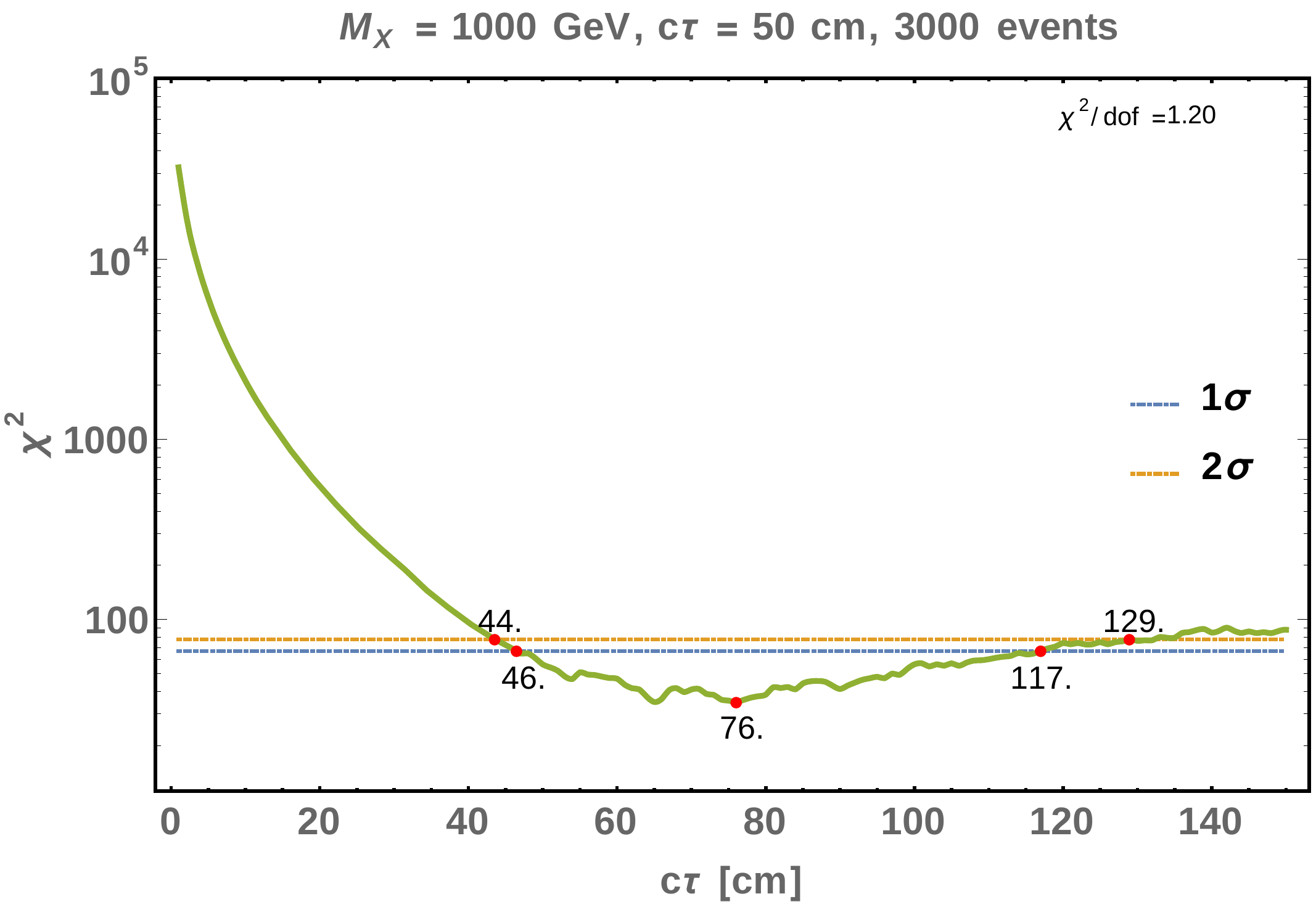}
    \caption{Model-independent $\chi^2$ as a function of the reconstructed decay length $c\tau$ for LLP masses of 100 GeV (top) and 1000 GeV (bottom) and decay lengths of 10\,cm (left) and 50\,cm (right). 
    }
    \label{fig:chi2_indep}
\end{figure}

\begin{table}
    \centering
    \begin{tabular}{|c|c|c||c||c|c||c|c|}
    \hline
    $M_X$ & DL & cross-section & Rec.\ DL & $1\sigma$ LL & $1\sigma$ UL & $2\sigma$ LL & $2\sigma$ UL \\
    (GeV) & (cm) & (fb) & (cm) & (cm) & (cm) & (cm) & (cm) \\
    \hline\hline
    \multirow{3}{*}{100} & \multirow{3}{*}{10} & 1 & 10 & 8 & 12 & 8 & 12\\
    & & 0.1 & 10 & 5 & 23 & 5 & 28\\
    & & 0.05 & 8 & 4 & 28 & 3 & 38\\    
    \hline
    \multirow{3}{*}{100} & \multirow{3}{*}{50} & 1 & 39 & 24 & 81 & 22 & 90\\
    & & 0.1 & - & 10 & $>$150 & 9 & $>$150 \\
    & & 0.05 & - & 6 & $>$150 & 5 & $>$150 \\
    \hline\hline
    \multirow{3}{*}{1000} & \multirow{3}{*}{10} & 1 & 10 & 9 & 12 & 9 & 12\\
    & & 0.1 & 10 & 6 & 16 & 6 & 18 \\
    & & 0.05 & 10 & 6 & 21 & 5 & 24 \\    
    \hline
    \multirow{3}{*}{1000} & \multirow{3}{*}{50} & 1 & 76 & 46 & 117 & 44 & 129\\
    & & 0.1 & - & 23 & $>$150 & 20 & $>$150 \\
    & & 0.05 & - & 10 & $>$150 & 9 & $>$150 \\    
    \hline
    \end{tabular}
    \caption{Lifetime estimates by model-independent $\chi^2$ fitting of the $d_T$ distribution (as shown in Figure \ref{fig:chi2_dep}) for the displaced leptons signature assuming different combinations of the LLP mass $M_X$ and decay length (DL). We display the reconstructed decay length as well as the corresponding lower (LL) and upper (UL) limits at the $1\sigma$ and $2\sigma$ confidence level, respectively. }
    \label{tab:chi2_indep}
\end{table}

Figure \ref{fig:chi2_indep} shows the $\chi^2$ distribution obtained through this method as a function of the reconstructed decay length for the same four benchmark configurations as in Sec.~\ref{sssec:chi2_dep}, each having 1 fb cross-section. Table \ref{tab:chi2_indep} shows the reconstructed decay length and the $1\sigma$ and $2\sigma$ lower and upper limits for each scenario for all the three assumed cross-sections. Similar to the model-dependent $\chi^2$ analysis presented previously, the number of degrees of freedom has been taken to be $dof = N-1 = 29$.

Our findings show that for LLPs characterized by relatively short lifetimes or heavy LLPs, through this method it is possible to reconstruct the mean decay length with a comparable precision as when knowledge of the underlying model is assumed. For longer lifetimes and lower masses, and within the lifetime interval that we considered, we could only infer a lower limit from this $\chi^2$ analysis as can seen in the upper right panel of Figure \ref{fig:chi2_indep}. Again with increasing mass, the lifetime estimation improves. These results are to be expected: as the mean proper lifetime and/or boost of the LLPs increases, the number of decays occurring within a radial distance of $30$ cm from the beam line decreases. This means that our sampling of the $\beta_T \gamma$ distribution carries larger uncertainties which, in turn, reflect upon our capacity to reconstruct the LLP lifetime. In a sense, a model-dependent fit corresponds to the limit at which the $\beta_T \gamma$ distribution is known with infinite precision. Note also that the estimated lifetime using this method tends to be on the higher side because the $\beta_T\gamma$ distribution is affected by the cuts, as shown in fig.\ \ref{fig:cuts_effect}, and is biased towards lower values. Consequently, larger $c\tau$ values are favoured in order to match the experimental $d_T$ distribution. 

Other estimators, like the binned likelihood and ML-based regressors, as discussed for the model-dependent approach, can also be employed in the model-independent case. We do not repeat the analysis here with these other estimators, since the procedure remains similar -- in this case, instead of comparing the $d_T$ distributions from a particular theoretical model with varying lifetimes with the experimentally observed $d_T$, one would compare the latter with $d_T$ distributions obtained from the product of experimentally observed $\beta_T\gamma$ and the exponential $c\tau$ distributions for various lifetimes.

\subsection{Displaced leptons with missing transverse energy}
\label{ssec:lep_met}

We now consider the 3-body decay of a neutral long-lived particle (LLP) $X$ into two leptons along with an invisible particle $Y$, 
\begin{equation*}
    X ~\rightarrow~ \ell^+ \ell^- Y.
\end{equation*}
The presence of more than one lepton implies that the position of the secondary vertex (SV) can be identified. But in this class of LLP decays, measuring the $\beta\gamma$ of $X$ is more challenging since not all the decay products can be reconstructed. Hence, we need a lifetime estimation method which does not rely on the knowledge of the $\beta\gamma$ information of the LLP. To the best that we can think of, the only option in order to reconstruct the LLP lifetime in this class of decay modes is a model-dependent analysis.

An additional complication arises due to the lack of knowledge concerning the LLP mass. One possibility is to employ the dilepton invariant mass edge, which is determined by the difference between the mass of the LLP and that of the invisible particle
\be
    M^{\rm edge}_{\ell\ell} ~=~ m_{X}-m_{Y} ~=~ \Delta
    \label{eq:mass_edge}
\ee
and to assume that particle $Y$ is massless. We can then indeed estimate the mass of $X$ from the edge of the dilepton invariant mass distribution. After determining the mass, we can follow the same procedure as in the case of displaced leptons. 

\begin{figure}
    \centering
    \includegraphics[scale=0.3]{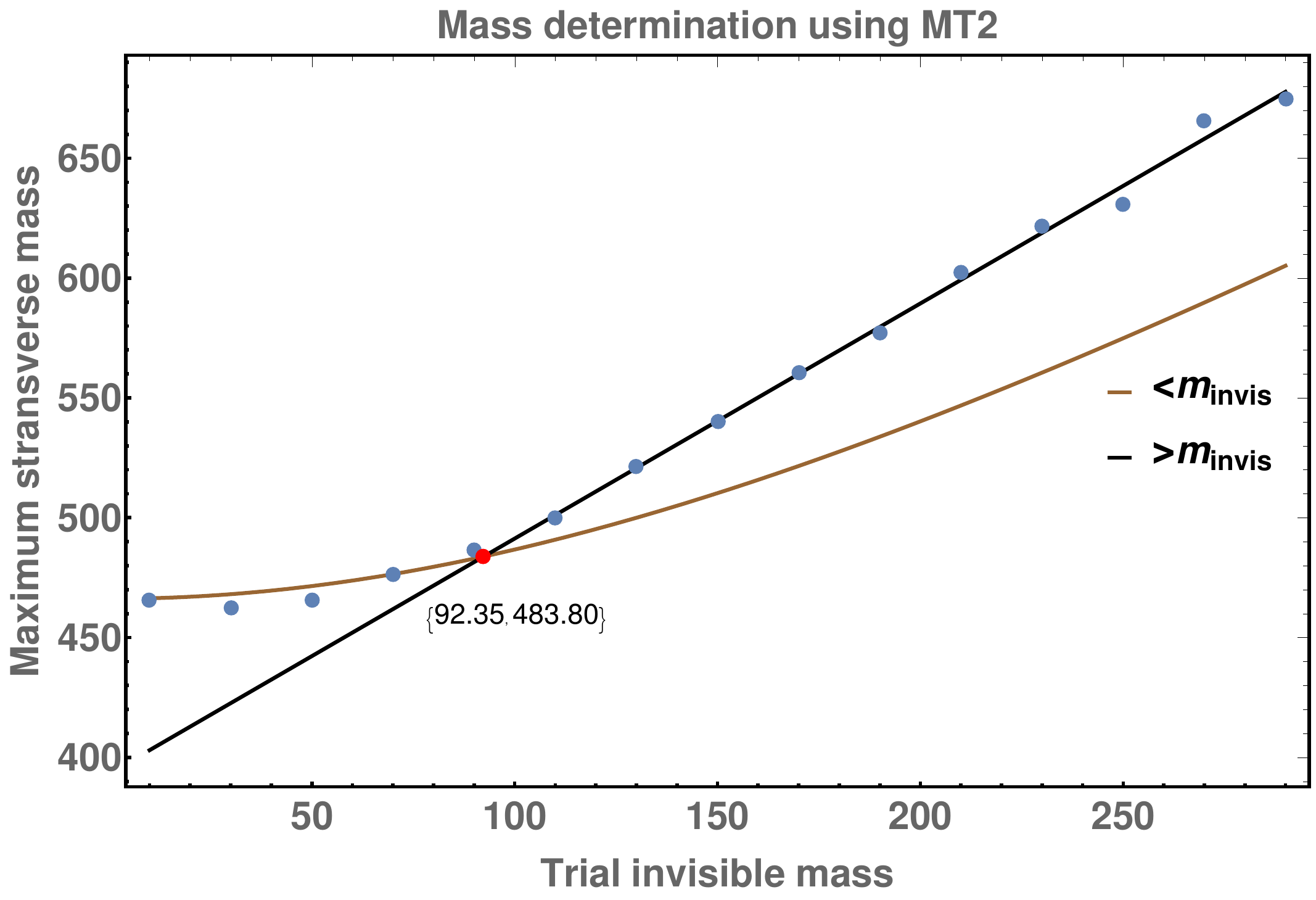}\\
    \includegraphics[scale=0.3]{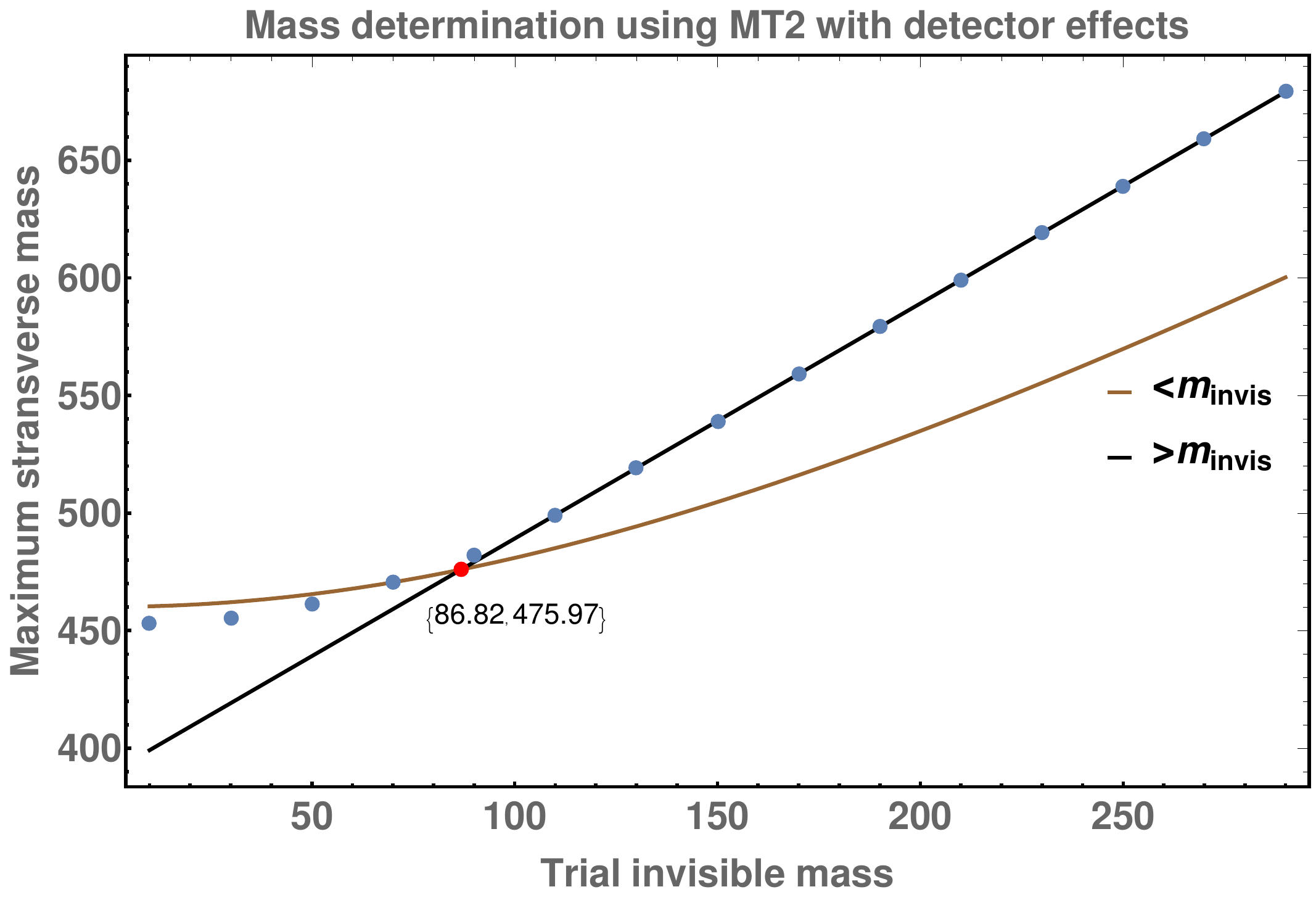}    
    \caption{Mass determination of the LLP and the invisible particle using the stransverse mass variable $-$ at the parton-level with 10\,000 events ({\it top}) and at the reconstruction level with 1000 events ({\it bottom}) for cases where the intermediate particle is off-shell.}
    \label{fig:mt2_mass}
\end{figure}

The massless invisible particle assumption can be avoided by employing the stransverse mass (MT2) variable to find out the masses of the mother particle as well as its invisible decay product as has been shown for the case of gluino decaying to neutralino and jets in Ref.\ \cite{Cho:2007qv}. The transverse mass of a gluino is given as
\begin{equation}
    m_T^2(m_{T,vis}, m_Y,\textbf{p}_T^{vis},\textbf{p}_T^Y) =m^2_{T,vis}+m^2_Y+2(E^{vis}_TE^Y_T-\textbf{p}^{vis}_T\cdot \textbf{p}^Y_T)
    \label{eq:transverse_mass}
\end{equation}
where $m_{T,vis}$ and $\textbf{p}^{vis}_T$ are the transverse invariant mass and transverse momentum of the visible system, respectively, while $m_Y$ and $\textbf{p}^Y_T$ are the assumed mass and transverse momentum of the invisible system, respectively. Each event will involve two such LLP decays and the stransverse mass variable (MT2) is defined as
\begin{equation}
    m^2_{T2} ~\equiv~ 
    \underset{\textbf{p}^{Y(1)}_T+\textbf{p}^{Y(2)}_T=\textbf{p}^{miss}_T}
    {\text{min}}\Big\{ \text{max}\big\{ m^{2(1)}_T, m^{2(2)}_T\big\} \Big\} \,,
    \label{eq:stransverse_mass}
\end{equation}
where the maximum transverse mass of the two LLPs in each event is minimised over all possible values of $\textbf{p}^{Y(1)}_T$ and $\textbf{p}^{Y(2)}_T$ such that they always satisfy $\textbf{p}^{Y(1)}_T + \textbf{p}^{Y(2)}_T = \textbf{p}^{miss}_T$. The edge of the transverse mass distribution (i.e.\ $m_{T2}^{max}$) gives the value of the LLP mass only if the correct mass of the invisible particle is used in Equation \eqref{eq:transverse_mass}. Otherwise, $m_{T2}^{max}$ has a different functional dependence on $m_Y$ depending on whether its value is smaller or greater than the actual invisible particle mass. The two functions, however, intersect at the invisible particle mass. As it has been shown in Ref.\ \cite{Cho:2007qv}, this feature can be used in order to deduce the mass of the LLP. In Figure \ref{fig:mt2_mass} we show the variation of the maximum stransverse mass with varying trial masses for the invisible particle fitted with two different functions and how the intersection of these functions can provide an estimate of both the LLP and invisible particle's masses. 

The reconstructed masses are $483.8\pm4.57~\text{GeV}$ and $92.35\pm8.62~\text{GeV}$ at parton-level with 10\,000 events for a $500~\text{GeV}$ LLP decaying into a $100~\text{GeV}$ invisible particle. 
Note that here we have performed a very simplistic analysis without considering any detector effects, just to illustrate that it is, indeed, possible to obtain at least a ballpark estimate of the particle masses even for signatures including missing transverse energy (even in the absence of timing). 
Still, in oder to obtain an estimate of the impact of detector effects, we repeat the analysis using \texttt{Delphes-3.4.2} with the CMS card. In the {\it bottom panel} of Fig.~\ref{fig:mt2_mass} we show the reconstruction of both the LLP and invisible particle mass using MT2 with the detector level missing transverse energy. The reconstructed masses are $475.97\pm5.13~\text{GeV}$ and $86.82\pm5.13~\text{GeV}$, assuming 1000 events. We find that even with this number of events, this method can provide a reasonable estimate of the two masses involved in the process at the reconstruction level.

With the LLP and invisible masses at hand, and within the framework of a specific model, we can then perform a model-dependent $\chi^2$ analysis to reconstruct the LLP lifetime. It should be noted that the same method can also be applied to the case in which the LLP decays to two leptons and an invisible particle through an on-shell intermediate particle. For further details of this analysis the reader is referred to Ref.\ \cite{Cho:2007qv}.

\begin{table}
    \centering
    \begin{tabular}{|c|c|c||c||c|c||c|c|}
    \hline
    $M_X$ & DL & cross-section & Rec.\ DL & $1\sigma$ LL & $1\sigma$ UL & $2\sigma$ LL & $2\sigma$ UL \\
    (GeV) & (cm) & (fb) & (cm) & (cm) & (cm) & (cm) & (cm) \\
    \hline\hline
    \multirow{3}{*}{100} & \multirow{3}{*}{10} & 1 & 9 & 6 & 14 & 6 & 15\\
    & & 0.1 & 9 & 3 & $>$150 & 3 & $>$150\\
    & & 0.05 & 5 & 2 & $>$150 & 2 & $>$150\\    
    \hline
    \multirow{3}{*}{100} & \multirow{3}{*}{50} & 1 & 34 & 11 & $>$150 & 10 & $>$150\\
    & & 0.1 & - & 2 & $>$150 & 2 & $>$150\\
    & & 0.05 & - & 1 & $>$150 & 1 & $>$150\\
    \hline\hline
    \multirow{3}{*}{1000} & \multirow{3}{*}{10} & 1 & 10 & 9 & 11 & 8 & 12\\
    & & 0.1 & 11 & 7 & 18 & 6 & 20\\
    & & 0.05 & 9 & 5 & 21 & 5 & 24\\    
    \hline
    \multirow{3}{*}{1000} & \multirow{3}{*}{50} & 1 & 47 & 32 & 88 & 30 & 93\\
    & & 0.1 & - & 18 & $>$150 & 16 & $>$150\\
    & & 0.05 & - & 11 & $>$150 & 9 & $>$150\\    
    \hline
    \end{tabular}
    \caption{Lifetime estimates by model-dependent $\chi^2$ fitting of $d_T$ distribution for displaced leptons plus $\met$ final state.}
    \label{tab:chi2_dep_lepsmet}
\end{table}

In Table \ref{tab:chi2_dep_lepsmet} we present the reconstructed lifetime values along with their $1\sigma$ ($68\%$ CL) and $2\sigma$ ($95\%$ CL) lower and upper limits for the GMSB model presented in Sec.~\ref{sec:diff_decay} for all the benchmarks considered with three different cross-sections. We observe that the true lifetime can be reconstructed with a precision of $40\%$ (at $68\%$ CL) for small masses and lifetimes, improving to roughly $15\%$ for heavier (\textit{i.e.} less boosted) LLPs. For longer lifetimes the latter number translates to roughly $40\%$, whereas for a light LLP with a longer lifetime we could only infer a lower limit on $c\tau$ within the considered interval.

\subsection{Charged LLP decaying into lepton and invisible particle}
\label{ssec:chargeLLP}

The last case we consider is that of a charged LLP decaying into a lepton and an invisible particle inside the tracker. If the mass difference between the charged LLP and the invisible particle is substantial, then the lepton will have sufficient transverse momentum and can be reconstructed, giving rise to a ``kinked'' track signature. If, on the other hand, the charged LLP and the invisible particle are degenerate in mass, the lepton will be too soft to be reconstructed, leading to a disappearing track in the Tracker. Here we will focus on the former case.

In the busy environment of the LHC, online triggering on a kinked track is challenging \cite{Evans:2016zau}, especially if the LLP decay occurs towards the outer parts of the Tracker system. However, as stated in Ref.~\cite{Evans:2016zau}, off-line reconstruction of this kink could be attempted, which would then provide the position of the SV with some uncertainty. Moreover, from the track of the charged LLP we can calculate its momentum, while the rate of energy loss due to ionisation (\textit{i.e.} the LLP's $dE/dx$) can be used to estimate its mass. Then, it is -- at least in principle -- possible to retrieve all the information that is necessary in order to reconstruct the lifetime in a similar manner as we did for displaced leptons, and we can use any of the alternatives to estimate the lifetime.

Note that until now we have not discussed issues related to the efficiency with which displaced objects can be detected. Given the exceptionally challenging nature of the kinked tracks, however, it is important to try and estimate, even in a crude manner, the efficiency of reconstructing such a signature in the first place. To this goal, in what follows we will assume that the probability to reconstruct a kinked track can be expressed as a convolution of three factors: first, the efficiency to identify the charged LLP track. This can be typically identified with about $95\%$ efficiency if the LLP travels a distance of at least 12 cm before decaying, as shown in Ref. \cite{ATL-PHYS-PUB-2019-011}. Secondly, the efficiency of identifying the (displaced) lepton track. We take this to be identical as for ordinary displaced leptons, and borrow it from Ref.\ \cite{ATL-PHYS-PUB-2017-014}. Finally, in order to be able to disentangle the two tracks, we also demand that the angular separation ($\Delta R$) between the LLP and the lepton should be greater than 0.1 radian, so that the kink is prominent.

\subsubsection{Model-dependent $\chi^2$ analysis}

With the previous remarks in mind, we first perform a model-dependent $\chi^2$ analysis. In Table \ref{tab:chi2_dep_kink} we show the reconstructed decay length values for our four benchmarks along with the $1\sigma$ and $2\sigma$ lower and upper limits on the LLP lifetime for each scenario.

\begin{table}
    \centering
    \begin{tabular}{|c|c|c||c||c|c||c|c|}
    \hline
    $M_X$ & DL & cross-section & Rec.\ DL & $1\sigma$ LL & $1\sigma$ UL & $2\sigma$ LL & $2\sigma$ UL \\
    (GeV) & (cm) & (fb) & (cm) & (cm) & (cm) & (cm) & (cm) \\
    \hline\hline
    \multirow{3}{*}{100} & \multirow{3}{*}{10} & 1 & 11 & 9 & 15 & 9 & 15\\
    & & 0.1 & 11 & 5 & 30 & 4 & 40\\
    & & 0.05 & 9 & 3 & 45 & 3 & 100\\    
    \hline
    \multirow{3}{*}{100} & \multirow{3}{*}{50} & 1 & 44 & 22 & $>$150 & 20 & $>$150\\
    & & 0.1 & 28 & 6 & $>$150 & 5 & $>$150\\
    & & 0.05 & - & 2 & $>$150 & 1 & $>$150\\
    \hline\hline
    \multirow{3}{*}{1000} & \multirow{3}{*}{10} & 1 & 10 & 8 & 12 & 8 & 12\\
    & & 0.1 & 8 & 4 & 14 & 4 & 15\\
    & & 0.05 & 8 & 3 & 19 & 3 & 23\\    
    \hline
    \multirow{3}{*}{1000} & \multirow{3}{*}{50} & 1 & 43 & 30 & 90 & 29 & 93\\
    & & 0.1 & 31 & 12 & $>$150 & 11 & $>$150\\
    & & 0.05 & - & 7 & $>$150 & 6 & $>$150\\    
    \hline
    \end{tabular}
    \caption{Lifetime reconstruction through a model-dependent $\chi^2$ fit of the $d_T$ distribution for kinked tracks.}
    \label{tab:chi2_dep_kink}
\end{table}

We see that a lifetime $c\tau = 10$ cm can be reconstructed with a precision of $\sim 15\%$ ($65\%$ CL) for a $100$ GeV LLP, which turns to $10\%$ for a $1$ TeV particle. As expected, the precision decreases as $c\tau$ increases but, for $c\tau = 50$ cm we can still obtain results within a rough factor of 2.

\subsubsection{Model independent $\chi^2$ analysis}

Let us now move to our model-independent analysis. One important point to note here is that the transverse decay length distribution as obtained from experiment is expected to be biased towards lower values because of the dependence of the displaced lepton track reconstruction efficiency on the decay length - the efficiency decreases as $d_T$ increases. Since, however, these efficiencies \textit{are} known, it should be possible to unfold the experimental $d_T$ distribution accordingly and then compare with the distributions that we obtain using the product of various $c\tau$ distributions with the fitted $\beta\gamma$ distribution\footnote{A similar remark also applies to the displaced leptons and displaced jets analyses of the previous Sections.}. However, for the unfolding to work in experiment, one needs to not only know the efficiencies associated with tracking, but also the uncertainties associated with these efficiencies very well.
Quantifying these experimentally is quite an arduous job and this will affect the lifetime estimates and the sensitivities. Since we do not know the uncertainties yet, we perform the analysis assuming that we know them precisely well.

In Table \ref{tab:chi2_indep_kink} we present the reconstructed lifetime value for each of our benchmarks, along with their $1\sigma$ ($68\%$) and $2\sigma$ ($95\%$) lower and upper limits.

\begin{table}
    \centering
    \begin{tabular}{|c|c|c||c||c|c||c|c|}
    \hline
    $M_X$ & DL & cross-section & Rec.\ DL & $1\sigma$ LL & $1\sigma$ UL & $2\sigma$ LL & $2\sigma$ UL \\
    (GeV) & (cm) & (fb) & (cm) & (cm) & (cm) & (cm) & (cm) \\
    \hline\hline
    \multirow{3}{*}{100} & \multirow{3}{*}{10} & 1 & 9 & 8 & 12 & 8 & 12\\
    & & 0.1 & 10 & 6 & 21 & 5 & 25\\
    & & 0.05 & 10 & 5 & 39 & 4 & 55\\    
    \hline
    \multirow{3}{*}{100} & \multirow{3}{*}{50} & 1 & 43 & 25 & 80 & 24 & 87\\
    & & 0.1 & 42 & 9 & $>$150 & 8 & $>$150\\
    & & 0.05 & 20 & 4 & $>$150 & 4 & $>$150\\
    \hline\hline
    \multirow{3}{*}{1000} & \multirow{3}{*}{10} & 1 & 10 & 9 & 11 & 9 & 11\\
    & & 0.1 & 9 & 6 & 16 & 6 & 18\\
    & & 0.05 & 8 & 4 & 18 & 4 & 21\\    
    \hline
    \multirow{3}{*}{1000} & \multirow{3}{*}{50} & 1 & 62 & 44 & 113 & 42 & 132\\
    & & 0.1 & - & 24 & $>$150 & 21 & $>$150\\
    & & 0.05 & - & 17 & $>$150 & 14 & $>$150\\    
    \hline
    \end{tabular}
    \caption{Lifetime reconstruction through a model-independent $\chi^2$ fit of the $d_T$ distribution for kinked tracks.}
    \label{tab:chi2_indep_kink}
\end{table}

We observe that for all of our benchmarks, it is possible to obtain a reasonable reconstruction of the LLP lifetime, with a precision that is comparable to the one obtained through the model-dependent analysis presented in the previous Section.

\subsection{Displaced jets}
\label{ssec:disp_jets}

Let us now move to the case of a neutral long-lived particle that decays into two quarks inside the tracker part of the detector. The observed LHC signature in this case consists of displaced jets. Since jets contain numerous charged particles, by extrapolating their tracks, it is possible to obtain the position of the secondary vertex quite accurately \cite{Chatrchyan:2014fea,ATL-PHYS-PUB-2019-013}. In our analysis, we will assume that the positions of the secondary vertices are known with high precision and we will study the reconstruction of the mother particle's, \textit{i.e.}\ the LLP's, $\beta\gamma$ from its decay products.

In the high PU scenario of HL-LHC, the displaced jets signature will get more affected than final states consisting of displaced leptons. The 140 vertices per bunch crossing can increase the jet multiplicity to very high values, even when a higher $p_T$ cut, and in this busy environment, it is difficult to identify the displaced jets coming from the LLPs. Considering narrow jets reduces the PU contribution and therefore, proves useful for identifying the LLP jets, since the latter deposits energy in smaller physical region, as has been discussed in Ref. \cite{Bhattacherjee:2020nno}. Also, the proposal to include the timing layer in the Phase-II upgrade of CMS (MTD) with a timing resolution of 30 ps, would help in bringing down the PU amount to the current PU amount, around 30-50 vertices per bunch crossing \cite{MTD}. Pile-up mitigation techniques, like PUPPI (PileUp Per Particle Identification), might also help, however, how these methods work in the HL-LHC environment and how well can one recover the displaced jets need separate studies, which is clearly beyond the scope of the present work. Here, we present the analysis assuming that complete removal of PU is possible. 
\texttt{Delphes}, by default, does not handle displaced objects properly as has been discussed in Ref. \cite{Bhattacherjee:2019fpt} due to the absence of the three-dimensional detector geometry and segmentation, and needs major modifications. Therefore, we have presented the analysis at \texttt{Pythia}-level here.

As in the displaced lepton case, we use \texttt{Pythia6} to generate events for pair-production of a long-lived particle $X$ and its eventual decay into quarks. We use the same set of cuts (EC) for the displaced jets as used for final states with displaced letpons, since these cuts are on the position of the secondary vertex to ensure that the displaced tracks can be reconstructed with good efficiency as motivated from the extent of large area tracking in ATLAS. We discuss the reconstruction of the boost of the LLP from displaced jets and show that the situation becomes less straightforward than in the case of displaced leptons due to several complications affecting jet reconstruction. First, the mismatch between the actual energy of the quarks and the one measured from the jet affects the reconstruction of the $\beta_T\gamma$ distribution of the LLP. Secondly, the reconstruction of jets as their displacement increases may introduce additional challenges at the LHC. Concerning the second issue, since we are restricting ourselves to decays occurring within 30 cm from the beamline, we don't expect much difficulty in reconstructing the displaced jets. However, the measured jet energy can be quite different than that of the initial quark coming from the LLP decay. This may, in particular, affect the model-independent analysis which crucially depends on the fitting of the transverse boost distribution of the LLP.

\subsubsection{Reconstructing $\beta_T\gamma$ of the LLP from displaced jets}
\label{sssec:reco_btg_jets}

\begin{figure}
    \centering
    \includegraphics[scale=0.5]{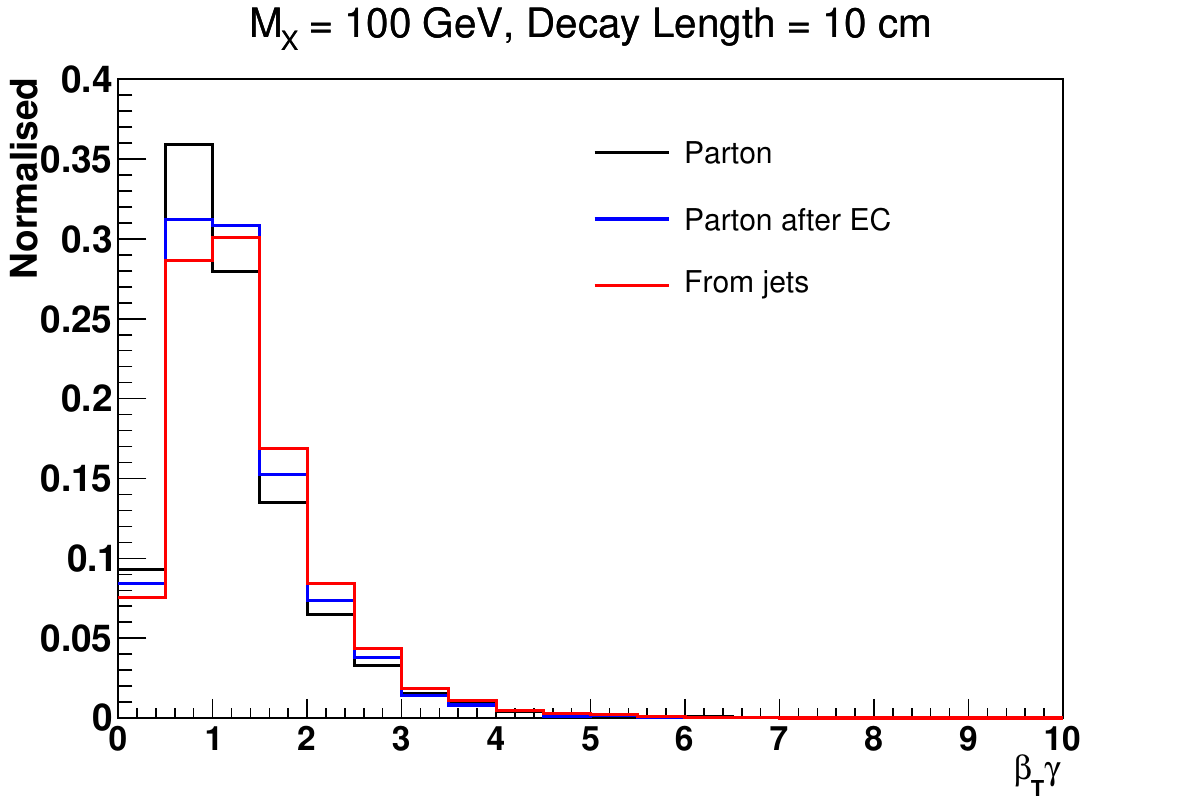}
    \caption{$\beta_T\gamma$ distribution of the LLP at the parton level before applying EC, after EC and that reconstructed from jets.}
    \label{fig:jets_btg}
\end{figure}

We cluster energy depositions with $p_T>5{\rm~GeV}$ and within pseudorapidity $|\eta|<4.9$ (taking into account both the barrel and the endcap regions) of charged particles which do not come from the primary vertex as well as all neutral particles\footnote{Charged particles coming from the primary vertex can be identified in the tracker and can, hence, be removed. Neutral particles coming from the primary vertex cannot be identified and hence can contribute to a jet's energy.} to build jets with a minimum transverse momentum of 20 GeV and $|\eta|<4.0$ (the HL-LHC pseudorapidity coverage of the tracker) using a cone of $R=0.4$~\footnote{$R=\sqrt{\eta^2+\phi^2}$}. We then identify charged particles coming from each secondary vertex and count how many of these particles are present in each jet. If a jet contains at least two particles from one secondary vertex and none from the other one, then that jet is associated with the former secondary vertex. If the number of jets associated with a particular secondary vertex is greater than two and if the invariant mass of all these jets falls within $40\%$ of the mass of the mother particle, which can be inferred from the peak of the invariant mass distribution, then we use all such jets to reconstruct the $\beta_T\gamma$ of the corresponding mother particle. In Fig.~\ref{fig:jets_btg} we show the $\beta_T\gamma$ distribution of the LLP at the parton level before and after applying the EC, along with the one reconstructed from the displaced jets. We can see that the shape of the reconstructed distribution is, indeed, modified, which we expect to affect the model-independent lifetime estimation. This effect is expected for jets due to many effects like out-of cone radiation, energy smearing and the segmentation in $\eta$-$\phi$ of the detector. In experiments, one usually applies jet energy correction factors to recover from these effects and get some matching between MC jets and detector jets, and this correction factor is a function of $p_T$ and $\eta$. However, for displaced jets, this correction factor will be a function of displacement in addition to the $p_T$ and $\eta$ of these jets. Moreover, the patterns of energy deposition for displaced and prompt jets also have some differences and again depend on displacement, as has been studied in Ref.~\cite{Banerjee:2017hmw,Bhattacherjee:2019fpt}. Once this is calibrated, and applied to the jets, we can expect to get closer to the correct $\beta_T\gamma$ distribution. In this work, we present our analysis without applying any jet energy correction.

We now move on to the LLP lifetime reconstruction, first within a model-dependent and afterwards within a model-independent framework.

\subsubsection{Reconstructing the lifetime: model-dependent $\chi^2$ analysis}
\label{sssec:reco_tau_dep}

Let us first assume that the underlying model is known\footnote{Relevant information can be inferred, for instance, by observing various distributions of the final state jets.}. Note also that the mass of the LLP can be reconstructed from the dijet invariant mass. Once we assume the model, we only need to know the position of the SV in order to perform a $\chi^2$ analysis as before. In the case of displaced jets, the SV can be identified more precisely than in the case of displaced leptons thanks to the greater number of charged particles involved in the process. 

\begin{table}
    \centering
    \begin{tabular}{|c|c|c||c||c|c||c|c|}
    \hline
    $M_X$ & DL & cross-section & Rec.\ DL & $1\sigma$ LL & $1\sigma$ UL & $2\sigma$ LL & $2\sigma$ UL \\
    (GeV) & (cm) & (fb) & (cm) & (cm) & (cm) & (cm) & (cm) \\
    \hline\hline
    \multirow{3}{*}{100} & \multirow{3}{*}{10} & 1 & 10 & 8 & 12 & 8 & 13\\
    & & 0.1 & 9 & 4 & 28 & 4 & 35\\
    & & 0.05 & 10 & 3 & $>$150 & 3 & $>$150\\    
    \hline
    \multirow{3}{*}{100} & \multirow{3}{*}{50} & 1 & 45 & 23 & $>$150 & 20 & $>$150\\
    & & 0.1 & - & 7 & $>$150 & 6 & $>$150\\
    & & 0.05 & - & 4 & $>$150 & 3 & $>$150\\
    \hline\hline
    \multirow{3}{*}{1000} & \multirow{3}{*}{10} & 1 & 10 & 8 & 11 & 8 & 11\\
    & & 0.1 & 8 & 6 & 15 & 5 & 16\\
    & & 0.05 & 8 & 4 & 16 & 4 & 18\\    
    \hline
    \multirow{3}{*}{1000} & \multirow{3}{*}{50} & 1 & 53 & 35 & 98 & 34 & 108\\
    & & 0.1 & 29 & 13 & $>$150 & 11 & $>$150\\
    & & 0.05 & - & 8 & $>$150 & 6 & $>$150\\    
    \hline
    \end{tabular}
    \caption{Lifetime estimates through a model-dependent $\chi^2$ fit of the $d_T$ distribution for displaced jets.}
    \label{tab:chi2_dep_jets}
\end{table}

Our results are presented in Table \ref{tab:chi2_dep_jets}, where we show the reconstructed decay lengths along with their $1\sigma$ and $2\sigma$ uncertainties for the same two LLP masses and decay lengths as in the displaced lepton case. We find that for a LLP of mass $100{\rm~GeV}$ and a decay length of $50{\rm~cm}$, through this type of analysis we can only infer a lower limit on the parent particle's proper decay length within the lifetime interval that we consider. For shorter decay lengths and/or heavier LLPs, however, the lifetime can be bounded both from above and from below with a precision of roughly $10\%-20\%$ at $1\sigma$.

\subsubsection{Reconstructing the lifetime: model-independent $\chi^2$ analysis}
\label{sssec:reco_tau_indep}

If the underlying model is unknown, much like in the displaced lepton case we need to assume that the $\beta_T\gamma$ distribution of the LLP can be measured experimentally. However, as we saw in Sec.~\ref{sssec:reco_btg_jets}, the $\beta_T\gamma$ distribution obtained from the final state displaced jets tends to deviate from the actual one. This will affect the lifetime estimation. In Table \ref{tab:chi2_indep_jets} we present the accuracy with which the LLP lifetime can be estimated in this framework.

\begin{table}
    \centering
    \begin{tabular}{|c|c|c||c||c|c||c|c|}
    \hline
    $M_X$ & DL & cross-section & Rec.\ DL & $1\sigma$ LL & $1\sigma$ UL & $2\sigma$ LL & $2\sigma$ UL \\
    (GeV) & (cm) & (fb) & (cm) & (cm) & (cm) & (cm) & (cm) \\
    \hline\hline
    \multirow{3}{*}{100} & \multirow{3}{*}{10} & 1 & 11 & 9 & 14 & 9 & 14\\
    & & 0.1 & 13 & 7 & 42 & 6 & 57\\
    & & 0.05 & 11 & 4 & 89 & 4 & $>$150\\    
    \hline
    \multirow{3}{*}{100} & \multirow{3}{*}{50} & 1 & 53 & 28 & $>$150 & 26 & $>$150\\
    & & 0.1 & 32 & 8 & $>$150 & 7 & $>$150\\
    & & 0.05 & - & 12 & $>$150 & 10 & $>$150\\
    \hline\hline
    \multirow{3}{*}{1000} & \multirow{3}{*}{10} & 1 & 13 & 12 & 15 & 12 & 16\\
    & & 0.1 & 11 & 8 & 17 & 7 & 19\\
    & & 0.05 & 11 & 6 & 21 & 6 & 24\\    
    \hline
    \multirow{3}{*}{1000} & \multirow{3}{*}{50} & 1 & - & 72 & $>$150 & 65 & $>$150\\
    & & 0.1 & - & 29 & $>$150 & 25 & $>$150\\
    & & 0.05 & - & 13 & $>$150 & 11 & $>$150\\    
    \hline
    \end{tabular}
    \caption{Lifetime estimates through a model-independent $\chi^2$ fit of the $d_T$ distribution for displaced jets.}
    \label{tab:chi2_indep_jets}
\end{table}

We find that the lifetime reconstruction is poor when we use the naive $\beta_T\gamma$ distribution from experiment, since the $2\sigma$ interval does not contain the actual decay length values, most of the time. A first way through which this situation could be improved would be by applying proper correction factors on the jets calibrated as a function of $p_T$, $\eta$ and $d_T$ of the jets as has been discussed in Sec.~\ref{sssec:reco_btg_jets}. An alternative idea could be to rely on timing information which will be available in the HL-LHC upgrade. The second option will be further discussed in Sec.~\ref{ssec:lifetime_time}.

\subsection{Summary of findings in the absence of timing information}
\label{sec:summarynotime}

Before moving on to discuss how the situation gets modified once timing information is taken into account, let us summarise our main findings so far. After some preliminary considerations related to existing limits on long-lived particles, experimental cuts, pile-up in the HL-LHC environment and some systematic uncertainties entering LLP-related measurements, we studied whether the LLP lifetime can be reconstructed at the HL-LHC in four different scenarios: LLPs decaying into diplaced leptons, into displaced leptons accompanied by missing energy, a charged LLP decaying into a lepton along with missing energy and a neutral LLP decaying into displaced jets. 

We considered two benchmark LLP masses, namely $100$ GeV and $1$ TeV and two different mean proper decay lengths of $10$ cm and $50$ cm for each case. For each of these four scenarios, we moreover assumed three different cross-section values for each process: $1$ fb, $0.1$ fb and $0.01$ fb. In all cases, we attempted to reconstruct the lifetime both assuming that the underlying model is known and assuming that it is unknown, but we have access to additional experimental information, in particular the LLP $\beta\gamma$ distribution.

We have found that in all four LLP decay scenarios, the lifetime can indeed be reconstructed if $\sim 3000$ events are available, with a precision ranging between $10$\% (in the case of a 1 TeV LLP decaying into displaced leptons with a mean proper decay length of 10 cm) up to a factor of a few (typically for lighter LLPs, longer lifetimes and decay modes involving missing energy). If the number of events is reduced by a factor 10, the situation becomes less clear, although in the most favourable of cases it is still possible to reconstuct the lifetime with a precision ranging in the $40$\% ballpark. We should, nonetheless, point out that some of the channels that we have considered, especially the ones involving MET, may allow for even larger cross-sections than the ones considered here. Perhaps even more interestingly we have, moreover, shown that prior knowledge of the underlying microscopic model is by no means necessary in order to access the LLP lifetime, at least in the most favourable scenarios. To the best of our knowledge, this has not been pointed out before in the literature.

\section{Including precise timing information}
\label{sec:timing}

The proposal for a MIP Timing Detector (MTD) \cite{MTD} for the phase II CMS upgrade, which will provide the timing information for MIPs (minimum ionizing particles), can provide additional information which is relevant for the lifetime reconstruction of LLPs. A few recent works have already studied aspects of the role that such a detector could play in LLP searches \cite{Liu:2018wte,Mason:2019okp}. It has also been shown in Refs.\ \cite{Meade:2010ji,Kang:2019ukr} that timing information can help in measuring the mass of the LLPs.

The proposed design for MTD indicates that it will be able to provide timing measurements for all charged particles having $p_T>0.7{\rm~GeV}$ in the barrel region ($|\eta|<1.5$) and $p>0.7{\rm~GeV}$ in the endcaps (up to $\eta=3$) with a time resolution of $30{\rm~ps}$. It will be placed at a radial distance of $1.161{\rm~m}$ from the beam axis \cite{MTD}: in the transition region from the tracker to the ECAL.

In what follows we will use the \texttt{ParticlePropagator} functionality of the \texttt{Delphes 3.4.1} \cite{deFavereau:2013fsa} package to get the timing of the charged particles coming from an LLP decay, restricting ourselves to the barrel region and applying a $30{\rm~ps}$ Gaussian smearing on the time obtained from \texttt{ParticlePropagator}. Our goal is not to repeat the analyses performed in the previous Sections. Instead, we will simply comment on the role that the MTD could play in the LLP mass reconstruction and in the estimation of its lifetime in the various scenarios we examined previously.

\subsection{Mass reconstruction}
\label{ssec:time_mass}

In Sec.\ \ref{ssec:lep_met}, we saw that when the LLP decay involves an invisible particle, the LLP mass can be reconstructed by employing the stransverse mass variable. In the case of a three-body decay of the LLP into two visible particles along with an invisible one, we can also get the mass difference of the LLP and the invisible particle from the edge of the invariant mass distribution of the two visible decay products. 

However, when the LLP decays into an invisible particle along with an \textit{on-shell} intermediate neutral particle (which, in turn, decays visibly), the invariant mass of the visible part of the decay will peak at the intermediate particle's mass and, therefore, we cannot rely on the mass edge to obtain the difference between the LLP and the invisible particle's mass.

In this Section, we discuss how the mass reconstruction could improve in these two cases once timing information from the MTD is taken into account.

\subsubsection{Two-body decay of the LLP involving an invisible particle}
\label{sssec:2body_mass}

Consider the following decay of an LLP $X$:
\begin{equation*}
X \rightarrow Z~Y \,,~Z \rightarrow l^+l^-
\end{equation*}
where $Y$ is an invisible particle. Here the dilepton invariant mass will peak at the SM $Z$ boson mass. However, using information from the MTD and the position of the SV, we can find out the boost of the LLP using the relation
\be
\frac{l_1}{\beta_X}+\frac{l_2}{\beta_{l}}=ct
\label{eq:timing}
\ee  
where $l_1$ is the distance travelled by the LLP from the PV to the SV where it decays and $l_2$ is the length traversed by the charged decay product (here, the lepton) from the SV to the point where it hits the MTD, $\beta_X$ and $\beta_l$ are the boosts of the LLP and the charged decay products respectively, $t$ is the time when the charged particle hits the MTD, and $c$ is the speed of light.

Once we know the LLP boost, we can boost back the leptons to the rest frame of the LLP. In this frame, the following relation holds 
\be 
m_X^2+m_{\rm vis}^2-m_{Y}^2 = 2 m_X E_{\rm vis}^{\rm rest}
\label{eq:rest_frame}
\ee
where $m_{\rm vis}$ and $E_{\rm vis}^{\rm rest}$ are the mass of the visible system (here, the dilepton invariant mass which peaks at $Z$ mass) and the total energy of the visible decay products in the LLP rest frame. If, now, we assume that the invisible particle is (quasi-)massless, then the mass of the LLP can be estimated using
\be 
m_X = E_{\rm vis}^{\rm rest} + p
\label{eq:2body}
\ee

\begin{figure}
    \centering
    \includegraphics[width=0.49\textwidth]{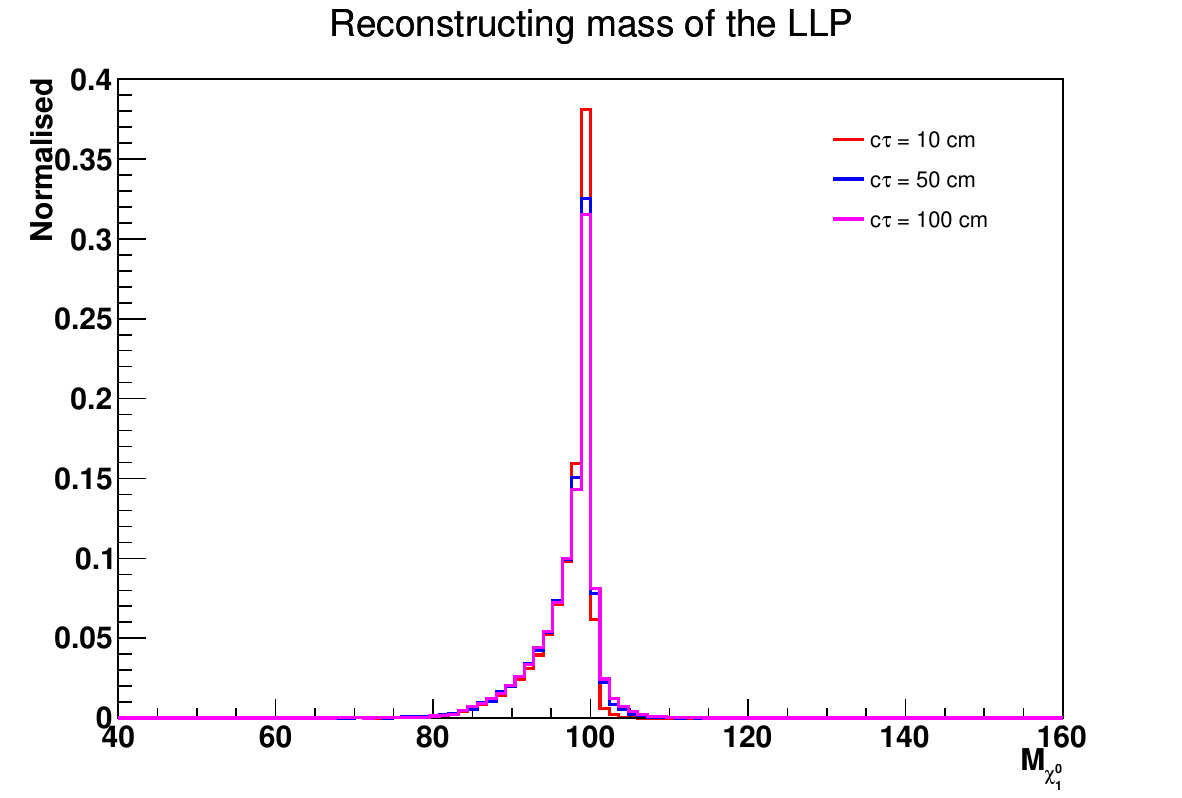}
    \includegraphics[width=0.49\textwidth]{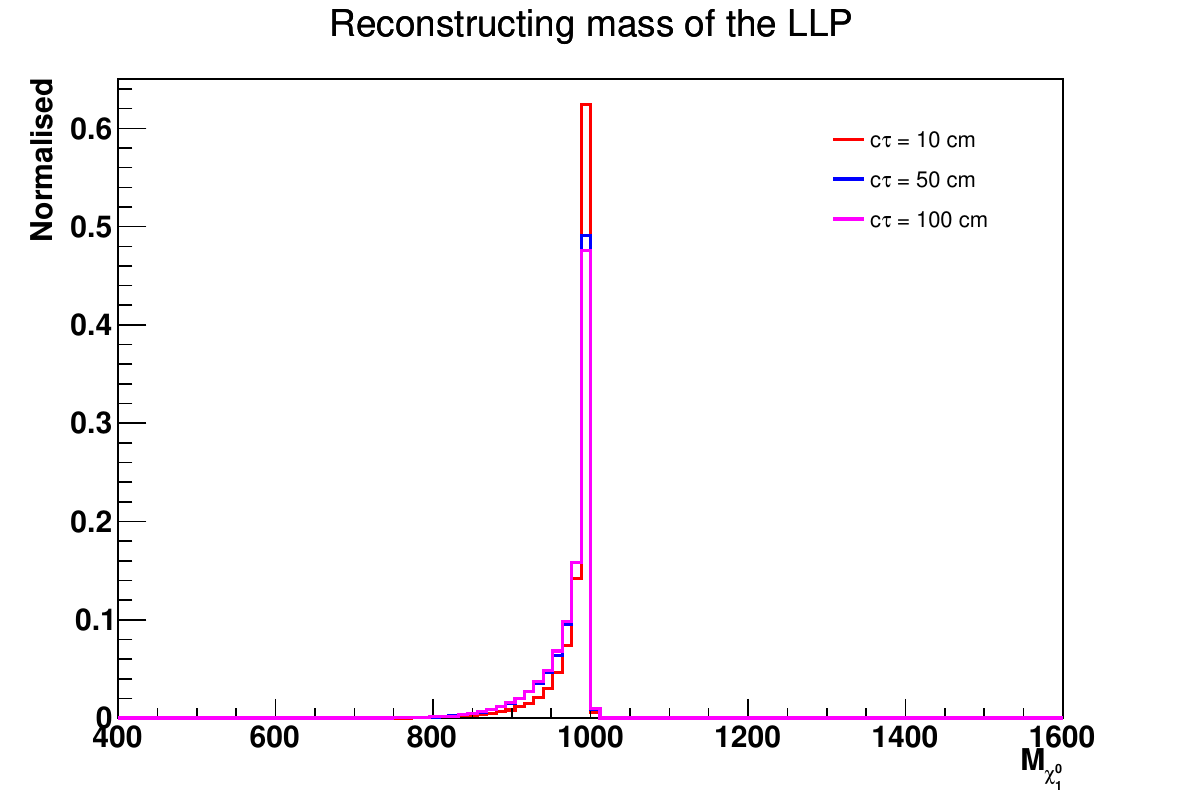}
    \caption{Reconstruction of the LLP mass for the decay of the LLP, neutralino, into $Z$ and gravitino and further decay of $Z$ into electrons for an LLP mass of $100{\rm~GeV}$  (left) and $1000{\rm~GeV}$ (right).}
    \label{fig:mchi10_timing}
\end{figure}

For concreteness, let us consider the decay of a neutralino (LLP) into a leptonically decaying SM $Z$ boson and a gravitino (invisible). We consider two different masses of the neutralino -- $100{\rm~GeV}$ and $1000{\rm~GeV}$. In Figure \ref{fig:mchi10_timing} we show the reconstruction of the LLP mass assuming the gravitino to be massless and using Eq.\ \eqref{eq:2body}.

The longer tail of the distribution towards lower masses is due to the mismatch between the azimuthal angle $\phi$ and pseudorapidity $\eta$ of the decay products measured at the collider and the actual $\eta-\phi$ value (which starts from the SV) of the displaced electrons. 

\begin{figure}
    \centering
    \includegraphics[width=0.49\textwidth]{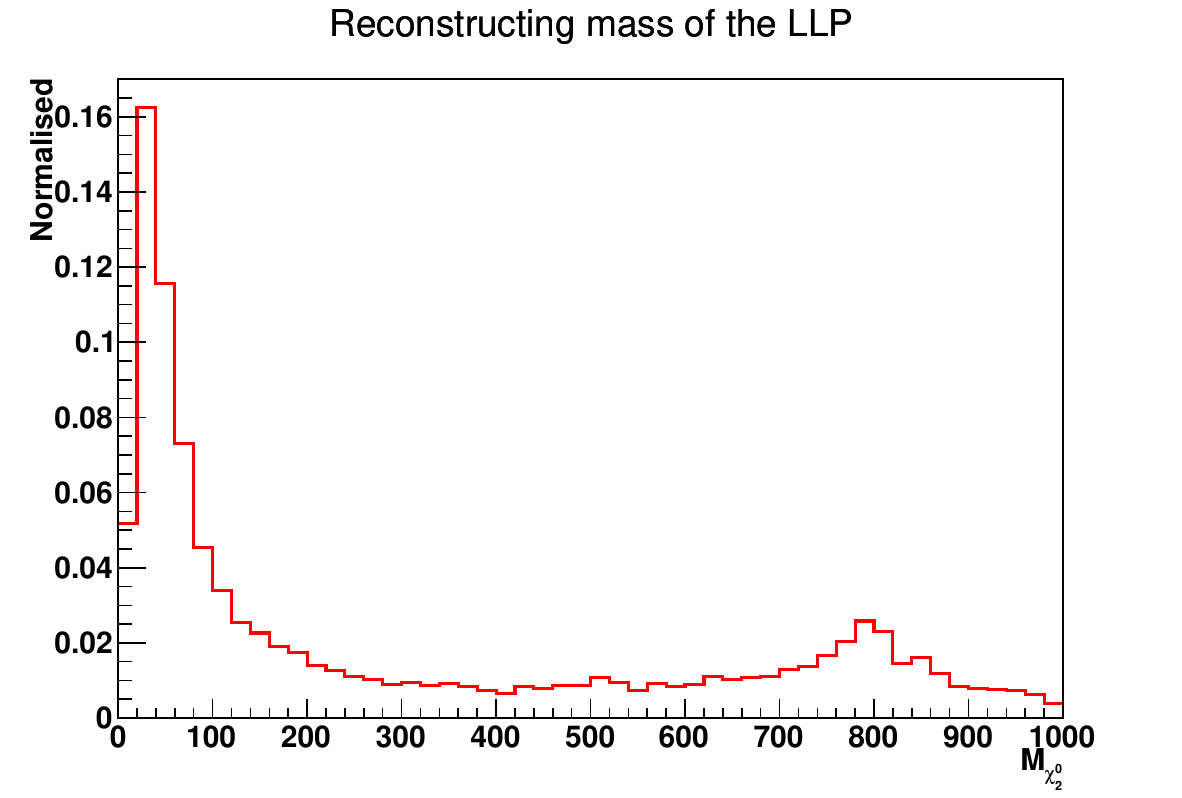}~
    \includegraphics[width=0.49\textwidth]{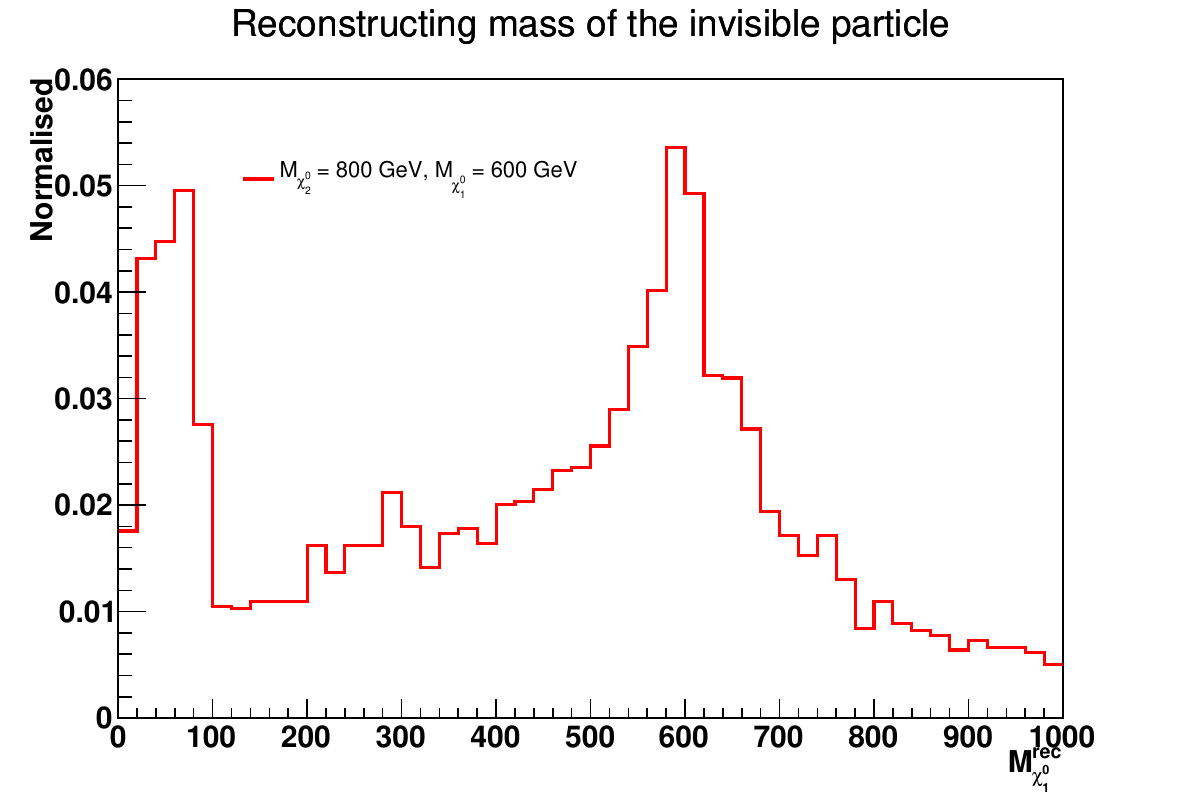}\\
    \includegraphics[width=0.49\textwidth]{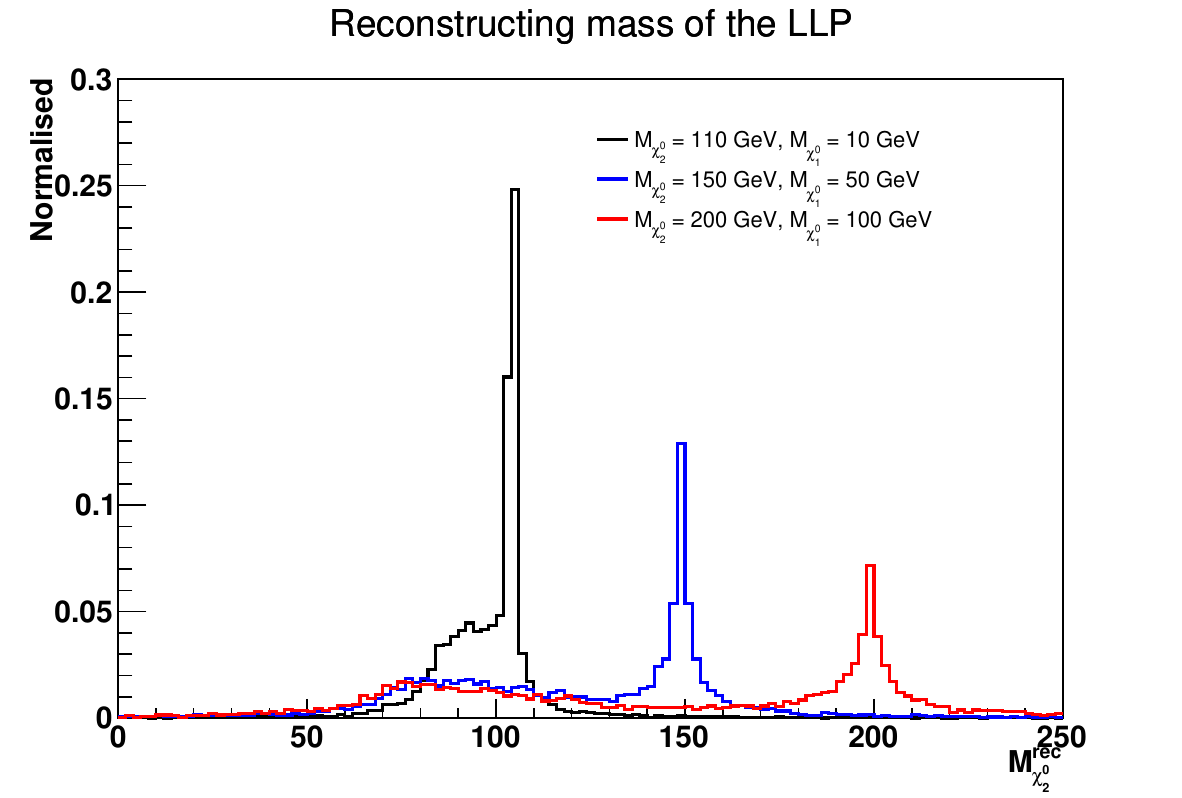}~
    \includegraphics[width=0.49\textwidth]{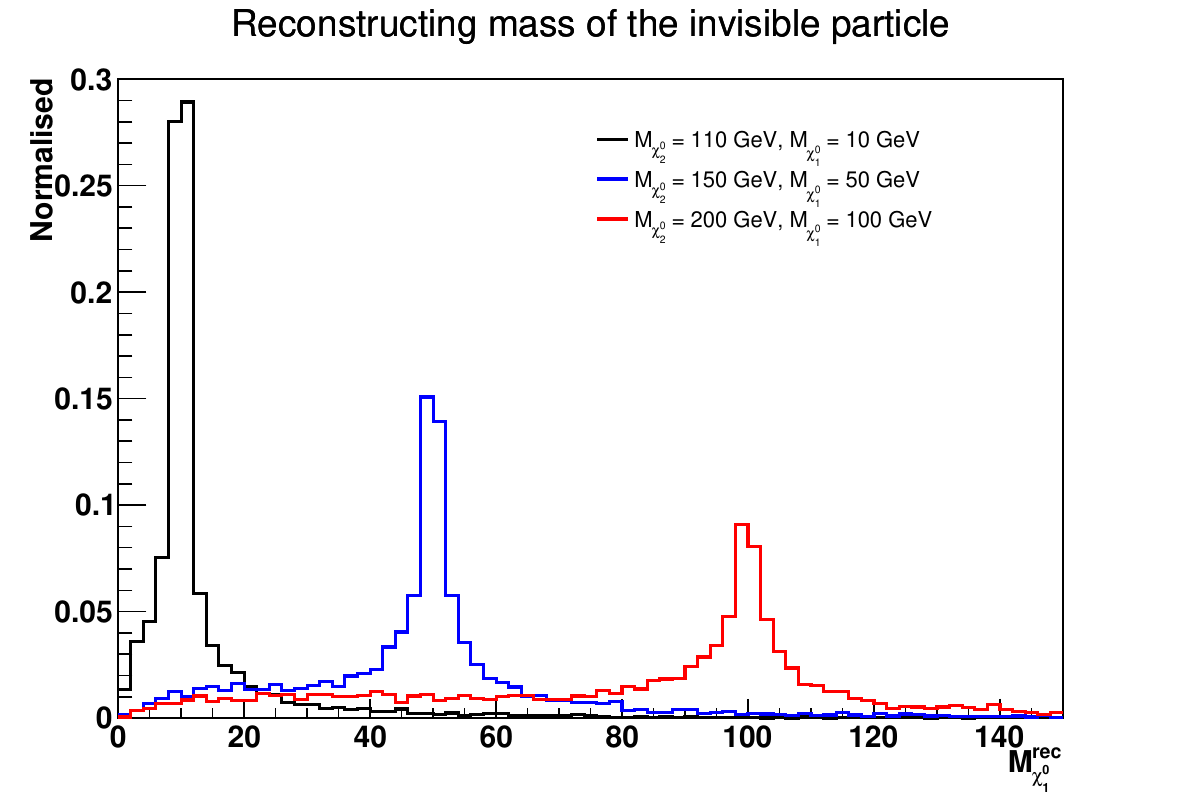}\\    
    \caption{{\it Top panel:} Reconstructing the LLP mass (left) and the invisible particle mass (right) by solving Eq.\ \eqref{eq:rest_frame} for every two observed events where the actual LLP and invisible particle masses are $800{\rm~GeV}$ and $600{\rm~GeV}$ respectively. {\it Bottom panel:} Reconstructing lower masses and the effect of FSR in these cases.}
    \label{fig:2body_mass}
\end{figure}

In order to circumvent the assumption that the invisible particle is massless, we can try to actually \textit{reconstruct} the two masses (LLP/invisible particle) by observing that Eq.\ \eqref{eq:rest_frame} holds on an event-by-event basis. Then, by solving it for different pairs of events, we can obtain an estimate of the two masses. We expect this method to work for cases where the intermediate particle has a non-vanishing decay width and, therefore, Eq.\ \eqref{eq:rest_frame} will involve slightly different values of $m_{\text{vis}}$ and $E^{\text{rest}}_{\text{vis}}$ for each event (otherwise we simply obtain the same equation each time). However, the addition of final state radiation and smearing will affect the solutions because these effects will bias the parameters of Eq.\ \eqref{eq:rest_frame} in different ways in each event. 

We demonstrate this method of mass reconstruction by simulating a LLP of mass 800 GeV decaying into two leptons and an invisible particle of mass 600 GeV.
In Fig.\ \ref{fig:2body_mass} we show the mass reconstruction of the LLP and the invisible particle by solving Eq.\ \eqref{eq:rest_frame} for every two observed events with final state radiation switched on and a smearing of $5\%$ on the transverse momenta of the leptons. We observe that the reconstruction is, indeed, affected but we deduce that the massless invisible particle assumption may be possible to drop. We do observe peaks around the actual LLP and invisible particle masses around 800 GeV and 600 GeV respectively. Note that the large number of solutions populating the low mass regions in both panels of Fig.\ \ref{fig:2body_mass} are a result of final state radiation which reduces the energy of the visible decay products. It is, then, interesting to examine whether in the case of lighter LLPs/invisible particles their mass peaks may overlap with the FSR-induced ones and, therefore, hamper their reconstruction. To this goal, we select three other benchmark points with lower LLP masses -- namely, 200 GeV, 150 GeV and 110 GeV, and in each case set the invisible particle mass to 100 GeV, 50 GeV and 10 GeV respectively. In the {\it bottom panel} of Fig.\ref{fig:2body_mass} we show the mass reconstruction for these three benchmarks. We find that even the lighter masses can be reconstructed reasonably well, and the FSR peaks do not affect the actual mass peaks.

\subsubsection{Three-body decay of the LLP involving invisible particle}
\label{sssec:3body_mass}

Let us now consider the three-body decay of an LLP $X$ as:
\begin{equation}
    X\rightarrow l^+~l^-~Y \,,
\end{equation}
where $Y$ is an invisible particle. In Sec.\ \ref{ssec:lep_met} we discussed one method through which both the mass of the LLP and that of the invisible particle can be reconstructed. By employing timing information, an alternative approach can be envisaged.
Namely, in this case, we have another equation (Eq.\ \eqref{eq:rest_frame}) in the rest frame of the LLP, which can be solved along with Eq.\ \eqref{eq:mass_edge} in order to obtain simple expressions for both masses as
\bea
    m_X ~=~ \frac{\Delta^2-M_{ll}^2}{2(\Delta-E_{\rm vis}^{\rm rest})} \,,\qquad\qquad
    m_Y ~=~ m_X - \Delta \,.
    \label{eq:3body}
\eea
As an example, let us consider the three-body decay of a neutral LLP into two muons along with an invisible particle. The mass of the LLP has been set to 200 GeV and the invisible particle's mass is set to 50 GeV. The decay length of the LLP in this case is set to 10 cm.

\begin{figure}
    \centering
    \includegraphics[width=0.68\textwidth]{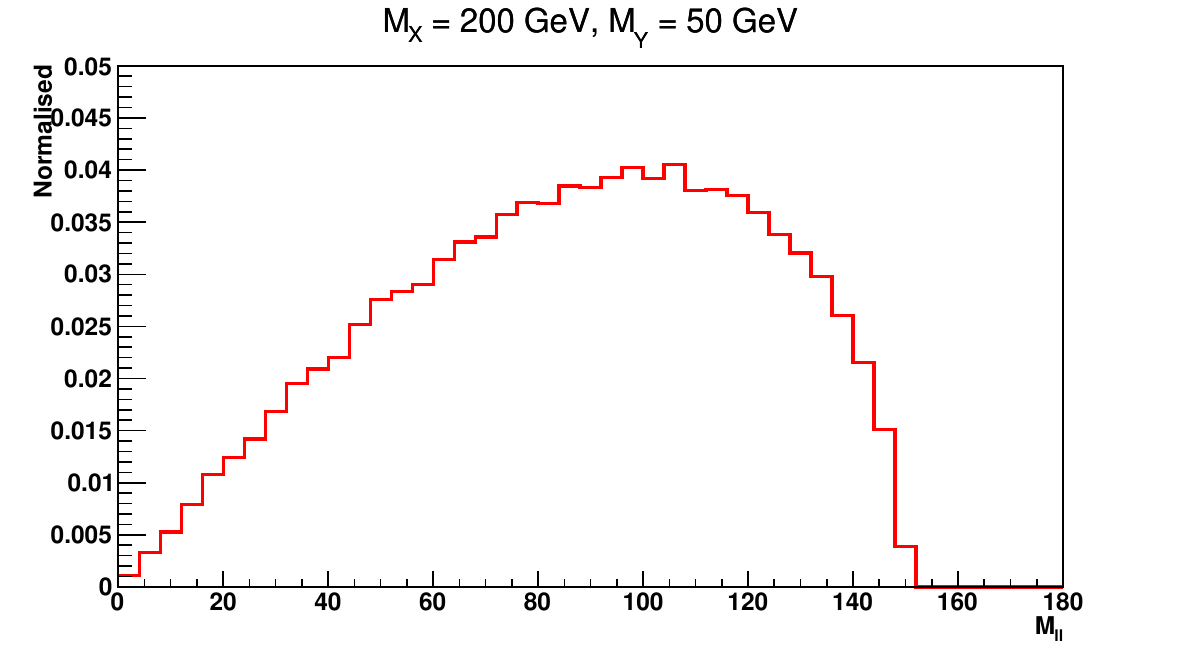}\\[3mm]
    \includegraphics[width=0.68\textwidth]{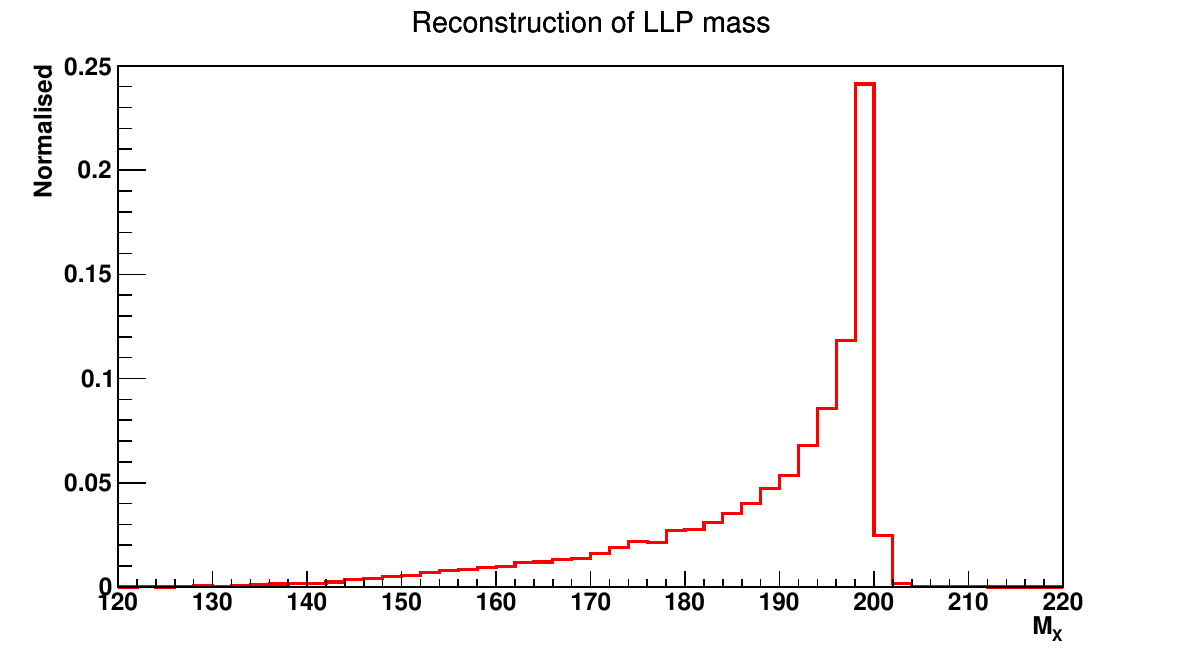}\\[3mm]
    \includegraphics[width=0.68\textwidth]{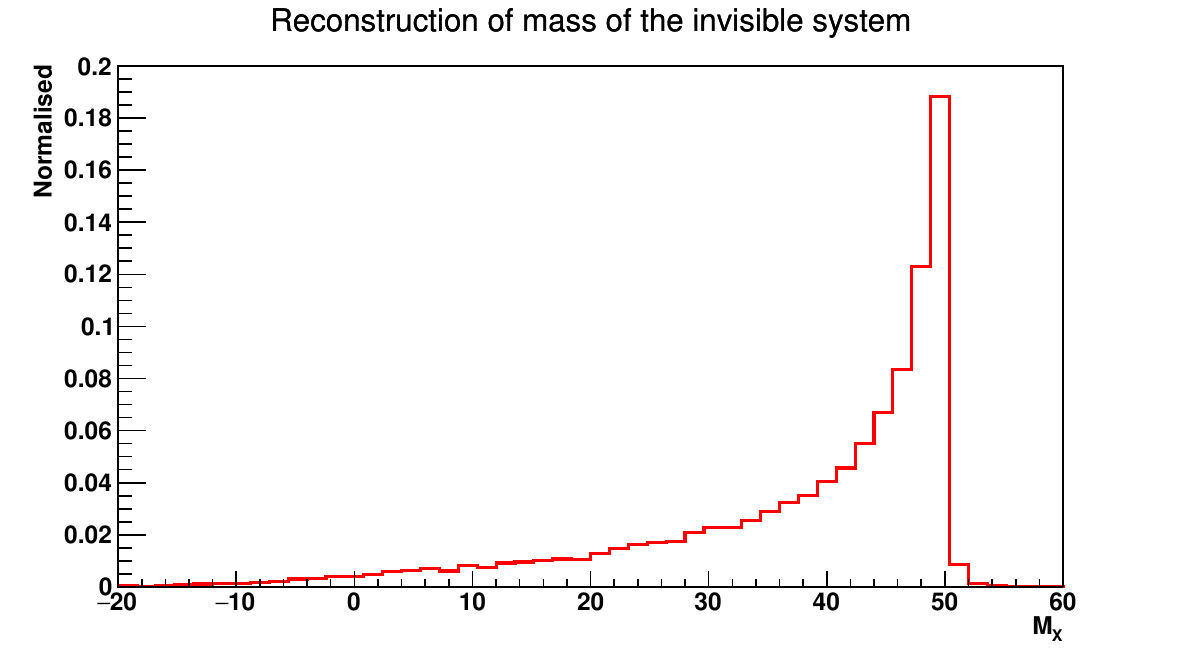}
    \caption{Estimation of the difference between LLP mass and the mass of the invisible system from dimuon invariant mass edge (top), reconstruction of the LLP mass (centre) and the invisible system's mass (bottom), respectively, using timing information.}
    \label{fig:3body_timing}
\end{figure}

In the top panel of Fig.\ \ref{fig:3body_timing}, we show the dimuon invariant mass distribution. The edge of the distribution corresponds to $\Delta$, which in this case is 150 GeV. Using this value of $\Delta$ and Eq.\ \eqref{eq:3body}, we can indeed calculate both $m_X$ and $m_Y$. In the lower panels of Fig.\ \ref{fig:3body_timing} we show the reconstructed mass of the LLP and that of the invisible system, respectively. Therefore, timing can help in mass reconstruction of LLPs as well as the invisible particle coming from the LLP decay, even if we do not use the missing transverse energy information
\footnote{If we have timing information available at ILC as well, given that its potential to reconstruct mother particle's masses to great precision as shown in~\cite{Li:2010mq} for processes like $e^+e^-\rightarrow\tilde{\chi}_1^+\tilde{\chi}_1^-\rightarrow\tilde{\chi}_1^0\tilde{\chi}_1^0W^+W^-$ and $e^+e^-\rightarrow\tilde{\chi}_2^0\tilde{\chi}_2^0\rightarrow\tilde{\chi}_1^0\tilde{\chi}_1^0ZZ$ for prompt decays ~\cite{Li:2010mq} can be extended to long-lived scenarios, we can then reconstruct the DM ($\tilde{\chi}_1^0$) mass at ILC even for the two body decays of the LLP using Eq. \eqref{eq:rest_frame}.}.

\subsection{Improving the lifetime estimation}
\label{ssec:lifetime_time}

The inclusion of timing information for charged particles enables us to calculate the boost of the LLP if the position of the secondary vertex is known, even when we cannot reconstruct all of its decay products. This is particularly useful in order to estimate the LLP lifetime when the final state particles' energy and momentum are significantly smeared, as in the case of displaced jets, or when there are invisible particles in the final state. In this Section we will compare the parton-level transverse boost factor ($\beta_T\gamma$) distributions with the ones obtained using MTD timing information in these two cases.

\subsubsection{LLPs decaying into jets}
\label{sssec:btg_jets}

The time taken by a jet to reach the MTD has no significance because it consists of multiple particles with varying momenta. Therefore, in order to compute the boost factor of the LLP, we will use the timing information of the fastest charged particle from the two jets associated with a SV and assume that this particle's $\beta=1$. This assumption introduces some error. The higher the $p_T$ of the particle, the lower will be the corresponding error. Note also that placing higher $p_T$ cuts makes the sample biased towards higher $\beta_T\gamma$ values.

\begin{figure}
    \centering
    \includegraphics[width=0.68\textwidth]{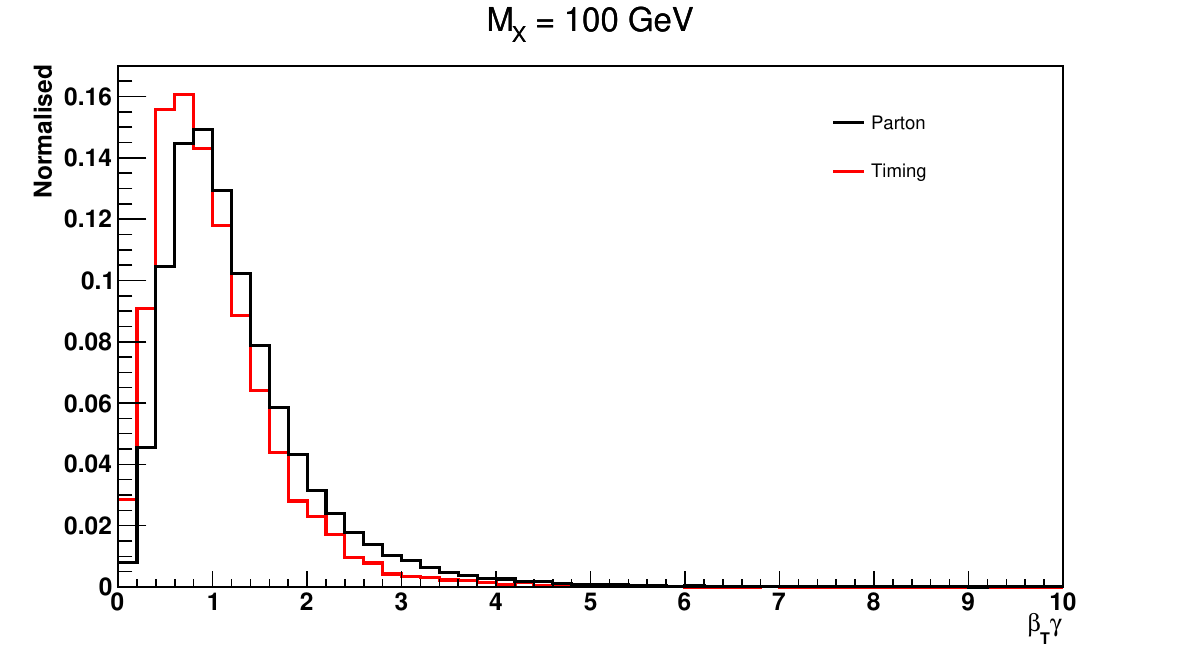}\\[3mm]
    \includegraphics[width=0.68\textwidth]{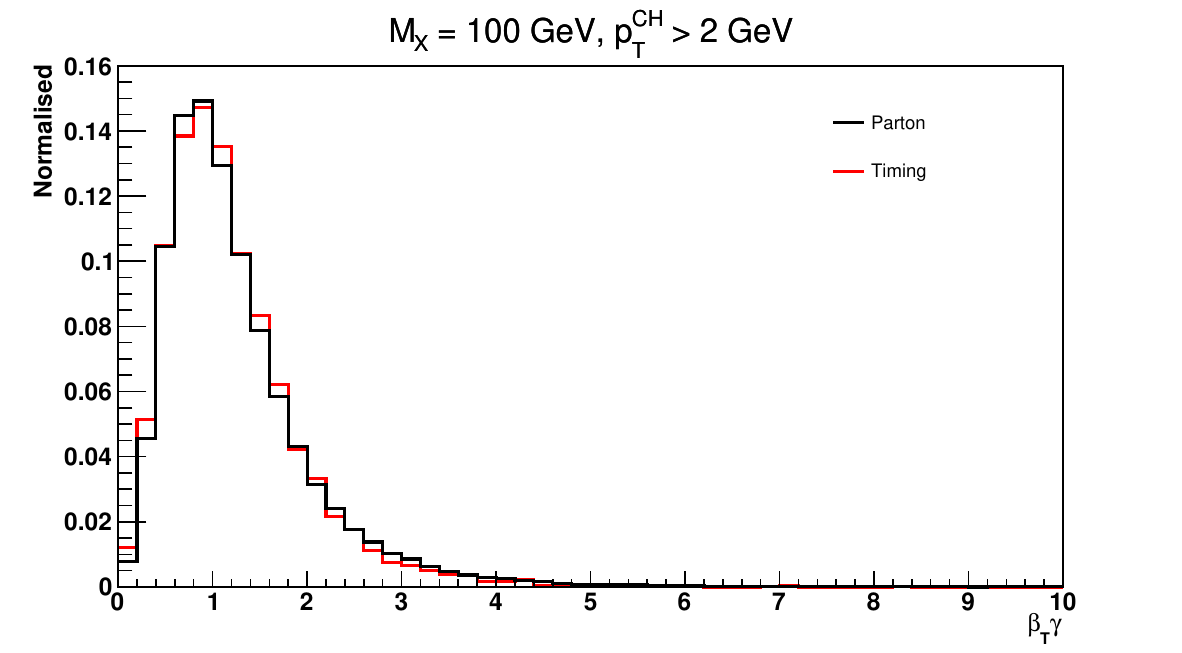}\\[3mm]
    \includegraphics[width=0.68\textwidth]{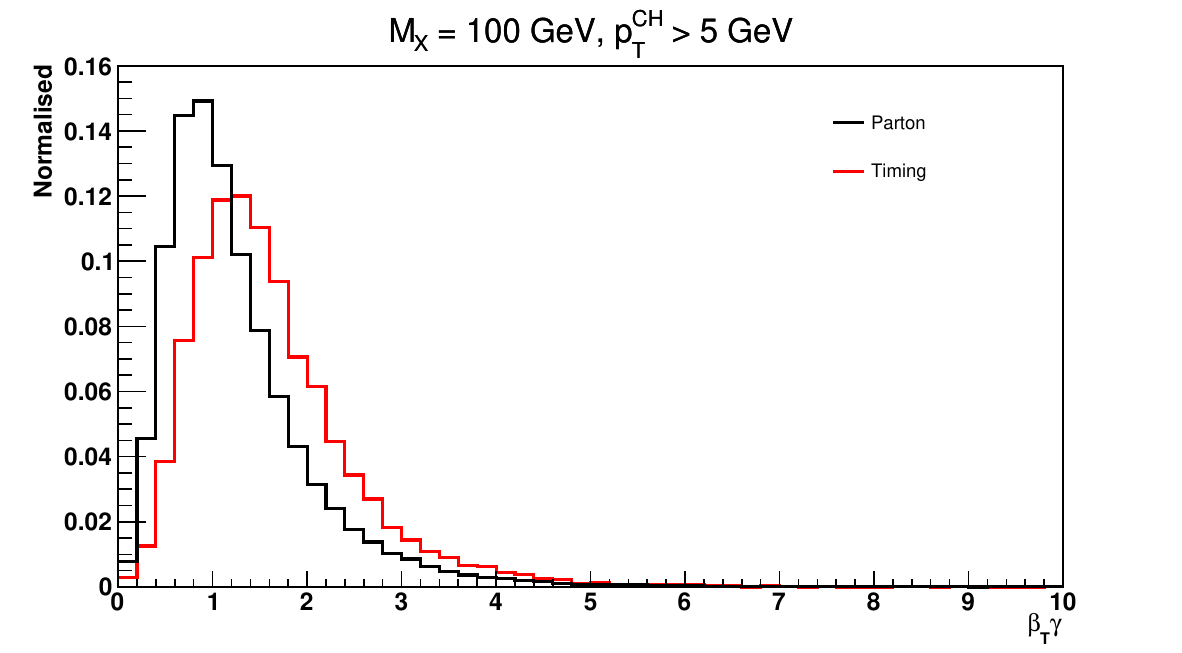}\\[3mm]
    \caption{$\beta_T\gamma$ distribution of an LLP of mass $100{\rm~GeV}$ at parton level and using timing information with no $p_T$ cut (top), $p_T>2{\rm~GeV}$ (centre), and $p_T>5{\rm~GeV}$ (bottom) for the fastest charged hadron coming from a SV.}
    \label{fig:btg_jets_wt}
\end{figure}

In Fig.\ \ref{fig:btg_jets_wt} we compare the boost factor distribution at the parton level with the one estimated using the timing of the fastest charged particle with no $p_T$ cut and lower $p_T$ cuts of $2{\rm~GeV}$ and $5{\rm~GeV}$ respectively. We find that for LLP of mass $100{\rm~GeV}$, if we place a cut of $2{\rm~GeV}$ on the fastest charged particle coming from the LLP decay, the assumption of this particle's $\beta=1$ works quite well, and we obtain a distribution from timing that is comparable to the parton-level distribution. For a LLP of mass $1000{\rm~GeV}$, higher $p_T$ cuts ($p_T>5{\rm~GeV}$) work better. 

Comparing Fig.\ \ref{fig:btg_jets_wt} (centre panel) with Fig.\ \ref{fig:jets_btg}, where the latter shows the comparison of the parton-level $\beta_T\gamma$ distribution $-$ without and with EC and that calculated from final state jet information, we find that timing improves the boost factor measurement. Hence, it will also improve the lifetime estimation.

\subsubsection{LLPs decays involving invisible particle}
\label{sssec:btg_invisible}

\begin{figure}
    \centering
    \includegraphics[width=0.68\textwidth]{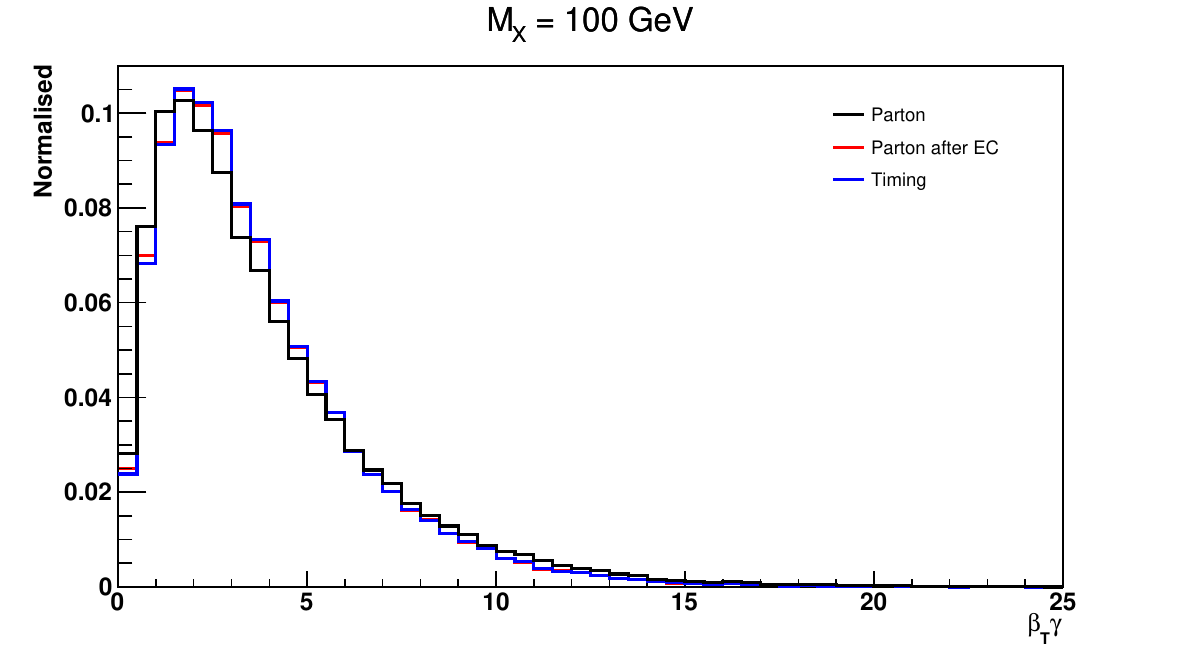} \\[3mm]
    \includegraphics[width=0.68\textwidth]{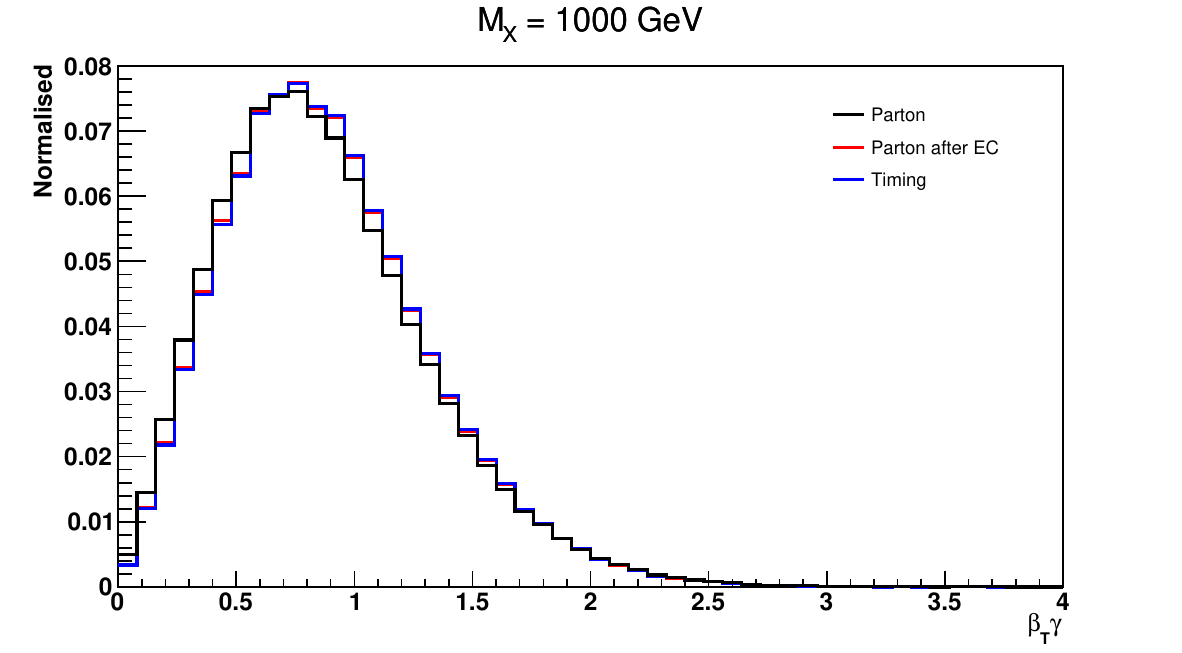}
    \caption{$\beta_T\gamma$ distribution for two-body decay of an LLP of mass $100{\rm~GeV}$ (top) and $1000{\rm~GeV}$ (bottom) at parton level without cuts, parton level with cuts and using timing information.}
    \label{fig:btg_timing_2}
\end{figure}

\begin{figure}
    \centering
    \includegraphics[width=0.68\textwidth]{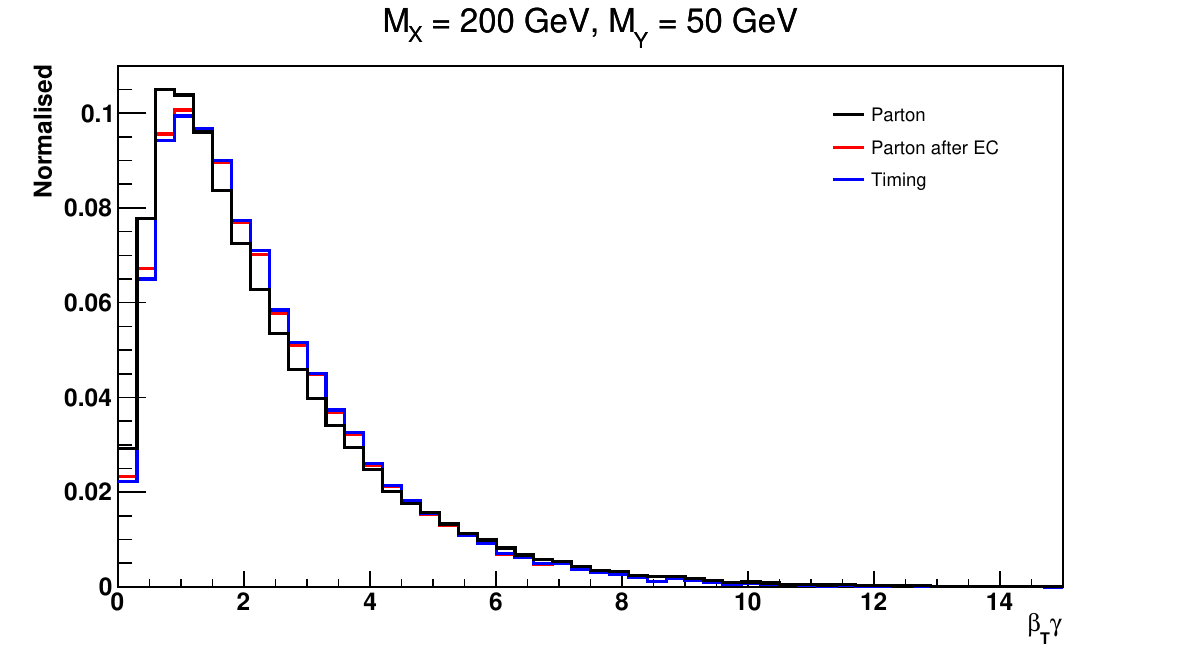}
    \caption{$\beta_T\gamma$ distribution for three-body decay of an LLP of mass $200{\rm~GeV}$ into two muons and an invisible particle of mass $50{\rm~GeV}$ at parton level without cuts, parton level with cuts and using timing information.}
    \label{fig:btg_timing_3}
\end{figure}

For LLPs decaying into leptons and missing particles, for both two-body and three-body decays, Figs.\ \ref{fig:btg_timing_2} and \ref{fig:btg_timing_3} show that the boost factor distribution obtained from timing of any one of the leptons coming from the LLP decay is comparable to the parton-level distribution. 

Therefore, timing information can help us to get the boost of the LLP if we know the SV and if any one of the decay products is a charged particle with $p_T>0.7{\rm~GeV}$ in the barrel and $p>0.7{\rm~GeV}$ in the endcaps, even when its decay products cannot be reconstructed entirely due to presence of invisible particles. We can then apply any one of the previous discussed methods in Sec.\ \ref{ssec:realistic} to estimate the lifetime of the LLP, especially the model-independent $\chi^2$ analysis which crucially needs the boost information of the LLP \footnote{Along the same lines, we can also expect to reconstruct the lifetime of LLPs decaying into a final state involving photons or even electrons or jets for which the track information (and thus the production location) is lost as in \cite{Sirunyan:2019gut}, by using the timing information of the ECAL. Although the time resolution of the ECAL is a few $100 {\rm~ps}$ \cite{CMS:2015sjc}, it can still be used to obtain a rough estimate of the lifetime.}.

\subsection{Challenges in use of timing for estimating boost of LLPs in high PU}
\label{ssec:time}

With the addition of 140 vertices per bunch crossing in the HL-LHC runs, identifying the LLP production vertex as the primary vertex is difficult, since the $\sum_{n_{trk}} p_T^2$ (or $\sum_{n_{trk}} p_T^2/n_{trk}$) calculated from the prompt tracks can be maximum for any of the PU vertices instead of the vertex at which the LLP is produced, as we have discussed in Sec.~\ref{ssec:PU}. This gives a wrong $l_1$ and $t$ in Eq.~\eqref{eq:timing} if the identified PV does not match with the LLP production vertex, since these are measured with respect to the PV. 
The LLP vertex could, in principle, be identified if all the decay products of the LLPs are properly reconstructed, however, then we need not use the timing information for finding out the boost of the LLP.

\begin{figure}
\centering
\includegraphics[width=0.75\textwidth]{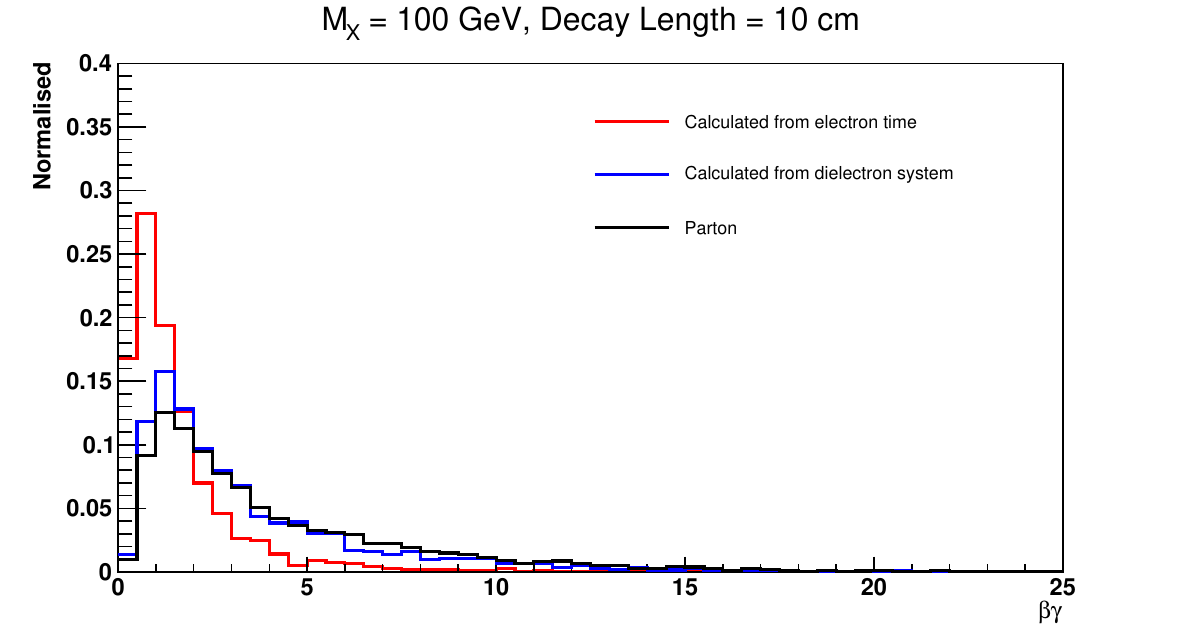}
\caption{Distribution of the boost factor ($\beta\gamma$) of the LLP at the parton level, and those calculated from the dielectron system and the electron timing at the MTD.}
\label{fig:timing_PU}
\end{figure}

For illustration, we use the displaced electrons signature, since in that scenario we can also calculate the $\beta\gamma$ from the dielectron system apart from the electron timing, which is useful for comparison.
Fig. \ref{fig:timing_PU}, we can see the difference in the $\beta\gamma$ distributions if they are calculated from timing, to the one which is obtained from the dielectron system, both compared to the parton level distribution. We find that the distribution calculated from electron time is shifted to lower values, and therefore, will provide a higher value of $c\tau$ for an observed $d$ distribution.

However, if the event has a high energy ISR jet, that can ensure that the LLP production vertex gets identified as the PV, and we can use eq. \ref{eq:timing} as discussed in the previous Sections. Also, having a hard ISR jet does not affect the model-dependent as well as model-independent methods for reconstructing the lifetime. The same discussion follows for reconstructing the boost of the LLP in the displaced jets scenario, where the time of the fastest track from the jet was used.

\section{Conclusions and outlook}
\label{sec:concl}

In this work we studied the capacity of the High-Luminosity LHC to reconstruct some key properties of long-lived particles, most notably their lifetime, in an optimistic scenario when such particles are observed. We examined a variety of different signatures, namely decays of neutral LLPs into pairs of displaced leptons, displaced jets or displaced leptons accompanied by missing transverse energy as well as the decay of an electrically charged LLP into missing energy along with a charged SM lepton. Perhaps unsurprisingly, in all cases we found that it is, indeed, possible to reconstruct the LLP lifetime if we assume that the underlying model is known. Going a step further, however, we showed that -- at least within the limitations of our study -- in most cases it is also possible to estimate the LLP lifetime in a model-independent manner, provided the LLP $\beta\gamma$ distribution can be experimentally accessed. We moreover commented upon how upgrades of the LHC detectors, and in particular the improvement of timing measurements, can be used in order to facilitate the determination of different quantities entering LLP-related measurements.

The present work is a theorist’s analysis of a topic which depends heavily on experimental information. There are numerous ways through which our study could be improved and/or extended. For instance, other than examining different final states that we have not considered here, it is clear that our treatment of experimental limitations related, \textit{e.g.}, to the determination of the secondary vertex location or, perhaps most crucially, to the measurement of the LLP $\beta\gamma$ distribution, can be improved. In this respect, we deem our results to be on the optimistic side. On the other side of the spectrum, nevertheless, in our analysis we mostly focused on information that can be inferred from the tracker systems of the LHC detectors. However, \textit{e.g.} in the case of kinked track signatures, calorimetric information can be used in the framework of an offline analysis in order to identify decay events in which the SM lepton track is too short to be reconstructed, but its presence can be inferred by energy depositions in the ECAL. Our hope is that this preliminary study will trigger more detailed analyses on the topic.

\acknowledgments
We would like to thank the organizers of the 2017 ``Les Houches -- Physics at TeV colliders'' workshop where this work was initiated. We thank Daniele Barducci for collaboration during the earlier stages of this work. We thank Swagata Mukherjee and Shilpi Jain for helpful discussions. The work of B.H.\ and S.B.\ are partially supported by {\it Investissements d'avenir}, Labex ENIGMASS, contrat ANR-11-LABX-0012. S.B.\ is also supported by a Durham Junior Research Fellowship COFUNDed by Durham University and the European Union. The works of S.B.\ and B.B.\ are partially supported by the CNRS LIA-THEP (Theoretical High Energy Physics) and the INFRE-HEPNET (IndoFrench Network on High Energy Physics) of CEFIPRA/IFCPAR (Indo-French Centre for the Promotion of Advanced Research). The work of S.B.\ is also partially supported by a Durham Junior Research Fellowship COFUNDed between Durham University and the European Union under grant agreement number 609412. The work of B.B.\ is also supported by the Department of Science and Technology, Government of India, under the Grant Agreement number IFA13-PH-75 (INSPIRE Faculty Award). A.G.\ would like to thank the Laboratoire de Physique Th\'eorique et Hautes \'Energies, where part of this work was performed, for warm hospitality and Dimitris Varouchas for illuminating discussions. The work of D.S.\ is supported by the National Science Foundation under Grant No.\ Grant No. PHY-1915147. R.S.\ would like to thank Rahool Kumar Barman and Amit Adhikary for useful discussions.



\providecommand{\href}[2]{#2}\begingroup\raggedright\endgroup

\end{document}